\documentclass[a4paper,11pt]{article}
\pdfoutput=1 

\usepackage{jheppub} 

\usepackage[T1]{fontenc} 
\usepackage{slashed}
\usepackage{tikz}
\usetikzlibrary{shapes,arrows,positioning,automata,backgrounds,calc,er,patterns}
\usepackage{tikz-feynman}
\tikzfeynmanset{compat=1.1.0}
\usepackage{xcolor}
\usepackage{makecell}
\usepackage{rotating}
\usepackage{pdflscape}
\usepackage{float}
\usepackage{multicol}
\usepackage{xspace}
\usepackage{booktabs}
\usepackage{xspace}
\usepackage[shortlabels]{enumitem}

\usepackage{tikz}
\usetikzlibrary{shapes.geometric, arrows, positioning}
\tikzstyle{fullspace} = [rectangle, minimum width=4cm, minimum height=1.5cm, text centered, text width=3cm, draw=black, fill=white]
\tikzstyle{subspace} = [rectangle, rounded corners, minimum width=4cm, minimum height=1.5cm,text centered, text width=3cm, draw=black, fill=white]
\tikzstyle{arrow} = [thick,->,>=stealth]
\tikzstyle{circlebox} = [circle,text centered, minimum width = 1.5 cm, minimum height = 1.5 cm, draw=black,fill=white]
\tikzstyle{recbox} = [rectangle,text centered, minimum width = 1.5 cm, minimum height = 1.5 cm, draw=black,fill=white]

\tikzstyle{startstop} = [rectangle, rounded corners, 
minimum width=4cm, 
minimum height=1cm,
text centered, 
draw=black, 
fill=red!30]

\tikzstyle{io} = [trapezium, 
trapezium stretches=true, 
trapezium left angle=70, 
trapezium right angle=110, 
minimum width=1.5cm, 
minimum height=1cm, text centered, 
draw=black, fill=blue!30]

\tikzstyle{process} = [rectangle, 
minimum width=1cm, 
minimum height=1cm, 
text centered, 
draw=black, 
fill=orange!30]

\tikzstyle{decision} = [diamond, 
minimum width=3cm, 
minimum height=1cm, 
text centered, 
draw=black, 
fill=green!30]

\newcommand{\Tr}[4]{\mathrm{Tr}(\slashed{#1}\slashed{#2} \slashed{#3} \slashed{#4})}
\newcommand{\ome}{(1-\epsilon)}
\newcommand{\e}{\epsilon}

\newcommand{\modsq}[1]{\ensuremath{\left\vert #1 \right\vert^2}}
\newcommand{\calM}{\ensuremath{\mathcal{M}}}
\newcommand{\order}[1]{\ensuremath{\mathcal{O}}\left( #1 \right)\xspace}

\newcommand{\fa}{a}
\newcommand{\fb}{b}
\newcommand{\fc}{\tilde{b}}
\newcommand{\fd}{c}
\newcommand{\fe}{\tilde{c}}
\newcommand{\ff}{d}
\newcommand{\fg}{\tilde{d}}

\newcommand{\NthreeLO}{N$^3$LO\xspace}

\newcommand{\Pqg}{P_{qg}}

\newcommand{\Pgg}{P_{gg}}
\newcommand{\Pqq}{P_{q\bar{q}}}

\newcommand{\PggS}{P_{gg}^\text{sub}}

\newcommand\nameggg{ggg \to g}

\newcommand\nameqgg{qgg \to q}
\newcommand\nameqpp{q\gamma\gamma \to q}
\newcommand\namegqbq{g\bar q q\to g}
\newcommand\namepqbq{q g \bar q \to \gamma}
\newcommand\nameqQQ{q\bar Q Q\to q}
\newcommand\nameqqq{q \bar q q\to q}

\renewcommand\nameggg{ggg}

\renewcommand\nameqgg{qgg}
\renewcommand\nameqpp{q\gamma\gamma}
\renewcommand\namegqbq{g\bar q q}
\renewcommand\namepqbq{q g \bar q }
\renewcommand\nameqQQ{q\bar Q Q}
\renewcommand\nameqqq{q \bar q q}

\newcommand{\Pggg}{P_{\nameggg}}

\newcommand{\Rgggsub}{R_{g(gg)}}
\newcommand{\Pqgg}{P_{\nameqgg}}
\newcommand{\Rqgg}{R_{\nameqgg}}
\newcommand{\Pqpp}{P_{\nameqpp}}
\newcommand{\Rqpp}{R_{\nameqpp}}
\newcommand{\Pgqbq}{P_{\namegqbq}}
\newcommand{\Rgqbq}{R_{\namegqbq}}
\newcommand{\Ppqbq}{P_{\namepqbq}}
\newcommand{\Rpqbq}{R_{\namepqbq}}
\newcommand{\PqQQ}{P_{\nameqQQ}}
\newcommand{\RqQQ}{R_{\nameqQQ}}
\newcommand{\Pqqq}{P_{\nameqqq}}
\newcommand{\Rqqq}{R_{\nameqqq}}

\newcommand{\softorder}{}
\newcommand{\Sgg}{S_{gg}^{\softorder}}
\newcommand{\Spp}{S_{\gamma\gamma}^{\softorder}}
\newcommand{\Sqq}{S_{q\qb}^{\softorder}}
\newcommand{\Sg}{S_{g}^{\softorder}}
\newcommand{\Sp}{S_{\gamma}^{\softorder}}
\newcommand{\Sq}{S_{q}^{\softorder}}
\newcommand{\Sqb}{S_{\bar{q}}^{\softorder}}
\newcommand{\Sb}{S_{b}^{\softorder}}

\newcommand{\PDSup}{\mathrm{\mathbf{DS}}^\uparrow\xspace}
\newcommand{\PTCup}{\mathrm{\mathbf{TC}}^\uparrow\xspace}

\newcommand{\PDCup}{\mathrm{\mathbf{DC}}^\uparrow\xspace}
\newcommand{\PSup}{\mathrm{\mathbf{S}}^\uparrow\xspace}
\newcommand{\PCup}{\mathrm{\mathbf{C}}^\uparrow\xspace}
\newcommand{\PDSdown}{\mathrm{\mathbf{DS}}^\downarrow\xspace}
\newcommand{\PTCdown}{\mathrm{\mathbf{TC}}^\downarrow\xspace}
\newcommand{\PSCdown}{\mathrm{\mathbf{SC}}^\downarrow\xspace}
\newcommand{\PDCdown}{\mathrm{\mathbf{DC}}^\downarrow\xspace}
\newcommand{\PSdown}{\mathrm{\mathbf{S}}^\downarrow\xspace}
\newcommand{\PCdown}{\mathrm{\mathbf{C}}^\downarrow\xspace}

\newcommand{\PPup}{\mathrm{\mathbf{P}}^\uparrow\xspace}
\newcommand{\PPdown}{\mathrm{\mathbf{P}}^\downarrow\xspace}

\newcommand{\X}{X_4^0}
\newcommand{\Xt}{\widetilde{X}_4^0}
\newcommand{\A}{A_4^0}
\newcommand{\B}{B_4^0}
\newcommand{\C}{C_4^0}
\newcommand{\D}{D_4^0}
\newcommand{\F}{F_4^0}
\newcommand{\E}{E_4^0}
\newcommand{\Ea}{E_4^{0}}
\newcommand{\Eb}{\overline{E}_4^{0}}
\newcommand{\G}{G_4^0}
\newcommand{\Ga}{G_4^{0}}
\newcommand{\Gb}{\overline{G}_4^{0}}
\renewcommand{\H}{H_4^0}

\newcommand{\At}{\widetilde{A}_4^0}
\newcommand{\Dt}{\widetilde{D}_4^0}
\newcommand{\Ft}{\widetilde{F}_4^0}
\newcommand{\Et}{\widetilde{E}_4^0}
\newcommand{\Gt}{\widetilde{G}_4^0}

\newcommand{\oldant}{\, \mathrm{OLD}}
\newcommand{\Xold}[1]{#1_3^{0,\oldant}}

\newcommand{\Aold}{A_4^{0,\oldant}}
\newcommand{\Atold}{\widetilde{A}_4^{0,\oldant}}
\newcommand{\Bold}{B_4^{0,\oldant}}
\newcommand{\Cold}{C_4^{0,\oldant}}
\newcommand{\Dold}{D_4^{0,\oldant}}
\newcommand{\Fold}{F_4^{0,\oldant}}
\newcommand{\Eold}{E_4^{0,\oldant}}
\newcommand{\Etold}{\widetilde{E}_4^{0,\oldant}}
\newcommand{\Gold}{G_4^{0,\oldant}}
\newcommand{\Gtold}{\widetilde{G}_4^{0,\oldant}}
\newcommand{\Hold}{H_4^{0,\oldant}}

\newcommand{\calX}{{\cal X}_4^0}

\newcommand{\calA}{{\cal A}_4^0}
\newcommand{\calAold}{{\cal A}_4^{0,\oldant}}
\newcommand{\calB}{{\cal B}_4^0}

\newcommand{\calC}{{\cal C}_4^0}
\newcommand{\calCold}{{\cal C}_4^{0,\oldant}}
\newcommand{\calD}{{\cal D}_4^0}
\newcommand{\calDold}{{\cal D}_4^{0,\oldant}}
\newcommand{\calF}{{\cal F}_4^0}
\newcommand{\calFold}{{\cal F}_4^{0,\oldant}}
\newcommand{\calEold}{{\cal E}_4^{0,\oldant}}
\newcommand{\calEa}{{\cal E}_4^{0}}
\newcommand{\calEb}{\overline{{\cal E}}_4^{0}}

\newcommand{\calGold}{{\cal G}_4^{0,\oldant}}
\newcommand{\calGa}{{\cal G}_4^{0}}
\newcommand{\calGb}{\overline{{\cal G}}_4^{0}}
\newcommand{\calH}{{\cal H}_4^0}
\newcommand{\calHold}{{\cal H}_4^{0,\oldant}}
\newcommand{\calAt}{\widetilde{{\cal A}}_4^0}
\newcommand{\calDt}{\widetilde{{\cal D}}_4^0}
\newcommand{\calFt}{\widetilde{{\cal F}}_4^0}
\newcommand{\calEt}{\widetilde{{\cal E}}_4^0}
\newcommand{\calGt}{\widetilde{{\cal G}}_4^0}
\newcommand{\calAtold}{\widetilde{{\cal A}}_4^{0,\oldant}}

\newcommand{\calEtold}{\widetilde{{\cal E}}_4^{0,\oldant}}
\newcommand{\calGtold}{\widetilde{{\cal G}}_4^{0,\oldant}}

\newcommand{\Ssoft}{\mathrm{Ssoft}}
\newcommand{\Scol}{\mathrm{Scol}}
\newcommand{\Dsoft}{\mathrm{Dsoft}}
\newcommand{\Tcol}{\mathrm{Tcol}}
\newcommand{\Dcol}{\mathrm{Dcol}}

\newcommand{\calSsoft}{\mathcal{S}\kern-0.05em{s\kern-0.05em o\kern-0.05em f\kern-0.05em t}}
\newcommand{\calScol}{\mathcal{S}\kern-0.05em{c\kern-0.05em o\kern-0.05em l}}
\newcommand{\calDsoft}{\mathcal{D}\kern-0.05em{s\kern-0.05em o\kern-0.05em f\kern-0.05em t}}
\newcommand{\calTcol}{\mathcal{T}\kern-0.05em{c\kern-0.05em o\kern-0.05em l}}
\newcommand{\calDcol}{\mathcal{D}\kern-0.05em{c\kern-0.05em o\kern-0.05em l}}

\newcommand{\qb}{\bar{q}}
\newcommand{\Qb}{\bar{Q}}

\newcommand{\omxi}{(1-x_i)}
\newcommand{\omxj}{(1-x_j)}
\newcommand{\omxk}{(1-x_k)}

\newcommand{\omyk}{(1-y_k)}
\newcommand{\sij}{s_{ij}}
\newcommand{\sjk}{s_{jk}}
\newcommand{\sik}{s_{ik}}
\newcommand{\sijk}{s_{ijk}}
\renewcommand{\xi}{x_{i}}
\newcommand{\xj}{x_{j}}
\newcommand{\xk}{x_{k}}

\newcommand{\yk}{y_{k}}

\title{\boldmath 
A general algorithm to build real-radiation antenna functions for higher-order calculations
}

\author[1]{Oscar Braun-White,}
\author[1]{Nigel Glover,}
\author[2]{Christian T Preuss}

\affiliation[1]{Institute for Particle Physics Phenomenology,\\Department of Physics, \\Durham University, Durham, DH1 3LE, UK}
\affiliation[2]{Institute for Theoretical Physics, ETH, CH-8093 Zürich, Switzerland}

\emailAdd{oscar.r.braun-white@durham.ac.uk}
\emailAdd{e.w.n.glover@durham.ac.uk}
\emailAdd{cpreuss@phys.ethz.ch}

\preprint{IPPP/23/7}

\abstract{The antenna subtraction method has been successfully applied to a wide range of processes relevant for the Large Hadron Collider at next-to-next-to-leading order in $\alpha_s$ (NNLO). We propose an algorithm for building antenna functions for any number of real emissions from an identified pair of hard radiator partons directly from a specified list of unresolved limits.  We use the algorithm to explicitly build all single- and double-real QCD antenna functions and compare them to the previous antenna functions, which were extracted from matrix elements. The improved antenna functions should be more easily applicable to NNLO subtraction terms. Finally, we match the integration of the antenna functions over the final-final unresolved phase space to the previous incarnation, serving as an independent check on our results.
}

\begin{document}
\maketitle
\flushbottom

\section{Introduction}

As experimental precision increases, phenomenological tools need to be refined to improve bounds on fundamental parameters and to detect potential deviations from theory. For many observables, percent-level accuracy cannot be reached without improvements in fixed-order calculations, parton distribution functions, parton showers and modelling of non-perturbative effects. Improvements are being made in all these areas.  
In the field of perturbative QCD, this typically requires fixed-order calculations to at least next-to-next-to-leading order (NNLO) in the strong-coupling expansion. 
Such higher-order calculations require particular attention due to the intricate interplay between real and virtual corrections across different multiplicity phase spaces~\cite{Kinoshita,LeeNauenberg}. Implicit infrared divergences are present because real emissions are unresolved (soft or collinear). These can only cancel against explicit poles in the dimensional regulator $\e$ (where $d=4-2\e$), that come from virtual graphs, after integration over the relevant unresolved phase space.
Subtraction schemes are currently considered the most elegant solution to overcome these subtleties.

At NLO, fully-differential calculations have been automated thanks to two general subtraction schemes known as Catani-Seymour dipole subtraction \cite{Catani:1996vz} and FKS subtraction \cite{Frixione:1995ms}.
The generality of these subtraction schemes has facilitated two fully-differential NLO matching schemes, known as MC@NLO \cite{Frixione:2002ik} and POWHEG \cite{Nason:2004rx,Frixione:2007vw} which systematically combine NLO fixed-order calculations with all-order parton-shower resummation.
These form the backbone of state-of-the-art multi-purpose event generators \cite{powheg:2010xd,Alwall:2014hca,Bellm:2019zci,Sherpa:2019gpd,Bierlich:2022pfr}, see Ref.~\cite{snowmass:2022qmc} for a recent summary.
The situation is inherently different for NNLO calculations, where subtraction and slicing schemes are much less automated, although a variety of methods have been developed \cite{Gehrmann-DeRidder:2005btv,Catani:2007vq,Czakon:2010td,Boughezal:2011jf,Gaunt:2015pea,Cacciari:2015jma,DelDuca:2016ily,Caola:2017dug,Magnea:2018hab,Herzog:2018ily}, see Ref.~\cite{TorresBobadilla:2020ekr} for a review.
These methods are generally implemented differently for each process due to the larger complexity at NNLO and do not scale straightforwardly to higher multiplicities.
Similarly, fully-differential NNLO matching techniques are so far still in their infancy, as they require higher-order corrections to be exponentiated in the shower algorithm \cite{Campbell:2021svd}. Progress on including such corrections in parton showers has been reported in \cite{Jadach:2011kc,Jadach:2013dfd,Hartgring:2013jma,Li:2016yez,Hoche:2017hno,Hoche:2017iem,Dulat:2018vuy,Gellersen:2021eci,Loschner:2021keu}.
At \NthreeLO, inclusive~\cite{Anastasiou:2015vya,Anastasiou:2016cez,Mistlberger:2018etf,Dreyer:2016oyx,Duhr:2019kwi,Duhr:2020kzd,Chen:2019lzz,Currie:2018fgr,Dreyer:2018qbw,Duhr:2020sdp,Duhr:2020seh} as well as more differential calculations have started to emerge~\cite{Dulat:2017prg,Dulat:2018bfe,Cieri:2018oms,Chen:2021isd,Chen:2021vtu,Billis:2021ecs,Chen:2022cgv,Neumann:2022lft,Camarda:2021ict,Chen:2022lwc,Baglio:2022wzu,Jakubcik:2022zdi}, the latter mainly for $2 \to 1$ processes via the use of the Projection-to-Born method~\cite{Cacciari:2015jma} or $k_T$-slicing techniques~\cite{Catani:2007vq} to promote known NNLO subtraction schemes to \NthreeLO.
Calculations for higher multiplicities are currently hindered by the lack of process-independent \NthreeLO subtraction schemes.

The antenna-subtraction scheme is one of the most successful methods for fully-differential NNLO calculations in QCD.
It was first proposed for perturbative QCD calculations with massless partons in electron-positron annihilation in Refs.~\cite{Gehrmann-DeRidder:2005btv,Gehrmann-DeRidder:2005alt,Gehrmann-DeRidder:2005svg}. It allowed the calculation of the NNLO corrections to 3-jet production and related event-shape observables in electron-positron annihilation \cite{Gehrmann-DeRidder:2007foh,Gehrmann-DeRidder:2007nzq,Gehrmann-DeRidder:2007vsv}. 
The extension of the scheme to the treatment of initial-state radiation relevant to processes with initial-state hadrons was established at NLO in \cite{Daleo:2006xa} and extended to NNLO in Refs.~\cite{Daleo:2009yj,Boughezal:2010mc,Gehrmann:2011wi,Gehrmann-DeRidder:2012too}. A cornerstone of the antenna-subtraction framework is that all of the integrals relevant for processes at NNLO with massless quarks are known analytically \cite{Daleo:2009yj,Boughezal:2010mc,Gehrmann:2011wi,Gehrmann-DeRidder:2012too}. The extension of antenna subtraction for the production of heavy particles at hadron colliders has been studied in Refs. \cite{Gehrmann-DeRidder:2009lyc,Abelof:2011ap,Bernreuther:2011jt,Abelof:2011jv,Abelof:2012bga,Abelof:2012rv,Bernreuther:2013uma,Dekkers:2014hna}. 
Besides its application in fixed-order calculations, the antenna framework has been utilised in antenna-shower algorithms \cite{Gustafson:1987rq,Lonnblad:1992tz,Giele:2007di,Giele:2011cb,Fischer:2016vfv,Brooks:2020upa}, where it enabled proof-of-concept frameworks for higher-order corrections \cite{Li:2016yez} and fully-differential NNLO matching \cite{Campbell:2021svd}.

This paper's focus is on improvements to the antenna-subtraction scheme at NNLO. In particular, we describe an algorithm to re-build antenna functions directly from the divergences we want them to contain. We note that some of the techniques employed in this paper have been used in other NNLO subtraction schemes such as~\cite{Caola:2017dug,Czakon:2010td,Magnea:2018hab}. While the antenna-subtraction formalism successfully enabled the calculation of many processes at NNLO and even at \NthreeLO, the complexity of the subtraction terms becomes increasingly difficult with growing particle multiplicity.
This is mainly due to two reasons. Firstly, double-real radiation antenna functions derived from matrix elements do not always identify which particles are the hard radiators. This is particularly the case for gluons. To get round this, so-called sub-antenna functions are introduced. The construction of sub-antenna functions at NNLO is extremely cumbersome and typically introduces unphysical denominators that make analytic integration difficult. Often analytic integrals are known only for the full antenna functions. This means that antenna-subtraction terms have to be assembled in such a way that sub-antenna functions recombine to full antenna functions before integration. 
Secondly, NNLO antenna functions can contain spurious singularities that have to be removed by explicit counter terms, which in turn can introduce further spurious singularities. In general, this can trigger an intricate chain of cross-dependent subtraction terms that have no relation to the actual singularity structure of the process at hand.

As a first step towards a refined antenna-subtraction scheme at NNLO, we construct a full set of improved single-real and double-real antenna functions. Instead of building these from physical matrix elements, as done originally \cite{Gehrmann-DeRidder:2005alt,Gehrmann-DeRidder:2005svg,Gehrmann-DeRidder:2005btv}, we build antenna functions directly from the relevant limits properly accounting for the overlap between different limits. The universal factorisation properties of multiparticle amplitudes when one or more particles are unresolved have been well studied in the literature and serve as an input to the algorithm.  The single-unresolved limits of tree amplitudes, where either one particle is soft, or two are collinear, are used to construct  NLO antennae, while the double-unresolved limits of tree amplitudes, with up to two soft particles, or three collinear particles~\cite{campbell,Catani:1998nv,Catani:1999ss,Kosower:2002su} are needed for the NNLO antennae.  We note that the triple-unresolved limits of tree amplitudes are available for the construction of real radiation \NthreeLO antennae~\cite{Catani:2019nqv,DelDuca:1999iql,DelDuca:2019ggv,DelDuca:2020vst,DelDuca:2022noh}. 
 
The paper is structured as follows. We outline the design principles for ideal real-radiation antenna functions in Section~\ref{sec:design-principles} before discussing the general construction algorithm in Section~\ref{sec:algorithm}.
To illustrate the algorithm, we explicitly construct a full set of single-real and double-real antenna functions for use in NLO and NNLO antenna subtraction in Sections~\ref{sec:single-real-antennae} and \ref{sec:double-real-antennae}, respectively.
We conclude and give an outlook on further work in Section~\ref{sec:outlook}.

\section{Design principles}
\label{sec:design-principles}
Within the antenna framework, subtraction terms are constructed from antenna functions which describe all unresolved partonic radiation (soft and collinear) between a hard pair of radiator partons.
In general, an antenna-subtraction term requires:
\begin{itemize}
\item antennae $X$, composed of two hard radiators that accurately reflect the infrared singularities of the $n$ unresolved partons radiated ``between'' them;
\item an on-shell momentum mapping $\mathcal{F}_{n+2\mapsto 2}$, clustering $n+2$ particles into $2$, while preserving the invariant mass of the radiators, used to define the ``reduced'' matrix element; and
\item a process- and antenna-dependent colour factor.
\end{itemize}
The calculation of colour factors strictly follows the non-abelian structure of QCD and, while in principle cumbersome, can be automated to all orders \cite{Sjodahl:2012nk,Sjodahl:2014opa,Gerwick:2014gya,Baberuxki:2019ifp}.
A general on-shell momentum mapping for multi-particle emissions in the antenna language has been derived in Ref.~\cite{Kosower:2002su} for massless particles. In the antenna framework, kinematic mappings are agnostic to the roles of the parent radiator partons and the transverse recoil of any additional emission is shared between them. Note that this is different to dipole-like kinematics, in which one of the parents is identified as the emitter and the other as the recoiler, whose role is solely to absorb the transverse recoil.

Antenna functions are used to subtract specific sets of unresolved singularities
\begin{equation} \label{eqn:subterm}
    X_{n+2}^\ell(i_1^h,i_3,\ldots,i_{n+2},i_2^h) 
    \modsq{\calM(\ldots,I_1^h,I_2^h,\ldots)} \, ,
\end{equation}
where $X_{n+2}^\ell$ represents an $\ell$-loop, $(n+2)$-particle antenna,
$i_1^h$ and $i_2^h$ represent the hard radiators, and $i_3$ to $i_{n+2}$ denote the unresolved particles. Particle $i$ carries a four-momentum $p_i^{\mu}$.
As the hard radiators may either be in the initial or in the final state, final-final (FF), initial-final (IF), and initial-initial (II) configurations need to be considered in general. $\calM$ is the reduced matrix element, with $n$ fewer particles and where $I_1^h$ and $I_2^h$ represent the particles obtained through an appropriate mapping,
\begin{align}
\{ p_{i_1},p_{i_3},\ldots,p_{i_2} \} \mapsto \{ p_{I_1}, p_{I_2} \}.  
\end{align}

Depending on the perturbative order of the calculation, antenna functions may contain a number of loops $\ell$ in addition to $n$ unresolved particles, so that a generic antenna function can be denoted by $X_{n+2}^\ell$.
At NLO, the number of loops $\ell$ is equal to 0 and the number of unresolved particles is equal to 1; these are the usual $X_3^0$ antenna functions.
At NNLO there are two additional cases; $\X$ corresponding to $\ell=0$ and $n=2$ with two unresolved particles, and $X_3^1$ corresponding to $\ell=1$ and $n=1$ with one unresolved particle.

Historically, antenna functions have been constructed directly from matrix elements which have the desired singularities. However, we note that these ``natural'' antenna functions do not share the same design principles we are describing -- for example many natural antenna functions do not have two identified hard radiators. This is the case for quark-antiquark antennae, but not for quark-gluon or gluon-gluon antennae because the matrix elements they are derived from will inevitably have a divergent limit when one of the gluonic radiators becomes soft.  
This makes the construction of the subtraction terms more complex, and correspondingly less automatable. 
Additionally, antenna momentum mappings require a clear identification of two hard radiators so that the mapping is appropriate for the limits the antenna is intended to describe. In order to counter these issues, so-called sub-antenna functions have been created by complicated use of supersymmetry relations, other antenna functions, and partial fractioning \cite{NigelGlover:2010kwr,Gehrmann-DeRidder:2007foh}. The momentum mapping is then different for each sub-antenna. These sub-antenna functions are in general difficult to integrate but combine in certain processes such that only the full antenna requires integration. In some cases, sub-antennae map onto different types of matrix elements and multiple subtraction terms cannot be combined easily. This means that a direct integration of the sub-antenna functions is required, which is not feasible. In these cases, intricate process-dependent combinations of other antennae have to be used to correct for over-subtraction of spurious limits. 

We aim to design antenna functions directly from their desired properties, with a uniform template, in a way that simplifies the construction of subtraction terms in general, while being straightforwardly integrable.
Specifically, we impose the following requirements on the new antenna functions:
\begin{enumerate}[I.]
\item each antenna function has exactly two hard particles (``radiators'') which cannot become unresolved;
\item each antenna function captures all single- and double-soft limits of its unresolved particles;
\item where appropriate (multi-)collinear and soft-collinear limits are decomposed over ``neighbouring'' antennae;
\item antenna functions do not contain any spurious (unphysical) limits;
\item antenna functions only contain singular factors corresponding to physical propagators; and
\item where appropriate, antenna functions obey physical symmetry relations (such as line reversal).
\end{enumerate}
We wish to emphasise again that the original NNLO antenna functions derived in \cite{Gehrmann-DeRidder:2005svg,Gehrmann-DeRidder:2005alt,Gehrmann-DeRidder:2005btv} do not obey these requirements, as they typically violate (some of) these principles, as alluded to above.
A subtlety is connected to (multi-)collinear and soft-collinear limits. For example, any gluon can become soft in a multi-gluon matrix element so a prescription has to be defined according to which the singularities are shared among antennae that identify each of the gluons as being hard.

In the following sections we will describe a general algorithm to construct real radiation antenna functions following strictly these design principles and apply it to the case of single-real and double-real radiation, required for NLO and NNLO calculations. 

\section{The algorithm}
\label{sec:algorithm}
The main goal of the algorithm is to construct (multiple-)real radiation antenna functions containing singular limits pertaining to exactly two hard radiators plus an, in principle arbritrary, number of additional particles that are allowed to become unresolved. To this end, each limit is defined by a ``target function'', which in the following we will denote by $L_i$. In general, there will be $N$ such limits, and we denote the ordered set of limits by $\{L_j\}$. The target functions have to capture the behaviour of the colour-ordered matrix element squared in the given unresolved limit and are taken as input to the algorithm. While the target functions may include process-dependent azimuthal terms, for the purposes of the present paper we will limit ourselves to azimuthally-averaged functions.

\begin{figure}[t]
    \centering
    \begin{tikzpicture}
    \node (x1) [fullspace] {Full Phase Space \\ $X^0_{n;i-1}$};
    \node (x2) [subspace, below=2cm of x1] {Subspace \\ $L_i$};
    \node (center) [below=1cm of x1] {};
    \node (AR) [right=0.001cm of x1] {};
    \node (AL) [left=0.001cm of x1] {};
    \node (BR) [right=0.001cm of x2] {};
    \node (BL) [left=0.001cm of x2] {};
    \node (a1) [left=2cm of center] {$\mathrm{\textbf{P}}^\downarrow_i$};
    \node (a2) [right=2cm of center] {$\mathrm{\textbf{P}}^\uparrow_i$};
    
    \draw [arrow] (AL) arc[start angle= 90,end angle=270,x radius=1,y radius=1.80];
    \draw [arrow] (BR) arc[start angle=270,end angle=450,x radius=1,y radius=1.80];
    \end{tikzpicture}
    \caption{Visual representation of the down- and up-projectors translating between the full phase space, on which antenna functions are defined, and the subspaces on which the target functions are defined. Here $X^0_{n;i-1}$ represents the accumulated antenna function having taken into account the limits $L_1, \ldots L_{i-1}$, which is projected into the subspace relevant for limit $L_i$ by $\PPdown_i$, subtracted from $L_i$ and the remainder projected back into the full phase space by $\PPup_i$. }
    \label{fig:spaces}
\end{figure}
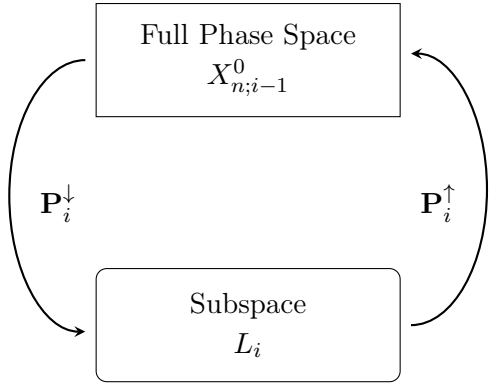

Antenna functions are defined over the full phase space appropriate to the respective antenna, whereas each singular limit lives on a restricted part of phase space, with one or more of the momenta being soft or collinear.
We relate the two by a ``down-projector'' $\PPdown$ that maps the invariants of the full phase space into the relevant subspace. 
An associated ``up-projector'' $\PPup$ restores the full phase space. That is, it re-expresses all variables valid in the sub-space in terms of invariants valid in the full phase space. This is illustrated in Fig.~\ref{fig:spaces}. 
It is to be emphasised that down-projectors $\PPdown$ and up-projectors $\PPup$ are typically not inverse to each other, as down-projectors destroy information about less-singular and finite pieces.

\begin{figure}[ht]
\centering
\begin{tikzpicture}[node distance=2.0cm]

\node (start) [startstop] {\Large $i=1, X_{n;0}^0 = 0, \{L_j\}, \{\PPdown_j\},\{\PPup_j\}$};
\node (Li) [io, below of=start, yshift=0.4cm] {\Large $L_i, X_{n;i-1}^0$};
\node (pro1) [process, below of=Li, yshift=0.5cm] {\Large $L_i$};
\node (pro2) [process, below of=pro1] {\Large $L_i -\PPdown_i X_{n;i-1}^0$};
\node (pro3) [process, below of=pro2] {\Large $\PPup_i (L_i -\PPdown_i X_{n;i-1}^0)$};
\node (pro4) [process, below of=pro3] {\Large $X_{n;i}^0 = \PPup_i (L_i -\PPdown_i X_{n;i-1}^0)+X_{n;i-1}^0$};
\node (dec1) [decision, below of=pro4, yshift=-1.5cm] {\Large $L_{i+1} \in \{L_j\}$?};

\node (pro2b) [process, right of=dec1, xshift=5cm] {\Large $i\rightarrow i+1$};

\node (stop) [io, below of=dec1, yshift=-1.5cm] {\Large $X_n^0 = X_{n;N}^0 $};
\node (end) [startstop, below of=stop, yshift=0.4cm] {\Large $X_n^0$ New Antenna};

\draw [arrow] (Li) -- (pro1);
\draw [arrow] (start) -- (Li);
\draw [arrow] (pro1) -- node[anchor=west] {\Large $-\PPdown_i X_{n;i-1}^0$} (pro2);
\draw [arrow] (pro2) -- node[anchor=west] {\Large $\PPup_i$} (pro3);
\draw [arrow] (pro3) -- node[anchor=west] {\Large $+X_{n;i-1}^0$} (pro4);
\draw [arrow] (pro4) -- (dec1);
\draw [arrow] (dec1) -- node[anchor=east] {\Large no} (stop);
\draw [arrow] (dec1) -- node[anchor=south] {\Large yes} (pro2b);
\draw [arrow] (pro2b) |- (Li);
\draw [arrow] (stop) -- (end);

\end{tikzpicture}
\caption{Flowchart representing the algorithm to construct a tree-level $n$-particle antenna $X_n^0$ using an ordered set of limits $\{L_j\}$ with associated down- and up-projectors. $X_{n;i}^0$ represents the accumulation of the contributions from the first $i$ limits.  The full antenna is obtained when all $N$ limits have been satisfied, $X_{n}^0 = X_{n:N}^0$.}
\label{fig:algorithm}
\end{figure}
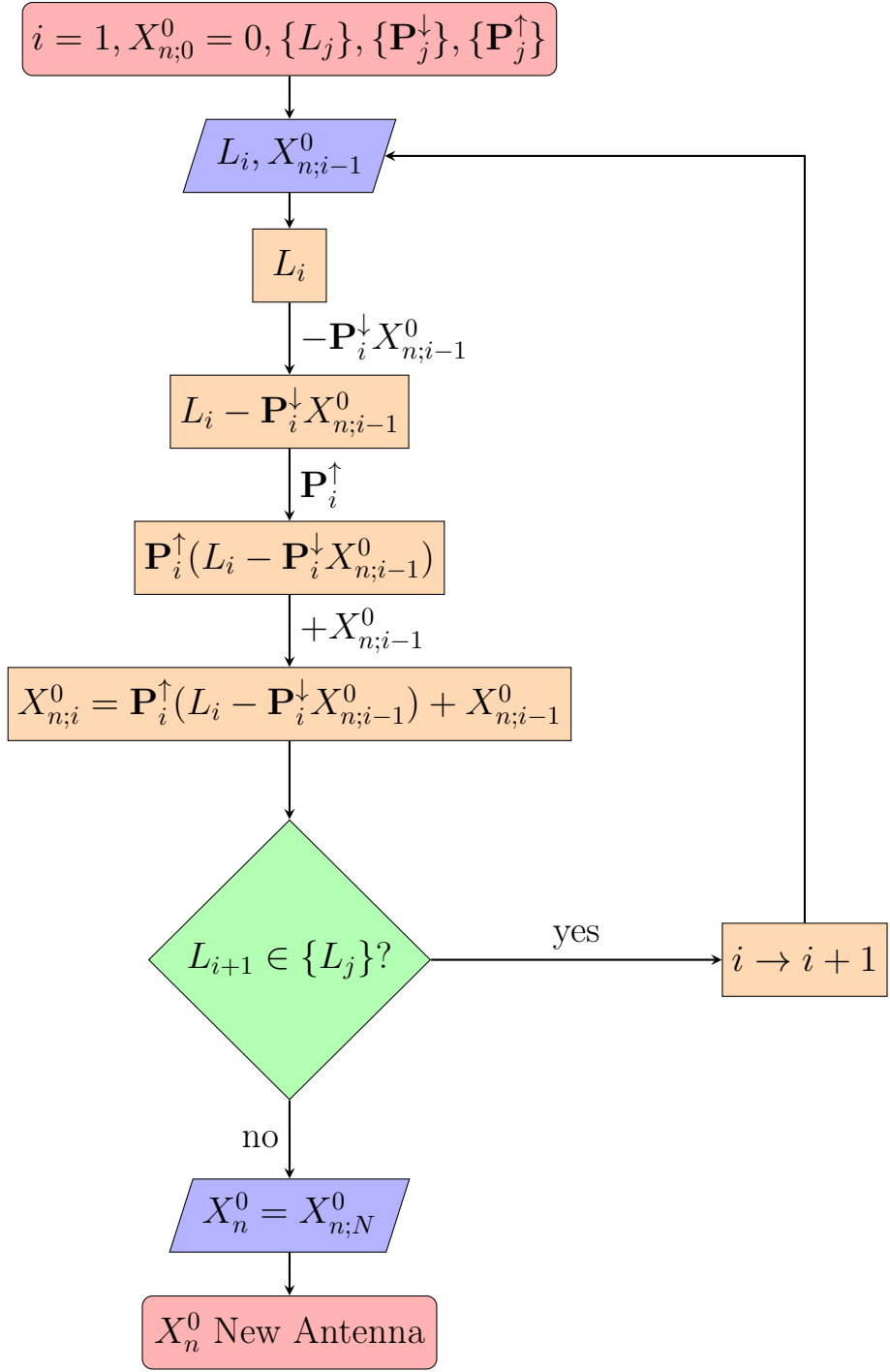

We construct antenna functions using an iterative process, which requires as input the full set of $N$ unresolved limits we aim to capture (defined in terms of target functions $L_i$, with $i$ running from 1 to $N$.), together with the appropriate set of ``down-'' and ``up-projectors'' pertaining to each of these limits.

The algorithm starts with the contribution corresponding to the deepest pole on the integrated level, and terminates upon reaching the level of finite corrections, i.e., terms that integrate to corrections of $\order{\e^0}$.
In each step of the iteration, we remove the overlap of the target function with all previously considered limits and accumulate the remainder.
To this end, we subtract the projection of the accumulated antenna function into the sub-space relevant to the target limit, $\PPdown_iX^0_{n;i-1}$, from the target function. The remainder $(L_i -\PPdown_i X^0_{n;i-1})$ is then restored to the full phase space via the associated up-projector $\PPup_i$ before adding it to the accumulated antenna function.
Schematically, this procedure is shown in Fig.~\ref{fig:algorithm} for $N$ target functions $L_i$ and can be written as 
\begin{equation}
\label{eq:algorithm}
  \begin{split}
    X^0_{n;1} &= \PPup_1 L_1 \, , \\
    X^0_{n;2} &= X^0_{n;1} + \PPup_2 (L_2 - \PPdown_2 X^0_{n;1}) \, ,\\
    & \vdots \\
    X^0_{n;N} &= X^0_{n;N-1} + \PPup_N (L_N-\PPdown_N X^0_{n;N-1}) \, ,
  \end{split}
\end{equation}
where $X^0_n \equiv X^0_{n;N}$. In particular, restoring the kinematics to the full phase space, in a judicious way, ensures that the antenna function can be expressed solely in terms of invariants corresponding to physical propagators. This specifically guarantees that the full antenna function can be integrated easily over its Lorentz-invariant antenna phase space.

Our algorithm resembles the ones in \cite{Czakon:2010td,Caola:2017dug,Magnea:2018hab} in spirit, in the sense that singular limits are considered subsequently, ordered according to the depth of the associated explicit pole. We wish to point out, however, that there are two important differences to the algorithms described in \cite{Czakon:2010td,Caola:2017dug,Magnea:2018hab}. Firstly, we do not use our algorithm to construct process-dependent subtraction terms (even though the construction of the counter terms is in principle process independent) but to construct universal antenna functions that are only assembled to process-dependent subtraction terms in a separate (automatable) step \cite{Gehrmann-DeRidder:2005btv,Currie:2013vh,Chen:2022ktf}. Secondly, we always reconstruct the kinematics of the subtraction term to the full phase space via up-projectors, a step that is not strictly necessary for the sake of the subtraction but vital to build a function in terms of multi-particle invariants that are valid in the full phase space.

The set of target functions unambiguously defines the behaviour of the antenna function in all unresolved limits pertaining to the antenna at hand.
In any unresolved limit, the full antenna function has to approach the target function in order to capture the correct singular behaviour of the respective squared matrix element.
In particular, the antenna function has to be finite in all limits not described explicitly by a target function.
This ensures that no spurious singularities enter, a feature not shared by antenna functions constructed directly from physical matrix elements.
As alluded to above, an important aspect to consider in the construction of antenna functions pertains to the presence of certain multi-collinear and soft-collinear limits, which are shared by ``neighbouring'' antennae in the antenna formalism.
In these cases, the correct multi-collinear or soft-collinear limit is only recovered in the sum over multiple antenna functions, each containing one of the involved partons as a hard radiator.
This means that each multi-collinear/soft-collinear splitting function has to be decomposed over all possibilities to identify one of the partons as the hard radiator before it can be used as a target function in our algorithm.
For simple-collinear splitting functions, this decomposition is simple and has been identified already in the original publications \cite{Altarelli:1977zs,Dokshitzer:1977sg}. A generalisation for triple-collinear splitting functions has been derived in Ref.~\cite{paper1}. 
As a byproduct of this procedure, soft-collinear limits do not have to be entered as explicit inputs into the algorithm at NNLO.

A core part of our algorithm is the definition of down-projectors into singular limits with corresponding up-projectors into the full phase space.
In each step of the construction, down-projectors are needed to identify the overlap of the so-far constructed antenna function with the target function of the respective unresolved limit, whereas up-projectors are required to re-express the subtracted target function in terms of antenna invariants.
In this way, the full (accumulated) antenna function can be expressed solely in terms of $n$-particle invariants and is therefore valid in the full phase space. By choosing the up-projectors judiciously, the antenna function can furthermore be expressed exclusively in terms of physical propagators.
As alluded to above, down-projectors $\PPdown$ and up-projectors $\PPup$ are not required to be inverse to each other.

The number of projectors depends directly on the perturbative order.
At NLO, only two types of down-projectors and their up-projectors into the full phase space are needed,
\begin{equation}
    \PSdown, \PCdown, \text{ and } \PSup, \PCup,
\end{equation}
corresponding to the single-soft and simple-collinear limits. 
These will be discussed in detail in Section~\ref{sec:single-real-antennae}.
At NNLO, three additional types of down-projectors are needed,
\begin{equation}
    \PDSdown, \PTCdown, \PDCdown,
\end{equation}
corresponding to the double-soft, triple-collinear, and double-collinear limits. The up-projectors into the full phase space are given by
\begin{equation}
    \PDSup, \PTCup, \text{ and } \PDCup,
\end{equation}
respectively. While not relevant to the construction of the antenna functions, for the sake of validating the correct singular behaviour of the constructed antenna functions, one can further define a down-projector related to the soft-collinear limit, $\PSCdown$.
Both down-projectors and up-projectors at NNLO will be discussed in Section~\ref{sec:double-real-antennae}.

Up to NNLO, we have automated our algorithm in a computer code based on MAPLE and FORM~\cite{Vermaseren:2000nd,Kuipers:2012rf}. 
To make the construction explicit, we work through the construction of all single-real antenna functions as an example in Section~\ref{sec:single-real-antennae} before constructing a full set of improved double-real antenna functions in Section~\ref{sec:double-real-antennae}.
The latter constitutes a first step towards a refined antenna-subtraction formalism at NNLO and beyond.

\section{Single-real radiation antennae}
\label{sec:single-real-antennae}
At NLO, we want to construct three-particle antenna functions $X_3^0(i^h_a,j_b,k^h_c)$, where the particle types are denoted by $a$, $b$, and $c$, which carry four-momenta $i$, $j$, and $k$ respectively. Particles $a$ and $c$ should be hard, and the antenna functions must have the correct limits when particle $b$ is unresolved. Frequently, we drop explicit reference to the particle labels in favour of a specific choice of $X$ according to Table~\ref{tab:X30}.

\begin{table}[t]
\centering
\begin{tabular}{ccc}
\underline{Quark-antiquark} & & \\
$qg\bar{q}$ & $X_3^0(i_q^h,j_g,k_{\bar{q}}^h)$ & $A_3^0(i^h,j,k^h)$ \\
\underline{Quark-gluon} & &  \\
$qgg$ & $X_3^0(i_q^h,j_g,k_g^h)$ & $D_3^0(i^h,j,k^h)$  \\
$q\bar{Q}Q$ & $X_3^0(i_q^h,j_{\bar{Q}},k_Q^h)$  & $E_3^0(i^h,j,k^h)$ \\ 
\underline{Gluon-gluon} & & \\
$ggg$ & $X_3^0(i_g^h,j_g,k_g^h)$ & $F_3^0(i^h,j,k^h)$  \\
$g\bar{Q}Q$ & $X_3^0(i_g^h,j_{\bar{Q}},k_Q^h)$ & $G_3^0(i^h,j,k^h)$  \\
\end{tabular}
\caption{Identification of $X_3^0$ antenna according to the particle type. These antennae only contain singular limits when particle $b$ (or equivalently momentum $j$) is unresolved. Antennae are classified as quark-antiquark, quark-gluon and gluon-gluon according to the particle type of the parents (i.e. after the antenna mapping). }
\label{tab:X30}
\end{table}

We systematically start from the most singular limit, and build the list of target functions from single-soft and simple-collinear limits.  For the particles of $X_3^0(i^h_a,j_b,k^h_c)$ there are three such limits, corresponding to particle $b$ becoming soft or particles $a$ and $b$ becoming collinear or particles $c$ and $b$ becoming collinear,
\begin{equation}
\begin{split}
    L_1(i^h,j,k^h) &= \Sb(i^h,j,k^h) \, , \\
    L_2(i^h,j;k) &= P_{ab}(i^h,j) \, , \\
    L_3(k^h,j;i) &= P_{cb}(k^h,j) \, .
\end{split}
\label{eq:NLOL}
\end{equation}
The tree-level soft factor $\Sb$ is given by the eikonal factor for particle $b$ radiated between two hard radiators,
\begin{align}
    \Sg(i^h,j_g,k^h) &= \Sp(i^h,j_{\gamma},k^h) =   \frac{2s_{ik}}{s_{ij}s_{jk}},  \\
    \Sq(i^h,j_q,k^h) &= \Sqb(i^h,j_{\bar{q}},k^h) =0,
\end{align}
where we use Lorentz-invariant momentum structures,
\begin{equation}
	s_{i,\ldots,n} \equiv (p_{i}+...+p_{n})^2.
\end{equation}
For massless quarks and gluons, $s_{ij} = 2p_i \cdot p_j = 2E_iE_j(1-\cos\theta_{ij})$, where $E_i$, $E_j$ are the energies of particles $i$, $j$ and $\theta_{ij}$ is the angle between them. The invariant $s_{ij}$ approaches zero if either particle is soft or they are collinear to each other. 
The splitting functions $P_{ab}(i^h,j)$ are {\em not singular} in the limit where the hard radiator $a$ becomes soft and are related to the usual spin-averaged splitting functions, cf.~\cite{Altarelli:1977zs,Dokshitzer:1977sg}, by, 
\begin{align} 
\label{eqn:Pqg}
\Pqg(i^h,j) &= \frac{1}{s_{ij}} \Pqg(\xj) \\
\Pqg(i,j^h) &= 0,\\
\label{eqn:Pqq}
\Pqq(i^h,j) &= \frac{1}{s_{ij}} \Pqq(\xj),\\
\Pqq(i,j^h) &= \frac{1}{s_{ij}} \Pqq\omxj,\\
\label{eqn:Pgg}
\Pgg(i^h,j) &= \frac{1}{s_{ij}} \PggS(\xj)\hfill\\
\Pgg(i,j^h) &= \frac{1}{s_{ij}}\PggS\omxj
\end{align}
with
\begin{eqnarray}
\Pqg(\xj) &=& \left(\frac{2\omxj}{\xj} + \ome \xj \right)\\
\Pqq(\xj) &=& \left( 1 -\frac{2\omxj\xj}{\ome} \right) = \Pqq\omxj\\
\label{eq:PggS}
\PggS(\xj)&=& \left( \frac{2\omxj}{\xj} + \xj \omxj \right) 
\end{eqnarray}
and
\begin{equation}
\PggS(\xj) + \PggS\omxj \equiv \Pgg(\xj).
\end{equation}
Here, the momentum fraction $\xj$ is defined with reference to the third particle in the antenna, $\xj = s_{jk}/(s_{ik}+s_{jk})$.

We define the soft down-projector by its action on invariants as
\begin{equation}
    \PSdown_j: \begin{cases} 
    s_{ij} \mapsto \lambda s_{ij}, 
    s_{jk} \mapsto \lambda s_{jk}, \\
    s_{ijk} \mapsto s_{ik},
    \end{cases}
\end{equation}
and keep only the terms proportional to $\lambda^{-2}$.
For the corresponding up-projector $\PSup_j$ we choose a trivial mapping which leaves all variables unchanged.
The collinear projector we define as,
\begin{equation}
    \PCdown_{ij}: \begin{cases}
    s_{ij} \mapsto \lambda s_{ij}, \\
    s_{ik} \mapsto \omxj (s_{ik}+s_{jk}),
    s_{jk} \mapsto \xj (s_{ik}+s_{jk}), 
    s_{ijk} \mapsto s_{ik} + s_{jk}.
    \end{cases}
\end{equation}
and keep only terms proportional to $\lambda^{-1}$, while its corresponding up-projector we define as,
\begin{equation}
    \PCup_{ij}: \begin{cases}
    \xj \mapsto s_{jk}/s_{ijk},
    \omxj \mapsto s_{ik}/s_{ijk} \\
    s_{ik}+s_{jk} \mapsto s_{ijk}
    \end{cases}
\end{equation}

For convenience, we define a general single-real radiation tree-level antenna function in terms of the contributions produced by the algorithm in Eq.~\eqref{eq:algorithm} as,
\begin{equation}
X_3^0(i^h,j,k^h) = \Ssoft(i^h,j,k^h) + \Scol(i^h,j;k^h) + \Scol(k^h,j;i^h) \, ,
\label{eq:X30def}
\end{equation}
where the individual pieces are given by
\begin{align}
  \Ssoft(i^h,j,k^h) &= \PSup_{j} L_1(i^h,j,k^h) = L_1(i^h,j,k^h), \\
  \Scol(i^h,j;k^h) &= \PCup_{ij}\left(L_2(i^h,j;k^h) - \PCdown_{ij} \Ssoft(i^h,j,k^h)\right), \\
  \Scol(k^h,j;i^h) &= \PCup_{kj}\left(L_3(k^h,j;i^h) - \PCdown_{kj} \left(\Ssoft(i^h,j,k^h) + \Scol(i^h,j;k^h)\right)\right) \nonumber, \\
                   &\equiv \PCup_{kj}\left(L_3(k^h,j;i^h) - \PCdown_{kj} \Ssoft(i^h,j,k^h)\right),
\end{align}
and we have used the fact that the overlap between the two collinear contributions is contained entirely in the projection of $\Ssoft(i^h,j,k^h)$ such that
\begin{equation}
    \PCdown_{kj} \Scol(i^h,j;k^h) = 0 \, .
\end{equation}
This algorithm guarantees that
\begin{align}
\PSdown_j X_3^0(i^h,j,k^h) &= L_1(i^h,j,k^h),\\
\PCdown_{ij} X_3^0(i^h,j,k^h) &= L_2(i^h,j,k^h),\\
\PCdown_{kj} X_3^0(i^h,j,k^h) &= L_3(i^h,j,k^h).
\end{align}

In the following subsections, we derive the single-real radiation antennae for pairs of quark-antiquark, quark-gluon and gluon-gluon parents. As a check, we also give the analytic form of the three-particle antennae integrated over the fully-inclusive $d$-dimensional antenna phase space, 
\begin{equation}
\label{eq:phijkFF}
{\cal X}_{3}^0(s_{ijk}) =
\left(8\pi^2\left(4\pi\right)^{-\e} e^{\e\gamma}\right)
\int {\rm d}\, \Phi_{X_{ijk}} X_{3}^0,
\end{equation}
with $d=4-2\e$.
As in Ref.~\cite{Gehrmann-DeRidder:2005btv}, we have included a normalisation factor to account for powers of the QCD coupling constant. 

\subsection{Quark-antiquark antennae}

Bulding the antenna iteratively according to Eq.~\eqref{eq:algorithm} using the list of limits in Eq.~\eqref{eq:NLOL}, we find that
the three-parton tree-level antenna function with quark-antiquark parents is given (to all orders in $\e$) by 
\begin{equation}
\label{eq:A30}
A_3^0(i_q^h,j_g,k_{\bar{q}}^h) =
\frac{2s_{ik}}{s_{ij}s_{jk}} 
+ \frac{\ome s_{jk}}{s_{ijk}s_{ij}}
+ \frac{\ome s_{ij}}{s_{ijk}s_{jk}}.
\end{equation}
Eq.~\eqref{eq:A30} differs from $\Xold{A}$, given in Eq.~(5.5) of Ref.~\cite{Gehrmann-DeRidder:2005btv} that was derived directly from the squared matrix element of $\gamma^* \to qg\bar{q}$, at $\order{\e}$. We note that in general $X_3^0$ can differ from $\Xold{X}$ at $\order{\e^0}$ as we only require that they have the same unresolved limits. Only in this specific case is the difference $\order{\e}$. 

Integrating over the final-final antenna phase space, Eq.~\eqref{eq:phijkFF}, yields 
\begin{eqnarray}
\label{eq:A30int}
{\cal A}_3^0(s_{ijk}) &=&
\left(s_{ijk}\right)^{-\e}
\left[ \frac{1}{\e^2} + \frac{3}{2\e}
+\frac{19}{4} -\frac{7\pi^2}{12}
+\left(\frac{113}{8}-\frac{7\pi^2}{8}-\frac{25\zeta_3}{3}\right)\e
\right. \nonumber \\
&& \left. \hspace{2cm}
+\left(\frac{675}{16}-\frac{133\pi^2}{48}-\frac{71\pi^4}{1440}-\frac{25\zeta_3}{2} \right)\e^2
+ \order{\e^3} \right], 
\end{eqnarray}
which, as expected, differs from the final-final integral of $\Xold{A}$, in Eq.~(5.6) of Ref.~\cite{Gehrmann-DeRidder:2005btv}, starting from $\order{\e}$. Integrals of the universal soft and collinear-remainder terms in Eq.~\eqref{eq:X30def} (and in this case Eq.~\eqref{eq:A30}) are given in Appendix~\ref{app:X30regionintegrations} for increased clarity. 

\subsection{Quark-gluon antennae}
Building the antenna iteratively according to Eq.~\eqref{eq:algorithm} and using the list of soft and collinear limits given in Eq.~\eqref{eq:NLOL} adapted to the particle content, we find that
\begin{equation}
\label{eq:D30}
D_3^0(i_q^h,j_g,k_g^h) =
\frac{2s_{ik}}{s_{ij}s_{jk}} 
+ \frac{\ome s_{jk}}{s_{ijk}s_{ij}}
+ \frac{s_{ij}s_{ik}}{s_{ijk}^2 s_{jk}}.
\end{equation}
Note that, owing to its origin in the neutralino-decay process, the $\Xold{D}$ antenna function given in Eq.~(4.3) and (4.9) of Ref.~\cite{Gehrmann-DeRidder:2005svg} contains two antennae, corresponding to the configurations: $j_g$ radiated between $i_q$ and $k_g$, and $k_g$ radiated between $i_q$ and $j_g$.
In Eq.~(6.13) of Ref.~\cite{Gehrmann-DeRidder:2005btv}, the full $\Xold{D}$ was decomposed into two subantennae, $\Xold{d}(i_q^h,j_g,k_g^h)$ and $\Xold{d}(i_q^h,k_g,j_g^h)$, each of which contains the soft limit of one of the gluons, and part of the collinear limit between the gluons.  $D_3^0(i_q^h,j_g,k_g^h)$ in Eq.~\eqref{eq:D30} is therefore to be compared with $\Xold{d}(i_q^h,j_g,k_g^h)$ given in Eq.~(6.13) of Ref.~\cite{Gehrmann-DeRidder:2005btv}. Eq.~\eqref{eq:D30} is more compact because it only  contains terms that contribute to the soft and collinear limits. The difference starts at $\order{\e^0}$ in the dimensional regularisation parameter $\e$. 
Integrating over the antenna phase space yields,
\begin{eqnarray}
\label{eq:D30int}
{\cal D}_3^0(s_{ijk}) &=&
\left(s_{ijk}\right)^{-\e}
\left[ \frac{1}{\e^2} + \frac{5}{3\e}
+\frac{61}{12} -\frac{7\pi^2}{12}
+\left(\frac{121}{8}-\frac{35\pi^2}{36}-\frac{25\zeta_3}{3}\right)\e
\right. \nonumber \\
&& \left. \hspace{2cm}
+\left(\frac{723}{16}-\frac{427\pi^2}{144}-\frac{71\pi^4}{1440}-\frac{125\zeta_3}{9} \right)\e^2
+ \order{\e^3} \right].
\end{eqnarray}
This differs from the final-final integral of $\Xold{d}(i_q^h,j_g,k_g^h)$, which is a half of Eq.~(6.9) in Ref.~\cite{Gehrmann-DeRidder:2005btv}, starting at $\order{\e^0}$.

Similarly, for the three-quark antenna, we find
\begin{equation}
\label{eq:E30}
E_3^0(i_q^h,j_{\bar{Q}},k_Q^h) =
\frac{1}{s_{jk}} 
-\frac{2s_{ij}s_{ik}}{s_{ijk}^2 s_{jk} \ome},
\end{equation}
which is to be compared with $\Xold{E}$ given in Eq.~(4.9) of Ref.~\cite{Gehrmann-DeRidder:2005svg} which contains terms that do not contribute to the quark-antiquark collinear limit starting at $\order{\e^0}$.
This has the consequence that the integrated antenna,
\begin{eqnarray}
\label{eq:E30int}
{\cal E}_3^0(s_{ijk}) &=&
\left(s_{ijk}\right)^{-\e}
\left[ - \frac{1}{3\e}
 -\frac{3}{4}
+\left(-\frac{15}{8}+\frac{7\pi^2}{36}\right)\e
\right. \nonumber \\
&& \left. \hspace{2cm}
+\left(-\frac{81}{16}+\frac{7\pi^2}{16}+\frac{25\zeta_3}{9} \right)\e^2
+ \order{\e^3} \right] ,
\end{eqnarray}
differs from the final-final integral of $\Xold{E}$given in Eq.~(6.15) of Ref.~\cite{Gehrmann-DeRidder:2005btv} starting at $\order{\e^0}$.

\subsection{Gluon-gluon antennae}
Directly constructing the antenna using Eq.~\eqref{eq:algorithm} with the list of soft and collinear limits given in Eq.~\eqref{eq:NLOL} adapted to the particle content, we find that
\begin{equation}
\label{eq:F30}
F_3^0(i_g^h,j_g,k_g^h) =
\frac{2s_{ik}}{s_{ij}s_{jk}} 
+ \frac{s_{ik}s_{jk}}{s_{ijk}^2 s_{ij}}
+ \frac{s_{ij}s_{ik}}{s_{ijk}^2 s_{jk}}.
\end{equation}
The $\Xold{F}$ antenna function, obtained from Higgs boson decay (Eq.~(4.3) of Ref.~\cite{Gehrmann-DeRidder:2005alt}), has limits when any of the three gluons are soft. For this reason, it was split into three sub-antenna functions $\Xold{f}(i_g^h,j_g,k_g^h)$ given by Eq.~(7.13) of Ref.~\cite{Gehrmann-DeRidder:2005btv} which differ from Eq.~\eqref{eq:F30} by terms that do not contribute in any of the unresolved limits.
Therefore, the integrated form, 
\begin{eqnarray}
\label{eq:F30int}
{\cal F}_3^0(s_{ijk}) &=&
\left(s_{ijk}\right)^{-\e}
\left[ \frac{1}{\e^2} + \frac{11}{6\e}
+\frac{65}{12} -\frac{7\pi^2}{12}
+\left(\frac{129}{8}-\frac{77\pi^2}{72}-\frac{25\zeta_3}{3}\right)\e
\right. \nonumber \\
&& \left. \hspace{2cm}
+\left(\frac{771}{16}-\frac{455\pi^2}{144}-\frac{71\pi^4}{1440}-\frac{275\zeta_3}{18} \right)\e^2
+ \order{\e^3} \right] ,
\end{eqnarray}
differs from the integrated form of $\Xold{f}(i_g^h,j_g,k_g^h)$, which is a third of Eq.~(7.9) in Ref.~\cite{Gehrmann-DeRidder:2005btv}, starting at $\order{\e^0}$.

The $g\bar{Q}Q$ antenna function is given by
\begin{equation}
\label{eq:G30}
G_3^0(i_g^h,j_{\bar{Q}},k_Q^h) =
\frac{1}{s_{jk}} 
-\frac{2s_{ij}s_{ik}}{s_{ijk}^2 s_{jk} \ome}.
\end{equation}
Note that $G_3^0(i_g^h,j_{\bar{Q}},k_Q^h)$ is identical to $E_3^0(i_q^h,j_{\bar{Q}},k_Q^h)$ because in both cases, the only limit that is required is the $g \to \bar{Q}Q$ collinear limit which is independent of the particle type of the other hard radiator.
The antenna derived from the Higgs-decay matrix element $\Xold{G}$ is given in Eq.~(7.14) of Ref.~\cite{Gehrmann-DeRidder:2005btv} and contains terms that do not contribute in the collinear limit. Integrating over the phase space, we find
\begin{eqnarray}
\label{eq:G30int}
{\cal G}_3^0(s_{ijk}) &\equiv& {\cal E}_3^0(s_{ijk})\\
&=&
\left(s_{ijk}\right)^{-\e}
\left[ - \frac{1}{3\e}
 -\frac{3}{4}
+\left(-\frac{15}{8}+\frac{7\pi^2}{36}\right)\e
\right. \nonumber \\
&& \left. \hspace{2cm}
+\left(-\frac{81}{16}+\frac{7\pi^2}{16}+\frac{25\zeta_3}{9} \right)\e^2
+ \order{\e^3} \right] ,
\end{eqnarray}
which differs from the integrated form of $\Xold{G}$, given in Eq.~(7.15) of Ref.~\cite{Gehrmann-DeRidder:2005btv}, at $\order{\e^0}$.

\section{Double-real radiation antennae}
\label{sec:double-real-antennae}
At NNLO, we want to construct four-particle antenna functions $\X(i^h_a,j_b,k_c,l^h_d)$, where the particle types are denoted by $a$, $b$, $c$, and $d$, which carry four-momenta $i$, $j$, $k$, and $l$ respectively. Particles $a$ and $d$ should be hard, and the antenna functions must have the correct limits when particles $b$ and $c$ are unresolved. Frequently, we again drop explicit reference to the particle labels in favour of a specific choice of $X$ according to Table~\ref{tab:X40}.

\begin{table}[t]
\centering
\begin{tabular}{ccc}
\underline{Quark-antiquark} & & \\
$qgg\bar{q}$ & $\X(i_q^h,j_g,k_g,l_{\bar{q}}^h)$ & $\A(i^h,j,k,l^h)$ \\
$q\gamma\gamma\bar{q}$ & $\Xt(i_q^h,j_\gamma,k_\gamma,l_{\bar{q}}^h)$ & $\At(i^h,j,k,l^h)$ \\
$q\bar{Q}{Q}\bar{q}$ & $\X(i_q^h,j_{\bar{Q}},k_Q,l_{\bar{q}}^h)$ & $\B(i^h,j,k,l^h)$ \\
$q\bar{q}{q}\bar{q}$ & $\X(i_q^h,j_{\bar{q}},k_q,l_{\bar{q}}^h)$ & $\C(i^h,j,k,l^h)$ \\
\underline{Quark-gluon} & &  \\
$qggg$ & $\X(i_q^h,j_g,k_g,l_g^h)$ & $\D(i^h,j,k,l^h)$  \\
       & $\Xt(i_q^h,j_g,k_g,l_g^h)$ & $\Dt(i^h,j,k,l^h)$  \\
$q\bar{Q}Qg$ & $\X(i_q^h,j_{\bar{Q}},k_Q,l_g^h)$  & $\Ea(i^h,j,k,l^h)$ \\ 
$qg\bar{Q}Q$ & $\X(i_q^h,j_g,k_{\bar{Q}},l_Q^h)$  & $\Eb(i^h,j,k,l^h)$ \\ 
$q\bar{Q}gQ$ & $\Xt(i_q^h,j_{\bar{Q}},k_g,l_Q^h)$  & $\Et(i^h,j,k,l^h)$ \\ 
\underline{Gluon-gluon} & & \\
$gggg$ & $\X(i_g^h,j_g,k_g,l_g^h)$ & $\F(i^h,j,k,l^h)$  \\
       & $\Xt(i_g^h,j_g,k_g,l_g^h)$ & $\Ft(i^h,j,k,l^h)$  \\
$g\bar{Q}Qg$ & $\X(i_g^h,j_{\bar{Q}},k_Q,l_g^h)$  & $\Ga(i^h,j,k,l^h)$ \\ 
$gg\bar{Q}Q$ & $\X(i_g^h,j_g,k_{\bar{Q}},l_Q^h)$  & $\Gb(i^h,j,k,l^h)$ \\ 
$g\bar{Q}gQ$ & $\Xt(i_g^h,j_{\bar{Q}},k_g,l_Q^h)$  & $\Gt(i^h,j,k,l^h)$ \\ 
$\bar{q}q\bar{Q}Q$ & $\X(i_{\bar{q}}^h,j_q,k_{\bar{Q}},l_Q^h)$ & $\H(i^h,j,k,l^h)$  \\
\end{tabular}
\caption{Identification of $\X$ antenna according to the particle type. These antennae only contain singular limits when one or both of particles $b$ and $c$ (or equivalently momenta $j$ and $k$) are unresolved. Antennae are classified as quark-antiquark, quark-gluon and gluon-gluon according to the particle type of the parents (i.e. after the antenna mapping)}
\label{tab:X40}
\end{table}

For double-real-radiation antenna functions, we have to distinguish between the case where the two unresolved particles are colour connected (which we denote by 
$\X(i^h_a,j_b,k_c,l^h_d)$) and the case where they are not (which we denote by $\Xt(i^h_a,j_b,k_c,l^h_d)$).
The list of limits included in each case is different, due to different possible double- and simple-collinear limits. Specifically, in $\Xt$ there are no $bc$-collinear limits, but there are $ac$- and $bd$-collinear limits, which are absent in $\X$.

As at NLO, we systematically start from the most singular limit and build the list of target functions using double- and single-unresolved limits. The list of limits for the 
$\X(i^h_a,j_b,k_c,l^h_d)$ double-real antenna function, from most singular to least singular, is given by
\begin{equation}
\begin{split}
    L_1(i^h,j,k,l^h) &= S_{bc}(i^h,j,k,l^h) \, , \\
    L_2(i^h,j,k;l^h) &= P_{abc}(i^h,j,k) \, , \\
    L_3(i^h,j,k,l^h) &= P_{dcb}(l^h,k,j) \, , \\
    L_4(i^h,j,k,l^h) &= P_{ab}(i^h,j) P_{dc}(l^h,k) \, ,  \\
    L_5(i^h,j,k,l^h) &= S_{b}(i^h,j,k) \, X_{3}^{0}(i^h,k,l^h) \, , \\
    L_6(i^h,j,k,l^h) &= S_{c}(j,k,l^h) \, X_{3}^{0}(i^h,j,l^h) \, , \\
    L_7(i^h,j,k,l^h) &= P_{ab}(i^h,j) \, X_{3}^{0}((i+j)^h,k,l^h) \, , \\
    L_8(i^h,j,k,l^h) &= P_{bc}(j,k) \, X_{3}^{0}(i^h,(j+k),l^h) \, , \\
    L_9(i^h,j,k,l^h) &= P_{dc}(l^h,k) \, X_{3}^{0}(i^h,j,(l+k)^h) \, ,
\end{split}
\label{eq:listX40}
\end{equation}
where for readability, we have suppressed the labels for the particle types in $X_3^0$.
Here, $L_1$ contains the double soft contribution, $L_2$ and $L_3$ the triple-collinear contributions, $L_4$ the double-collinear contribution, $L_5$ and $L_6$ the single-soft limits, and $L_7$, $L_8$ and $L_9$ the simple-collinear limits. The specific choice of $X_3^0$ in $L_5$ -- $L_9$ is fixed by the flavour structure of the relevant single-unresolved limit.
Because the iterative procedure is organised such that there are no overlaps between contributions of the same level, the ordering of limits of the same type is not important -- e.g. between $L_2$ and $L_3$.
For the 
$\Xt(i^h_a,j_b,k_c,l^h_d)$ the list of limits is given by
\begin{equation}
\begin{split}
    \tilde{L}_1(i^h,j,k,l^h) &= S_{bc}(i^h,j,k,l^h) \, , \\
    \tilde{L}_2(i^h,j,k;l^h) &= P_{abc}(i^h,j,k) \, , \\
    \tilde{L}_3(i^h,j,k,l^h) &= P_{dcb}(l^h,k,j) \, , \\
    \tilde{L}_4(i^h,j,k,l^h) &= P_{ab}(i^h,j) P_{dc}(l^h,k) \, , \\
    \tilde{L}_5(i^h,j,k,l^h) &= P_{ac}(i^h,k) P_{db}(l^h,j) \, , \\
    \tilde{L}_6(i^h,j,k,l^h) &= S_{b}(i^h,j,l^h) \, X_{3}^{0}(i^h,k,l^h) \, , \\
    \tilde{L}_7(i^h,j,k,l^h) &= S_{c}(i^h,k,l^h) \, X_{3}^{0}(i^h,j,l^h) \, , \\
    \tilde{L}_8(i^h,j,k,l^h) &= P_{ab}(i^h,j) X_{3}^{0}((i+j)^h,k,l^h) \, , \\
    \tilde{L}_9(i^h,j,k,l^h) &= P_{ac}(i^h,k) X_{3}^{0}((i+k)^h,j,l^h) \, , \\
    \tilde{L}_{10}(i^h,j,k,l^h) &= P_{dc}(l^h,k) X_{3}^{0}(i^h,j,(l+k)^h) \, , \\
    \tilde{L}_{11}(i^h,j,k,l^h) &= P_{db}(l^h,j) X_{3}^{0}(i^h,k,(l+j)^h) \, .
\end{split}
\label{eq:listX40t}
\end{equation}
Note that: 
\begin{itemize}
\item Neither Eq.~\eqref{eq:listX40} nor Eq.~\eqref{eq:listX40t} includes an explicit soft-collinear limit. These limits are present (and are verified after $\X$ construction) but arise naturally from the combination of double-soft and triple-collinear limits.
\item Because the antenna functions are built from spin-averaged splitting functions, there are no azimuthal correlations in the constructed $\X$.
\end{itemize}

The tree-level double-soft factors are given by 
\begin{eqnarray}
\label{eq:Sggtwo}
\Sgg(i^h,j,k,l^h) &=&
\frac{2 s_{ik}}{s_{ij}s_{jk}} 
\frac{2 s_{il}}{s_{ijk}s_{jkl}}
+
\frac{2 s_{jl}}{s_{kl}s_{jk}} 
\frac{2 s_{il}}{s_{ijk}s_{jkl}}
\nonumber \\
&&+ \frac{2s_{il} \Tr{i}{l}{j}{k} }{s_{ij}s_{jk}s_{kl}s_{ijk}s_{jkl}}
+ \frac{2\ome 
\left(s_{ij}s_{kl}-s_{ik}s_{jl}\right)^2}
{s_{jk}^2s_{ijk}^2s_{jkl}^2}, \\
\label{eq:Spp}    
\Spp(i^h,j,k,l^h) &=& \frac{2 s_{il}}{s_{ij}s_{jl}}
\frac{2 s_{il}}{s_{ik}s_{kl}}, \\
\label{eq:Sqq}
\Sqq(i^h,j,k,l^h) &=&
\frac{2 s_{il}}{s_{jk}s_{ijk}s_{jkl}} 
-\frac{2
\left(s_{ij}s_{kl}-s_{ik}s_{jl}\right)^2}{s_{jk}^2s_{ijk}^2s_{jkl}^2},
\end{eqnarray}
and zero otherwise.
Here we write the two-gluon soft factor in a suggestive manner such that the first two terms are iterations of the single-soft-gluon eikonal factor and
\begin{equation}
\Tr{i}{l}{j}{k} = s_{il}s_{jk}-s_{ij}s_{lk}+s_{ik}s_{lj} \, .
\end{equation}

The triple-collinear splitting functions $P_{abc}(i^h,j,k)$, when particle $a$ is hard, are organised according to the notation of Ref.~\cite{paper1}.  We exploit the decomposition into strongly-ordered iterated contributions (which are products of the usual spin-averaged two-particle splitting functions) and a remainder function $R_{abc\to P} (i,j,k)$ that is finite when any pair of $\{i,j,k\}$ are collinear. Here we write the splitting functions for which the organisation is not trivial while the others are given in Appendix~\ref{app:Pabc}. The three-gluon splitting function is split into three functions corresponding to one of the gluons being hard, 
\begin{align}
\label{eq:Pgggdecomp1}
P_{ggg}(i,j,k) &= P_{ggg}(i^h,j,k)
+P_{ggg}(i,j^h,k)
+P_{ggg}(i,j,k^h),
\end{align}
with 
\begin{align}
\Pggg(i^h,j,k) &= 
\frac{\PggS(x_k)}{s_{ijk}}  
\frac{\PggS\left(\frac{x_j}{1-x_k} \right)}{s_{ij}} +
\frac{\PggS(1-x_i)}{s_{ijk}}  
\frac{\Pgg\left(\frac{x_j}{1-x_i} \right)}{s_{jk}} \nonumber \\ 
& + 
\frac{1}{s_{ijk}^2} \Rgggsub(i,j,k) \, ,\\
\Pggg(i,j^h,k) &= 
\frac{\PggS(x_k)}{s_{ijk}}  
\frac{\PggS\left(\frac{x_i}{1-x_k} \right)}{s_{ij}}
+
\frac{\PggS(x_i)}{s_{ijk}}  
\frac{\PggS\left(\frac{x_k}{1-x_i} \right)}{s_{jk}} \, ,\\ \label{eq:Pgpp}
\Pggg(i,j,k^h) &= \Pggg(k^h,j,i) \, ,
\end{align}
where the momentum fraction is defined with respect to the fourth particle in the antenna,
\begin{equation}
x_k = \frac{s_{kl}}{s_{il}+s_{jl}+s_{kl}} \, ,
\end{equation}
and $\Rgggsub(i,j,k)$ is given in Ref.~\cite{paper1} and Eq.~\eqref{eq:Rggg} while $\PggS$ is defined in \eqref{eq:PggS}.
There are also two distinct splitting functions representing the clustering of a gluon with a quark-antiquark pair into a parent gluon. 
The splitting function for a gluon collinear with a quark-antiquark pair, when the gluon is colour-connected to the antiquark, is given by
\begin{align}
\label{eq:Pgqbqdecomp1}
\Pgqbq(i,j,k) &= \Pgqbq(i^h,j,k)
+\Pgqbq(i,j^h,k)
+\Pgqbq(i,j,k^h),
\end{align}
with
\begin{align}
\Pgqbq(i^h,j,k) &=
\frac{\PggS(1-x_i)}{s_{ijk}}
\frac{\Pqq\left(\frac{x_k}{1-x_i}\right)}{s_{jk}} + \frac{1}{s_{ijk}^2} \Rgqbq(i,j,k) \, , \\
\Pgqbq(i,j^h,k) &= 0 \, , \\
\Pgqbq(i,j,k^h) &= \frac{\Pqq(x_k)}{s_{ijk}} 
\frac{\Pqg\left(\frac{x_i}{1-x_k}\right)}{s_{ij}} 
+ \frac{\PggS(x_i)}{s_{ijk}} 
\frac{\Pqq\left(\frac{x_k}{1-x_i}\right)}{s_{jk}} \, ,
\label{eq:Pgqbqh}
\end{align}
where $\Rgqbq (i,j,k)$ is given in Ref.~\cite{paper1} and Eq.~\eqref{eq:Rgqbq}. Additionally, the projections of the double-soft factors in the triple-collinear phase space are given in Appendix~\ref{app:TCprojectionsofDS} and they indicate the overlap between the two limits.

The three-particle antennae appearing in Eqs.~\eqref{eq:listX40} and \eqref{eq:listX40t} are those discussed in Section~\ref{sec:single-real-antennae}.  For example, $X_3^0(i+j,k,l)$ denotes the antenna with particle types $(ab)$, $c$ and $d$ according to Table~\ref{tab:X30} carrying momenta $i+j$, $k$ and $l$ respectively. In Appendix~\ref{app:X40limits}, we list in full the limits for each $\X$, for convenient reference. 

Together with the list of limits, we need to have a procedure for mapping the antenna into the particular limit subspace, and then returning to the full phase space.
We define the double-soft down-projector for particles $j$ and $k$ soft as
\begin{equation}
    \PDSdown_{jk}: \begin{cases}
        s_{jk} \mapsto \lambda^2 s_{jk}, \\
        s_{ij} \mapsto \lambda s_{ij}, \,
        s_{ik} \mapsto \lambda s_{ik}, \,
        s_{jl} \mapsto \lambda s_{jl}, \,
        s_{kl} \mapsto \lambda s_{kl}, \\
        s_{ijk} \mapsto \lambda s_{ijk}, \,
        s_{jkl} \mapsto \lambda s_{jkl}, \\
        s_{ijl} \mapsto s_{il}, \,
        s_{ikl} \mapsto s_{il}, \,
        s_{ijkl} \mapsto s_{il},
    \end{cases}
\end{equation}
and we keep only terms proportional to $\lambda^{-4}$. As at NLO, for the corresponding up-projector $\PDSup_{jk}$ we choose a trivial mapping which leaves all variables unchanged.

We define the triple-collinear down-projector for collinear particles $i$, $j$, and $k$ as
\begin{equation}
    \PTCdown_{ijk}: \begin{cases}
        s_{ij} \mapsto \lambda s_{ij}, \,
        s_{ik} \mapsto \lambda s_{ik}, \,
        s_{jk} \mapsto \lambda s_{jk}, \,
        s_{ijk} \mapsto \lambda s_{ijk}, \\
        s_{ijkl} \mapsto s_{il}+s_{jl}+s_{kl}, \\
        s_{il} \mapsto \xi \left (s_{il}+s_{jl}+s_{kl}\right ),\,
        s_{jkl} \mapsto \omxi \left (s_{il}+s_{jl}+s_{kl}\right ), \\
        s_{jl} \mapsto \xj \left (s_{il}+s_{jl}+s_{kl}\right ),\,
        s_{ikl} \mapsto \omxj\left (s_{il}+s_{jl}+s_{kl}\right ), \\
        s_{kl} \mapsto \xk \left (s_{il}+s_{jl}+s_{kl}\right ),
        s_{ijl} \mapsto \omxk\left (s_{il}+s_{jl}+s_{kl}\right ),
    \end{cases}
\end{equation}
with $\xi+\xj+\xk = 1$ and we keep only terms proportional to $\lambda^{-2}$. Note that when $\omxi$ appears in the numerator, expressions are simplified according to $\omxi = \xj + \xk$, and so on. 
The corresponding up-projector we define as
\begin{equation}
    \PTCup_{ijk}: \begin{cases}
        \xi  \mapsto s_{il}/(s_{il}+s_{jl}+s_{kl}),\\ 
        \omxi  \mapsto s_{jkl}/(s_{il}+s_{jl}+s_{kl}),\\
        \xj  \mapsto s_{jl}/(s_{il}+s_{jl}+s_{kl}),\\ 
        \omxj  \mapsto s_{ikl}/(s_{il}+s_{jl}+s_{kl}),\\
        \xk  \mapsto s_{kl}/(s_{il}+s_{jl}+s_{kl}),\\ 
        \omxk  \mapsto s_{ijl}/(s_{il}+s_{jl}+s_{kl}),\\
        s_{il}+s_{jl}+s_{kl} \mapsto s_{ijkl}. \\
    \end{cases}
\end{equation}

The double-collinear down-projector for particles $i||j$ and $k||l$ we choose as
\begin{equation}
    \PDCdown_{ij;kl}: \begin{cases}
        s_{ij} \mapsto \lambda s_{ij}, \,
        s_{kl} \mapsto \mu s_{kl}, \\
        s_{il} \mapsto \omxj\omyk s_{ijkl}, \,
        s_{jl} \mapsto \xj\omyk s_{ijkl}, \\
        s_{ik} \mapsto \omxj\yk s_{ijkl}, \,
        s_{jk} \mapsto \xj\yk s_{ijkl}, \\
        s_{ijk} \mapsto \yk s_{ijkl} , \,
        s_{ijl} \mapsto \omyk s_{ijkl} , \,
        s_{ikl} \mapsto \omxj s_{ijkl}, \,
        s_{jkl} \mapsto \xj s_{ijkl}, \\
        s_{ijkl} \mapsto s_{ik}+s_{jk}+s_{il}+s_{jl}, \\
    \end{cases}
\end{equation}
and we keep only terms proportional to $\lambda^{-1}\mu^{-1}$ in order to count the divergences of both collinear limits separately.
The corresponding up projector is
\begin{equation}
    \PDCup_{ij;kl}: \begin{cases}
        \xj\yk \mapsto  s_{jk}/s_{ijkl}, \\
        \xj \mapsto (s_{jk}+s_{jl})/s_{ijkl}, \\
        \yk \mapsto (s_{ik}+s_{jk})/s_{ijkl}, \\
        1/\yk \mapsto  s_{ijkl}/s_{ijk} , \,
        1/\omyk \mapsto  s_{ijkl}/s_{ijl}  , \\
        1/\xj \mapsto s_{ijkl}/s_{jkl} , \,
        1/\omxj \mapsto  s_{ijkl}/s_{ikl}.
    \end{cases}
\end{equation}

The single-soft down-projector acts on invariants as
\begin{equation}
    \PSdown_j: \begin{cases} 
    s_{ij} \mapsto \lambda s_{ij}, \,
    s_{jk} \mapsto \lambda s_{jk}, \,
    s_{jl} \mapsto \lambda s_{jl}, \\
    s_{ijk} \mapsto s_{ik}, \,
    s_{ijl} \mapsto s_{il}, \,
    s_{jkl} \mapsto s_{kl}, \\
    s_{ijkl} \mapsto s_{ikl},
    \end{cases}
\end{equation}
and keeps only terms proportional to $\lambda^{-2}$.
For the corresponding up-projector $\PSup_j$, we again choose a trivial mapping which leaves all variables unchanged, in line with the choice for the single-soft up-projector in Section~\ref{sec:single-real-antennae}.

The simple-collinear projector we choose as,
\begin{equation}
    \PCdown_{ij}: \begin{cases}
    s_{ij} \mapsto \lambda s_{ij}, \\
    s_{ik} \mapsto \omxj s_{ijk}, \,
    s_{jk} \mapsto \xj s_{ijk}, \\
    s_{il} \mapsto \omxj s_{ijl}, \,
    s_{jl} \mapsto \xj s_{ijl}, \\    
    s_{ijk} \mapsto s_{ik} + s_{jk}, \\
    s_{ijl} \mapsto s_{il} + s_{jl}, \\
    \end{cases}
\end{equation}
and keep only terms proportional to $\lambda^{-1}$, while its corresponding up-projector we define as,
\begin{equation}
    \PCup_{ij}: \begin{cases}
    \xj (s_{ik}+s_{jk}) \mapsto s_{jk}, \,\,
    \omxj (s_{ik}+s_{jk}) \mapsto s_{ik} \\
    \xj (s_{il}+s_{jl}) \mapsto s_{jl}, \,\,
    \omxj (s_{il}+s_{jl}) \mapsto s_{il} \\
    s_{ik}+s_{jk} \mapsto s_{ijk}, \\
    s_{il}+s_{jl} \mapsto s_{ijl}, \\
    \xj \omxj \mapsto s_{jk}s_{il}/(s_{ijk} s_{ijl}), \\
    \xj \mapsto s_{jk}/s_{ijk}.
    \end{cases}
\end{equation}
Note that there is an ambiguity on how to write $\xj$ and $\omxj$ in terms of invariants.  In particular, we use the following identities,
\begin{eqnarray}
\xj^2 &=& \xj -\xj\omxj, \\
\omxj^2 &=& 1-\xj -\xj\omxj,\\
\xj\omxj &\mapsto& \frac{s_{jk}s_{il}}{s_{ijk}s_{ijl}} \text{ or } \frac{s_{ik}s_{jl}}{s_{ijk}s_{ijl}},
\end{eqnarray}
to avoid repeated powers of triple invariants.  Additional identities are imposed to preserve the symmetry of the antenna. 

Note that even after defining all projectors, some ambiguity remains.
For example, if we consider what happens to a term like $s_{ijl}/s_{ijl}$ in the triple-collinear $ijk$ space, then 
\begin{alignat*}{2}
1 \equiv \frac{s_{il}+s_{jl}+s_{ij}}{s_{ijl}} 
&\stackrel{\PTCdown_{ijk}}{\longrightarrow} \frac{\xi+\xj}{\omxk} \equiv 1 &&\stackrel{\PTCup_{ijk}}{\longrightarrow} 1,
\end{alignat*}
while applying the process to the individual terms does not give back unity upon summation in the full phase space,
\begin{alignat*}{2}
\frac{s_{il}}{s_{ijl}} 
&\stackrel{\PTCdown_{ijk}}{\longrightarrow} \frac{\xi}{\omxk} &&\stackrel{\PTCup_{ijk}}{\longrightarrow} \frac{s_{il}}{s_{ijl}},\\
\frac{s_{jl}}{s_{ijl}} 
&\stackrel{\PTCdown_{ijk}}{\longrightarrow} \frac{\xj}{\omxk} &&\stackrel{\PTCup_{ijk}}{\longrightarrow} \frac{s_{jl}}{s_{ijl}},\\
\frac{s_{ij}}{s_{ijl}} 
&\stackrel{\PTCdown_{ijk}}{\longrightarrow} 0 
&&\stackrel{\PTCup_{ijk}}{\longrightarrow} 0.
\end{alignat*}
It is clear that the sum of the last three lines is equivalent to unity in the triple-collinear subspace. Exactly how one returns to the full space can introduce differences there.   However, these differences do not affect the limit in question, but do influence less singular limits. These are corrected by the iterative process which systematically moves from more singular to less singular limit and which guarantees that each limit is correctly described.  However, after the iterative construction of the antenna function is complete, there can be differences in finite terms that do not contribute to any limit.

For antennae where $i,j,k$ and $l$ are colour connected to adjacent particles, i.e. $i$ to $j$ to $k$ to $l$, we define the general antenna function as
\begin{equation}
\begin{split}
\X(i^h,j,k,l^h) &= \Dsoft(i^h,j,k,l^h) \\
    &\quad + \Tcol(i^h,j,k;l^h) + \Tcol(l^h,k,j;i^h) \\
    &\quad + \Dcol(i^h,j; k,l^h) \\
    &\quad + \Ssoft(i^h,j,k;l^h) + \Ssoft(l^h,k,j;i^h) \\
    &\quad + \Scol(i^h,j;k,l^h) + \Scol(j,k;i^h,l^h) + \Scol(l^h,k;j,i^h) \, ,
\end{split}
\label{eq:X40def}
\end{equation}
where the individual pieces are given by
\begin{align}
\label{eq:Xpieces}
  \Dsoft(i^h,j,k,l^h) &= \PSup_{jk} L_1(i^h,j,k,l^h) = L_1(i^h,j,k,l^h) \, , \\
  \Tcol(i^h,j,k;l^h) &= \PTCup_{ijk}\left(L_2(i^h,j,k,l^h) 
  - \PTCdown_{ijk} \Dsoft(i^h,j,k,l^h)\right) \, , \\
  \Tcol(l^h,k,j;i^h) &= \PTCup_{lkj}\left(L_3(i^h,j,k,l^h) 
  - \PTCdown_{lkj} \left(\Dsoft(i^h,j,k,l^h) + \Tcol(i^h,j,k;l^h)\right)\right) \nonumber \\
                   &\equiv \PTCup_{lkj}\left(L_3(i^h,j,k,l^h) - \PTCdown_{lkj} \Dsoft(i^h,j,k,l^h)\right) \, , \\
  \Dcol(i^h,j; k,l^h) &= \PDCup_{ij;kl}\left(L_4(i^h,j,k,l^h) - \PDCdown_{ij;kl} X_{4;3}^{0}(i^h,j,k,l^h)\right) \, , \\
  \Ssoft(i^h,j,k;l^h) &= \PSup_{j}\left(L_5(i^h,j,k,l^h) - \PSdown_{j} X_{4;4}^{0}(i^h,j,k,l^h))\right) \, , \\
  \Ssoft(j,k,l^h;i^h) &= \PSup_{k}\left(L_6(i^h,j,k,l^h) - \PSdown_{k} \left(X_{4;4}^{0}(i^h,j,k,l^h)+\Ssoft(i^h,j,k;l^h)\right)\right) \nonumber \\
  &\equiv \PSup_{k}\left(L_6(i^h,j,k,l^h) - \PSdown_{k} X_{4;4}^{0}(i^h,j,k,l^h)\right) \, , \\
  \Scol(i^h,j;k,l^h) &= \PCup_{ij}\left(L_7(i^h,j,k,l^h) - \PCdown_{ij} X_{4;6}^{0}(i^h,j,k,l^h)\right) \, , \\
  \Scol(j,k;i^h,l^h) &= \PCup_{jk}\left(L_8(i^h,j,k,l^h) - \PCdown_{jk} \left (X_{4;6}^{0}(i^h,j,k,l^h)+ \Scol(i^h,j;k,l^h)\right)\right) \nonumber \\
  &\equiv \PCup_{jk}\left(L_8(i^h,j,k,l^h) - \PCdown_{jk} X_{4;6}^{0}(i^h,j,k,l^h)\right) \, , \\
  \Scol(l^h,k;j,i^h) &= \PCup_{kl}\left(L_9(i^h,j,k,l^h) - \PCdown_{kl} \left( X_{4;6}^{0}(i^h,j,k,l^h)+\Scol(i^h,j;k,l^h)\right.\right. \nonumber \\
  & \hspace{5cm}\left.\left. + \Scol(j,k;i^h,l^h)\right)\right) \nonumber \\
  &\equiv \PCup_{kl}\left(L_9(i^h,j,k,l^h) - \PCdown_{kl} X_{4;6}^{0}(i^h,j,k,l^h)\right) \, ,
\end{align}
where we have used
\begin{equation}
\begin{split}
\PTCdown_{lkj} \Tcol(i^h,j,k;l^h) &=0 \, , \\
\PSdown_{k} \Ssoft(i^h,j,k;l^h) &=0 \, , \\
\PCdown_{jk} \Scol(i^h,j;k,l^h) &=0 \, , \\
\PCdown_{kl} \Scol(i^h,j;k,l^h) &=0 \, , \\
\PCdown_{kl} \Scol(j,k;i^h,l^h) &=0 \, ,
\end{split}
\label{eq:Xpiecerelations1}
\end{equation}
together with
\begin{equation}
\begin{split}
X_{4;3}^{0}(i^h,j,k,l^h) &= \Dsoft(i^h,j,k,l^h) + \Tcol(i^h,j,k;l^h) + \Tcol(l^h,k,j;i^h) \, , \\
X_{4;4}^{0}(i^h,j,k,l^h) &= X_{4:3}^{0}(i^h,j,k,l^h) + \Dcol(i^h,j; k,l^h), \, , \\
X_{4;6}^{0}(i^h,j,k,l^h) &= X_{4:4}^{0}(i^h,j,k,l^h) + \Ssoft(i^h,j,k;l^h) + 
\Ssoft(j,k,l^h;i^h) \, .
\end{split}
\label{eq:Xpiecerelations2}
\end{equation}
An example of this type of antenna would be the leading-colour antenna with a quark and an antiquark as hard radiators, emitting two gluons. As for the $X_3^0$ antennae, the algorithm ensures that for all limits
\begin{equation}
    \PPdown_i \X(i^h,j,k,l^h) = L_i(i^h,j,k,l^h).
\end{equation}
In addition, $\X(i^h,j,k,l^h)$ has the correct soft-collinear limits. 

For subleading-colour antennae, where $j$ and $k$ are each colour connected to the hard radiators, and not to each other,  we define a general antenna function $\Xt$ as
\begin{equation}
\begin{split}
\Xt(i^h,j,k,l^h) & = \Dsoft(i^h,j,k,l^h) \\
    &\quad + \Tcol(i^h,{j},k;l^h) + \Tcol(l^h,k,j;i^h) \\
    &\quad + \Dcol(i^h,{j};k,l^h) + \Dcol(i^h,k; j,l^h) \\
    &\quad + \Ssoft(i^h,{j},l^h;k) + \Ssoft(i^h,k,l^h;j) \\
    &\quad + \Scol(i^h,{j};k,l^h) + \Scol(i^h,k;j,l^h) \\
    &\quad + \Scol(l^h,j;k,i^h) + \Scol(l^h,k;j,i^h).
\end{split}
\label{eq:X40tdef}
\end{equation}
where $\Dsoft$, $\Tcol$, $\Dcol$, $\Ssoft$ and $\Scol$ are defined analogously to Eq.~\eqref{eq:Xpieces} with a similar absence of overlaps between terms at the same level as in Eq.~\eqref{eq:Xpiecerelations1}.
An example of this type of antenna would be one with quark and antiquark hard radiators, emitting two photons. 
Again, the algorithm ensures that $\Xt$ satisfies
\begin{equation}
    \PPdown_i \Xt(i^h,j,k,l^h) = \tilde{L}_i(i^h,j,k,l^h)
\end{equation}
and has the correct soft-collinear limits.

We also give the analytic form of the four-particle antennae integrated over the fully inclusive $d$-dimensional antenna phase space~\cite{Gehrmann-DeRidder:2005btv,Gehrmann-DeRidder:2003pne}, 
\begin{equation}
\label{eq:phijklFF}
\calX(s_{ijkl}) =
\left(8\pi^2\left(4\pi\right)^{-\e} e^{\e\gamma}\right)^2
\int {\rm d}\, \Phi_{X_{ijkl}} \X,
\end{equation}
including a normalisation factor to account for powers of the QCD coupling constant.

\subsection{Quark-antiquark antennae}
As shown in Table~\ref{tab:X40}, there are four tree-level four-parton antennae with quark-antiquark parents that describe the emission of
\begin{enumerate}[(a)]
    \item two colour-connected gluons ($\A$),
    \item two photons, or equivalently two gluons that are not colour connected ($\At$),
    \item the emission of a quark-antiquark pair of different flavour to the radiators ($\B$),
    \item the emission of a quark-antiquark pair of the same flavour as the radiators ($\C$).
\end{enumerate}
These antenna functions can straightforwardly be obtained from matrix elements for $\gamma^*\to 4~\text{partons}$~\cite{Gehrmann-DeRidder:2005btv} and the antenna functions constructed here are directly related to the antenna functions given in Ref.~\cite{Gehrmann-DeRidder:2005btv} by
\begin{align}
\A(i_q^h, j_g, k_g, l_{\bar{q}}^h) &\sim
\Aold(i_q^h, j_g, k_g, l_{\bar{q}}^h),\\
\At(i_q^h, j_g, k_g, l_{\bar{q}}^h) &\sim
\Atold(i_q^h, j_g, k_g, l_{\bar{q}}^h),\\
\B(i_q^h, j_{\bar{Q}}, k_Q, l_{\bar{q}}^h) &\sim
\Bold(i_q^h, j_{\bar{Q}}, k_Q, l_{\bar{q}}^h),\\
\C(i_q^h, j_{\bar{q}}, k_q, l_{\bar{q}}^h) &\sim
\Cold(i_q^h, j_{\bar{q}}, k_q, l_{\bar{q}}^h),
\end{align}
in the sense that the new antennae should have the same double unresolved limits as the original antenna (as indicated by the $\sim$ symbol).

\begin{figure}[t]
    \centering
    \includegraphics[height=0.22\textheight,page=2]{A40n.pdf}
    \includegraphics[height=0.22\textheight,page=10]{A40n.pdf} 
    \includegraphics[height=0.22\textheight,page=6]{A40n.pdf} 
    \includegraphics[height=0.22\textheight,page=1]{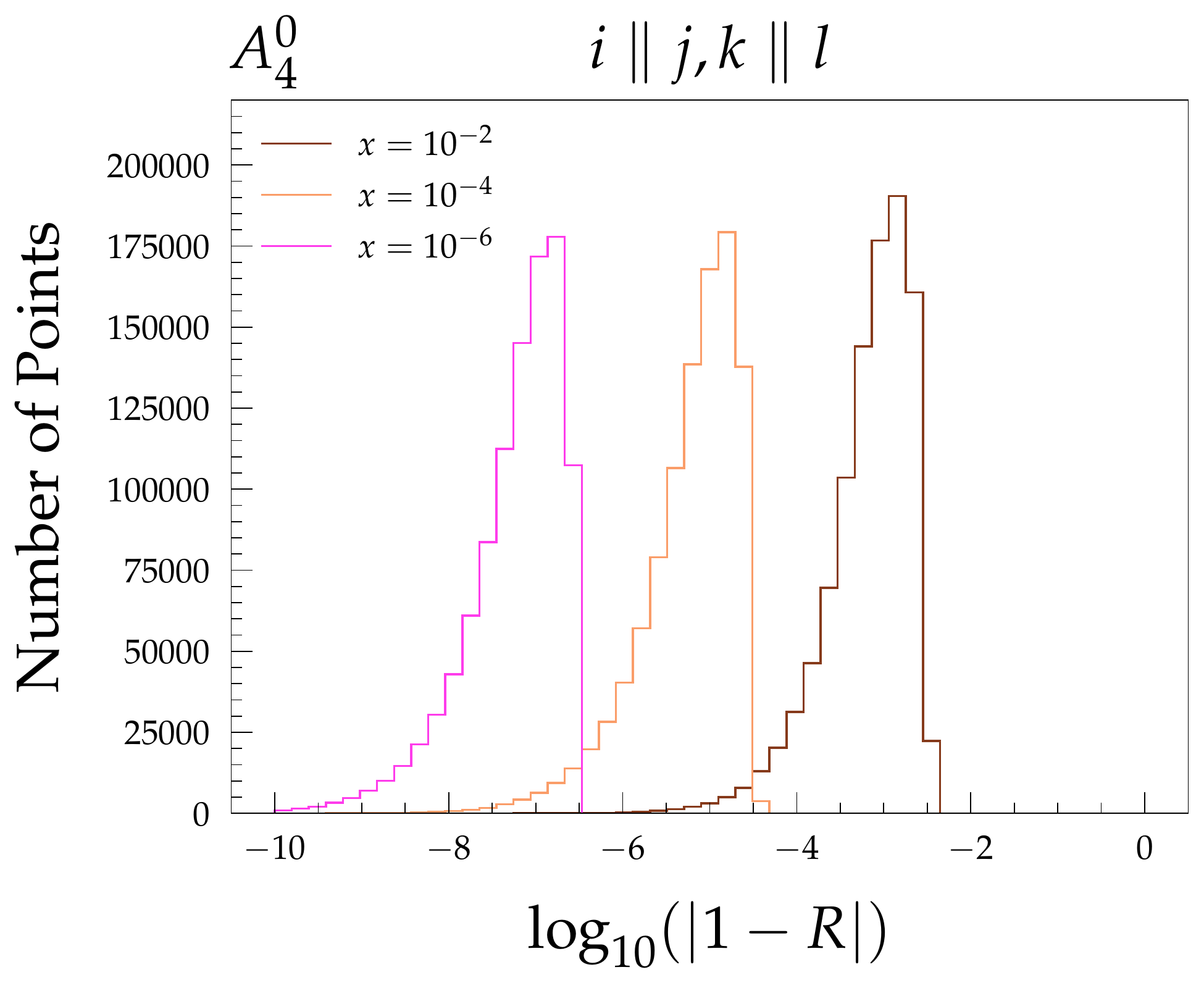}  
    \includegraphics[height=0.22\textheight,page=3]{A40n.pdf} 
    \includegraphics[height=0.22\textheight,page=4]{A40n.pdf} 
    \includegraphics[height=0.22\textheight,page=8]{A40n.pdf}    
    \caption{Numerical tests of the new $\A$ against $\Aold$ in all relevant singular limits. For three different values of the scaling parameter $x$, the relative disagreement of the ratio $R = \A/\Aold$ is shown on a logarithmic axis.}
    \label{fig:spiketests_A40}
\end{figure}

To build the $\A$ antenna using the algorithm described in Section~\ref{sec:algorithm}, we simply identify the particles in the list of required limits given in Eq.~\eqref{eq:listX40} -- $b$ and $c$ are gluons, while $a$ ($d$) are the quark (antiquark) hard radiators. In this case, the three-particle antennae are of type $A_3^0$.
The resulting expression for $\A$ is included in the auxiliary materials. It satisfies the line reversal property,
\begin{equation}
\A(i^h,j,k,l^h) = \A(l^h,k,j,i^h).
\end{equation}
As an example of the numerical behaviour of the new antenna functions, in Fig.~\ref{fig:spiketests_A40} we show numerical tests of the the new $\A$ against the original $\Aold$. 
We follow Ref. \cite{Pires:2010jv} and build trajectories into unresolved limits by scaling the relevant invariants by a fraction $x$ relative to the antenna invariant mass, $s_{ij\ldots} = x s_{ijkl}$. Due to the absence of azimuthal terms in our antenna functions, we combine phase-space points that are correlated by angular rotations about the collinear direction in every (multi-)collinear and soft-collinear limit.
Each histogram shows the relative agreement of $\A$ and $\Aold$ in digits,
\begin{equation}
    \log_{10}\left(\left\vert 1-R\right\vert\right) \text{ with } R = \frac{\A}{\Aold} \, .
\end{equation}
We wish to point out that due to the explicit line-reversal symmetry of $\A$, we only show representative examples for each limit. 
In all unresolved limits, we find percent-level agreement already for $x=10^{-2}$ and increasing agreement for smaller values of $x$.
Taking into account the different scaling behaviour of the double- and single-unresolved limits, it is evident that the antenna function develops quantitatively similar behaviour in approaching the singularity.

Integrating over the antenna phase space, we find 
\begin{eqnarray}
\label{eq:A40int}
\calA (s_{ijkl}) &=& \left( s_{ijkl} \right)^{-2\e}\Biggl [
+\frac{3}{4\e^4}
+\frac{65}{24\e^3}
+\frac{1}{\e^2} \left(
\frac{217}{18}
-\frac{13}{12}\pi^2
\right)
+\frac{1}{\e} \left(
\frac{44087}{864}
-\frac{589}{144}\pi^2
-\frac{71}{4}\zeta_3
\right)
\nonumber \\&& \hspace{2cm}
 + \left(
\frac{1134551}{5184}
-\frac{8117}{432}\pi^2
-\frac{1327}{18}\zeta_3
+\frac{373}{1440}\pi^4
\right)
 + \order{\e}\Biggr],
\end{eqnarray} 
which differs from $\calAold$ in Eq.~(5.31) of Ref.~\cite{Gehrmann-DeRidder:2005btv}, starting from the rational part at $\order{1/\e}$. This is completely understood and is simply because the $A_3^0$ given in Eq.~\eqref{eq:A30} differs at $\order{\e}$ from $\Xold{A}$ given in Eq.~(5.5) of Ref.~\cite{Gehrmann-DeRidder:2005btv}. 
The choice of $A_3^0$ impacts the algorithm at the stage of the single-soft limits, $L_{5,6} \sim \Sg A_3^0$. Integrating the  difference $2 \Sg(\Xold{A}-A_3^0)$ over the antenna phase space yields precisely the observed discrepancy of $1/\e$. Integrated forms of the four-particle antennae using the original set of three-particle antennae, $\Xold{X}$, derived directly from the squared matrix elements are listed in Appendix~\ref{app:intX4oldX3}. The pole structure of Eq.~\eqref{eq:A40intoldX3} agrees precisely with Eq.~(5.31) of Ref.~\cite{Gehrmann-DeRidder:2005btv} and differs only at $\order{\e^0}$. 
Integrals of the universal double-unresolved contributions in Eq.~\eqref{eq:X40def} for all $\X$ are given in Appendix~\ref{app:X40regionintegrations} for increased clarity. 

To build the $\At$ antenna, we identify $a$ ($d$) as the quark (antiquark) hard radiators and $b$ and $c$ as photons (or abelianised gluons). The resulting expression for $\At$ is included in the auxiliary materials. It is symmetric under both line reversal
\begin{equation}
\At(i^h,j,k,l^h) = \At(l^h,k,j,i^h)
\end{equation}
and exchange of the two gluons,
\begin{equation}
\At(i^h,j,k,l^h) = \At(i^h,k,j,l^h).
\end{equation}
Integrating over the antenna phase space, we find
\begin{eqnarray}
\label{eq:A40tint}
\calAt (s_{ijkl}) &=& \left( s_{ijkl} \right)^{-2\e}\Biggl [
+\frac{1}{\e^4}
+\frac{3}{\e^3}
+\frac{1}{\e^2} \left(
13
-\frac{3}{2}\pi^2
\right)
+\frac{1}{\e} \left(
\frac{861}{16}
-\frac{9}{2}\pi^2
-\frac{80}{3}\zeta_3
\right)
\nonumber \\&& \hspace{2cm}
 + \left(
\frac{7105}{32}
-\frac{39}{2}\pi^2
-80\zeta_3
+\frac{29}{120}\pi^4
\right)
 + \order{\e}\Biggr],
\end{eqnarray}
which differs from $\calAtold$ in Eq.~(5.32) of Ref.~\cite{Gehrmann-DeRidder:2005btv}, starting from the rational part at $\order{1/\e}$. The integrated form of $\At$ using the original set of three-particle antenna functions, $\Xold{X}$, is given in Eq.~\eqref{eq:A40tintoldX3}, where the pole structure agrees precisely with $\calAtold$ in Eq.~(5.32) of Ref.~\cite{Gehrmann-DeRidder:2005btv} and differs only at $\order{\e^0}$.

To construct $\B$ we identify $b$ and $c$ as a quark-antiquark pair (that is of different flavour to the hard radiators $a$ and $d$).  
The resulting expression for $\B$ is included in the auxiliary materials. It is symmetric under line reversal, 
\begin{equation}
\B(i^h,j,k,l^h) = \B(l^h,k,j,i^h),
\end{equation}
and exchange of the quark-antiquark pair,
\begin{equation}
\B(i^h,j,k,l^h) = \B(i^h,k,j,l^h).
\end{equation}
Integrating over the antenna phase space, we find
\begin{eqnarray}
\label{eq:B40int}
\calB (s_{ijkl}) &=& \left( s_{ijkl} \right)^{-2\e}\Biggl [
-\frac{1}{12\e^3}
-\frac{7}{18\e^2}
+\frac{1}{\e} \left(
-\frac{407}{216}
+\frac{11}{72}\pi^2
\right)
\nonumber \\&& \hspace{2cm}
+ \left(
-\frac{5917}{648}
+\frac{145}{216}\pi^2
+\frac{67}{18}\zeta_3
\right)
+ \order{\e}\Biggr],
\end{eqnarray}
which differs from Eq.~(5.39) of Ref.~\cite{Gehrmann-DeRidder:2005btv}, starting from the rational part of $\order{\e^0}$.

The $\C$ antenna has no double-soft or single-soft limits. It is therefore considerably simpler than the other quark-antiquark antennae. By construction, the only limit it contains is when particles $b$, $c$ and $d$ are collinear -- the $\bar{q}q\bar{q}$ triple-collinear limit for identical quarks.  
The resulting expression for $\C$ is included in the auxiliary materials. It is symmetric under the exchange of the two identical antiquarks in the triple collinear group, 
\begin{equation}
\C(i^h,j,k,l^h) = \C(i^h,l,k,j^h).
\end{equation}
This is as a direct result of the symmetry in the $\Pqqq$ splitting function (given in Appendix~\ref{app:Pabc}). 
Integrating over the antenna phase space, we find
\begin{eqnarray}
\label{eq:C40int}
\calC (s_{ijkl}) &=& \left( s_{ijkl} \right)^{-2\e}\Biggl [
+\frac{1}{\e} \left(
-\frac{13}{32}
+\frac{1}{16}\pi^2
-\frac{1}{4}\zeta_3
\right)
\nonumber \\&& \hspace{2cm}
 + \left(
-\frac{73}{16}
+\frac{23}{96}\pi^2
+\frac{23}{8}\zeta_3
-\frac{1}{45}\pi^4
\right)
 + \order{\e}\Biggr],
\end{eqnarray}
which differs from $\calCold$ in Eq.~(5.44) of Ref.~\cite{Gehrmann-DeRidder:2005btv}, starting from $\order{\e^0}$.

\subsection{Quark-gluon antennae}
The antenna functions for quark-gluon parents have been systematically derived from an effective Lagrangian describing heavy neutralino decay~\cite{Gehrmann-DeRidder:2005svg}.  There are three different configurations corresponding to the tree level processes $\widetilde{\chi} \to \widetilde{g}ggg$ (labelled $\Dold$) and $\widetilde{\chi} \to \widetilde{g}q\bar{q}g$ (with leading-colour and sub-leading colour antennae labelled $\Eold$ and $\Etold$ respectively).  Because these were based on matrix elements, the $\Dold$ and $\Eold$ antennae did not strictly follow the design principles laid out in Section~\ref{sec:design-principles}.  In particular, the antennae did not clearly specify which particles should be hard radiators and over-included unresolved limits that are not desirable. In Ref.~\cite{Gehrmann-DeRidder:2007foh} work was done to divide both $\D$ and $\E$ into sub-antennae with better properties, however yielding functions that were not analytically integrable.
Here we derive antennae that contain specific limits using the algorithm.
As indicated in Table~\ref{tab:X40}, there are five antennae in total:
\begin{itemize}
\item Two types of antennae describe the $qggg$ system:
$\D(i_q^h,j_g, k_g, l_g^h)$ and $\Dt(i_q^h,j_g, k_g, l_g^h)$.
In $\D$, the two unresolved gluons are colour-connected, while in $\Dt$ they are disconnected.  
In terms of the antennae of Ref.~\cite{Gehrmann-DeRidder:2005btv}, 
\begin{equation}
\Dold(i_q,j_g,k_g,l_g) \sim
  \D(i^h,j,k,l^h)
+ \D(i^h,l,k,j^h)
+ \Dt(i^h,j,l,k^h) \, .
\label{eq:Doldtonew}
\end{equation}
\item Three types of antennae describe the $q\bar{Q}{Q}g$ system:
$\Ea(i_q^h,j_{\bar{Q}},k_{Q},l_g^h)$,
$\Eb(i_q^h,j_g,k_{\bar{Q}},l_{Q}^h)$
and
$\Et(i_q^h,j_{\bar{Q}},k_g,l_{Q}^h)$.
At leading colour, two configurations are necessary: $\Ea$ in which the $Q\bar{Q}$ pair can be soft and the gluon is a hard radiator, and $\Eb$ where the gluon can be soft.
The soft-gluon singularities are therefore all contained in $\Eb$, but the triple-collinear $gQ\bar{Q}$ singularities are distributed between $\Ea$ and $\Eb$ as in Eq.~\ref{eq:Pgqbqdecomp1}.
At sub-leading colour, only one antenna is needed. 
These antennae are related to the antennae of Ref.~\cite{Gehrmann-DeRidder:2005btv} by,
\begin{align}
\Eold(i_q,j_{\bar Q},k_Q,l_g) &\sim
  \Ea(i^h,j,k,l^h)
+ \Eb(i^h,l,k,j^h) \\
\Etold(i_q,j_{\bar Q},k_Q,l_g) & \sim
  \Et(i^h,j,l,k^h).
\end{align}
\end{itemize}
To build the antennae using the algorithm, we simply identify the particles in the list of required limits given in Eq.~\eqref{eq:listX40}. For the $\D$ and $\Dt$ antennae, $a$ is a quark and $b$, $c$, and $d$ are gluons. For $\D$ the double-soft limit is $\Sgg$ while for $\Dt$ the double-soft limit is $\Spp$.
The colour-connected triple-gluon collinear limit is shared between the three antennae in Eq.~\eqref{eq:Doldtonew} according to the decomposition in Eq.~\eqref{eq:Pgggdecomp1}. The three-particle antennae that appear in the single-unresolved limits can be either of type $A_3^0$ or type $D_3^0$.

The resulting expressions for $\D$ and $\Dt$ are included in the auxiliary materials. $\D$ has no symmetries, while $\Dt$ is symmetric under exchange of the unresolved abelianised gluons,
\begin{equation}
\Dt(i^h,j,k,l^h) = \Dt(i^h,k,j,l^h).
\end{equation}
Note, in particular, that $\Dt(i^h,j,k,l^h)$ encapsulates the triple-collinear limits $\Pqpp(i^h,j,k)$ and $\Pggg(j,l^h,k)$, as given in Eq.~\eqref{app:Pqpp} and Eq.~\eqref{eq:Pgpp} respectively.
Integrating over the antenna phase space, we find
\begin{eqnarray}
\label{eq:D40int}
\calD (s_{ijkl}) &=& \left( s_{ijkl} \right)^{-2\e}\Biggl [
+\frac{3}{4\e^4}
+\frac{71}{24\e^3}
+\frac{1}{\e^2} \left(
\frac{118}{9}
-\frac{13}{12}\pi^2
\right)
+\frac{1}{\e} \left(
\frac{11849}{216}
-\frac{35}{8}\pi^2
-\frac{35}{2}\zeta_3
\right)
\nonumber \\&& \hspace{2cm}
 + \left(
\frac{74369}{324}
-\frac{8579}{432}\pi^2
-\frac{5473}{72}\zeta_3
+\frac{9}{32}\pi^4
\right)
 + \order{\e}\Biggr], \\
\label{eq:D40tint}
\calDt (s_{ijkl}) &=& \left( s_{ijkl} \right)^{-2\e}\Biggl [
+\frac{1}{\e^4}
+\frac{10}{3\e^3}
+\frac{1}{\e^2} \left(
\frac{29}{2}
-\frac{3}{2}\pi^2
\right)
+\frac{1}{\e} \left(
\frac{26749}{432}
-5\pi^2
-\frac{83}{3}\zeta_3
\right)
\nonumber \\&& \hspace{2cm}
 + \left(
\frac{113227}{432}
-\frac{1045}{48}\pi^2
-\frac{818}{9}\zeta_3
+\frac{19}{120}\pi^4
\right)
 + \order{\e}\Biggr].
\end{eqnarray}
The combination $2 \calD + \calDt$ agrees with the pole structure for $\calDold$ (given in Eq.~(6.45) of Ref.~\cite{Gehrmann-DeRidder:2005btv}), up to the rational part at $\order{1/\e^2}$.  This is because the $D_3^0$ antenna differs at $\order{\e^0}$ from the three-particle antenna, $\Xold{d}$, used in Ref.~\cite{Gehrmann-DeRidder:2005btv}.  The integrated forms of $\D$ and $\Dt$ using the original set of three-particle antenna, given in Appendix~\ref{app:intX4oldX3}, restore the agreement with the pole structure of Eq.~(6.45) of Ref.~\cite{Gehrmann-DeRidder:2005btv} through to $\order{\e^0}$.

\begingroup
The expressions for $\Ea$ and $\Eb$ are given in the auxiliary materials and have no symmetries. Integrating over the antenna phase space yields
\begin{eqnarray}
\label{eq:E40aint}
\calEa (s_{ijkl}) &=& \left( s_{ijkl} \right)^{-2\e}\Biggl [
-\frac{1}{12\e^3}
-\frac{5}{12\e^2}
+\frac{1}{\e} \left(
-\frac{1463}{864}
+\frac{1}{8}\pi^2
\right)
\nonumber \\&& \hspace{2cm}
+ \left(
-\frac{38401}{5184}
+\frac{77}{108}\pi^2
+\frac{20}{9}\zeta_3
\right)
 + \order{\e}\Biggr], \\
\label{eq:E40bint}
\calEb (s_{ijkl}) &=& \left( s_{ijkl} \right)^{-2\e}\Biggl [
-\frac{1}{3\e^3}
-\frac{35}{24\e^2}
+\frac{1}{\e} \left(
-\frac{5537}{864}
+\frac{1}{2}\pi^2
\right)
\nonumber \\&& \hspace{2cm}
+ \left(
-\frac{47345}{1728}
+\frac{35}{16}\pi^2
+\frac{80}{9}\zeta_3
\right)
 + \order{\e}\Biggr].
\end{eqnarray}
The combination $\calEa + \calEb$ agrees with the pole structure of $\calEold$ (given in Eq.~(6.51) of Ref.~\cite{Gehrmann-DeRidder:2005btv}), up to the rational part at $\order{1/\e^2}$. This is because the $E_3^0$ antenna differs at $\order{\e^0}$ from the three-particle antenna, $\Xold{E}$, used in Ref.~\cite{Gehrmann-DeRidder:2005btv}. Using the original set of three-particle antenna functions leads to the integrated four-particle antenna listed in Appendix~\ref{app:intX4oldX3}, which restores the agreement with the pole structure of $\calEold$ in Eq.~(6.51) of Ref.~\cite{Gehrmann-DeRidder:2005btv} through to $\order{\e^0}$.
\endgroup

The subleading-colour antenna $\Et (i_q^h,j_{\bar{Q}},k_g,l_{Q}^h)$ only contains singularities associated with the $\bar{Q}gQ$ cluster -- namely the triple-collinear limit, the soft-gluon limit, and the collinear limits $\bar{Q}g$ and $gQ$.  These limits are independent of the particle type of the first hard radiator $a$. The expression for $\Et$ is included in the auxiliary materials. $\Et$ is symmetric under exchange of the quark-antiquark pair,
\begin{equation}
\Et(i^h,j,k,l^h) = \Et(i^h,l,k,j^h).
\end{equation}
Integrating over the antenna phase space, we find
\begin{eqnarray}
\label{eq:E40tint}
\calEt (s_{ijkl}) &=& \left( s_{ijkl} \right)^{-2\e}\Biggl [
-\frac{1}{6\e^3}
-\frac{13}{18\e^2}
+\frac{1}{\e} \left(
-\frac{80}{27}
+\frac{1}{4}\pi^2
\right)
\nonumber \\&& \hspace{2cm}
+ \left(
-\frac{7501}{648}
+\frac{13}{12}\pi^2
+\frac{40}{9}\zeta_3
\right)
 + \order{\e}\Biggr].
\end{eqnarray}
The pole structure for $\calEt$ agrees with $\calEtold$ (given in Eq.~(6.52) of Ref.~\cite{Gehrmann-DeRidder:2005btv}), up to the rational part at $\order{1/\e^2}$. This is because the $E_3^0$ antenna differs at $\order{\e^0}$ from the three-particle antenna, $\Xold{E}$, used in Ref.~\cite{Gehrmann-DeRidder:2005btv}. Using as input the original three-particle antenna, $\Xold{E}$, restores the agreement through to $\order{\e^0}$.

\subsection{Gluon-gluon antennae}
Antenna functions for gluon-gluon parents have been systematically derived from the effective Lagrangian describing Higgs boson decay into gluons~\cite{Gehrmann-DeRidder:2005alt}. There are four possibilities, 
$ H \to gggg$ (labelled $\Fold$),
$H \to gg Q\bar{Q}$ (with leading-colour and sub-leading colour antennae labelled $\Gold$ and $\Gtold$ respectively)
and 
$H \to q\bar q Q \bar Q$ (labelled $\Hold$). The $\Fold$ and $\Gold$ antennae also did not follow the design principles laid out in Section~\ref{sec:design-principles}, by not clearly specifying which particles should be hard radiators and/or by over-including unresolved limits that are not desirable. Ref.~\cite{NigelGlover:2010kwr} reorganised the $\Fold$ antenna into $F^{4, \,a}_0$ and $F^{4, \,b}_0$ sub-antennae that had better properties, but at the cost of introducing composite denominators. Similar work was also carried out for the $\Gold$ antenna. 

Using our algorithm, we can build gluon-gluon antennae that do satisfy the design properties. As shown in Table~\ref{tab:X40} there are six antennae in total:
\begin{itemize}
\item Two types of antenna describe the $gggg$ system: $\F(i_g^h,j_g,k_g,l_g^h)$ and $\Ft(i_g^h,j_g,k_g,l_g^h)$. They are related to the antenna of Ref.~\cite{Gehrmann-DeRidder:2005btv} by,
\begin{align}
\label{eq:Foldtonew}
\Fold(i,j,k,l) &\sim
\F(i^h,j,k,l^h)
+ 3~{\rm cyclic~permutations} \nonumber \\
&+\Ft(i^h,j,l,k^h) 
+\Ft(l^h,i,k,j^h)
\end{align}
In $\F$ the two unresolved gluons are colour-connected, while in $\Ft$ they are disconnected. This means that the relevant double-soft factors are $\Sgg$ for $\F$ and $\Spp$ for $\Ft$. Each $ggg$ colour-connected triple-collinear limit is shared between three antennae in Eq.~\eqref{eq:Foldtonew} (two $\F$ and one $\Ft$) according to the decomposition in Eq.~\eqref{eq:Pgggdecomp1}. 
\item Three types of antenna describe the $g Q\bar{Q}g$ system:
$\Ga(i_g^h,j_Q,k_{\bar{Q}},l_g^h)$, 
$\Gb(i_g^h,j_g,k_Q,l_{\bar{Q}}^h)$ and
$\Gt(i_g^h,j_Q,k_g,l_{\bar{Q}}^h)$. 
At leading colour, two configurations are necessary: $\Ga$ in which $Q\bar{Q}$ pair can be soft and both gluons are hard radiators, and $\Gb$ where one of the gluons can be soft (and the other is a hard radiator).
The soft-gluon singularities are therefore all contained in $\Gb$, but the triple-collinear $gQ\bar{Q}$ singularities are distributed between $\Ga$ and $\Gb$. 
They are related to the antenna of Ref.~\cite{Gehrmann-DeRidder:2005btv} by,  
\begin{align}
\Gold(i_g,j_Q,k_{\bar{Q}},l_g) &\sim
\Ga(i^h,j,k,l^h) \nonumber \\
& + \Gb(l^h,i,j,k^h)
+ \Gb(i^h,l,k,j^h) \, .
\end{align}
At sub-leading colour, only one antenna, $\Gt (i_g^h,j_{\bar{Q}},k_g,l_Q^h)$,  is needed. There are no double-soft limits and only one triple collinear limit describing the $\bar{Q} g Q$ cluster. 
$\Gt$ is related to the antenna of Ref.~\cite{Gehrmann-DeRidder:2005btv} by,  
\begin{equation}
\Gtold(i_g,j_Q,k_{\bar{Q}},l_g) \sim
  \Gt(i^h,j,l,k^h)
+ \Gt(l^h,k,i,j^h) \, .
\end{equation}
\item Finally, one gluon-gluon antenna is needed to describe the $q\bar q Q\bar{Q}$ final state, called $\H(i_q^h,j_{\bar{q}},k_{Q},l_{\bar{Q}}^h)$, which is directly related to the analogous antenna in Ref.~\cite{Gehrmann-DeRidder:2005btv},
\begin{equation}
\Hold(i_{\bar{q}},j_{q},k_{\bar{Q}},l_{Q}) \sim
\H(i^h,j,k,l^h) \, .
\end{equation}
Note that only different quark flavours need to be considered, since the identical-flavour contribution to this final state is finite.
\end{itemize}
As for the previous antennae, we simply identify the particles in the list of required limits given in Eq.~\eqref{eq:listX40}. For the $\F$ and $\Ft$ antennae, all particles are gluons. For $\F$ the double-soft limit is $\Sgg$ while for $\Ft$ the double-soft limit is $\Spp$. The assignment of the triple-gluon splitting function to $\F$ and $\Ft$ exactly parallels the division for $\D$ and $\Dt$ and the triple-gluon collinear limit is shared between the two antennae according to the decomposition in Eq.~\eqref{eq:Pgggdecomp1}. The three-particle antennae that appear in the single-unresolved limits are all of type $F_3^0$.

The resulting expressions for $\F$ and $\Ft$ are included in the auxiliary materials. Both $\F$ and $\Ft$ satisfy the line reversal property,
\begin{align}
\F(i^h,j,k,l^h)  &= \F(l^h,k,j,i^h) \, ,\\
\Ft(i^h,j,k,l^h) &= \Ft(l^h,k,j,i^h) \, ,
\end{align}
while $\Ft$ is symmetric under the exchange of the two unresolved gluons,
\begin{equation}
\Ft(i^h,j,k,l^h) = \Ft(i^h,k,j,l^h) \, .
\end{equation}
After integration over the final-final antenna phase space, Eq.~\eqref{eq:phijkFF}, we find the following infrared pole structure, 
\begin{eqnarray}
\label{eq:F40int}
\calF (s_{ijkl}) &=& \left( s_{ijkl} \right)^{-2\e}\Biggl [
+\frac{3}{4\e^4}
+\frac{77}{24\e^3}
+\frac{1}{\e^2} \left(
\frac{511}{36}
-\frac{13}{12}\pi^2
\right)
+\frac{1}{\e} \left(
\frac{50801}{864}
-\frac{671}{144}\pi^2
-\frac{69}{4}\zeta_3
\right)
\nonumber \\&& \hspace{2cm}
 + \left(
\frac{415927}{1728}
-\frac{9059}{432}\pi^2
-\frac{2819}{36}\zeta_3
+\frac{437}{1440}\pi^4
\right)
 + \order{\e}\Biggr],\\
\label{eq:F40tint}
\calFt (s_{ijkl}) &=& \left( s_{ijkl} \right)^{-2\e}\Biggl [
+\frac{1}{\e^4}
+\frac{11}{3\e^3}
+\frac{1}{\e^2} \left(
\frac{289}{18}
-\frac{3}{2}\pi^2
\right)
+\frac{1}{\e} \left(
\frac{30347}{432}
-\frac{11}{2}\pi^2
-\frac{86}{3}\zeta_3
\right)
\nonumber \\&& \hspace{2cm}
 + \left(
\frac{785743}{2592}
-\frac{193}{8}\pi^2
-\frac{916}{9}\zeta_3
+\frac{3}{40}\pi^4
\right)
 + \order{\e}\Biggr].
\end{eqnarray}
The combination $4 \calF + 2\calFt$ agrees with the pole structure for $\calFold$ (given in Eq.~(7.45) of Ref.~\cite{Gehrmann-DeRidder:2005btv}), up to the rational part at $\order{1/\e^2}$. This is because the $F_3^0$ antenna differs at $\order{\e^0}$ from the three-particle antenna, $\Xold{f}$, used in Ref.~\cite{Gehrmann-DeRidder:2005btv}. Using as input the original three-particle antenna, $\Xold{f}$, restores the agreement through to $\order{\e^0}$.

The resulting expressions for $\G$ and $\Gb$ are included in the auxiliary materials. $\Gb$ has no symmetries, while $\Ga$ satisfies the line reversal property
\begin{eqnarray}
\Ga(i^h,j,k,l^h) &=& \Ga(l^h,k,j,i^h).
\end{eqnarray}
After integration over the antenna phase space, we find
\begin{eqnarray}
\label{eq:G40aint}
\calGa (s_{ijkl}) &=& \left( s_{ijkl} \right)^{-2\e}\Biggl [
-\frac{1}{12\e^3}
-\frac{4}{9\e^2}
+\frac{1}{\e} \left(
-\frac{649}{432}
+\frac{7}{72}\pi^2
\right)
\nonumber \\&& \hspace{2cm}
+ \left(
-\frac{1637}{288}
+\frac{163}{216}\pi^2
+\frac{13}{18}\zeta_3
\right)
 + \order{\e}\Biggr],\\
\label{eq:G40bint}
\calGb (s_{ijkl}) &=& \left( s_{ijkl} \right)^{-2\e}\Biggl [
-\frac{1}{3\e^3}
-\frac{109}{72\e^2}
+\frac{1}{\e} \left(
-\frac{5741}{864}
+\frac{1}{2}\pi^2
\right)
\nonumber \\&& \hspace{2cm}
+ \left(
-\frac{146651}{5184}
+\frac{109}{48}\pi^2
+\frac{80}{9}\zeta_3
\right)
 + \order{\e}\Biggr].
\end{eqnarray}
The combination $ \calGa + 2\calGb$ agrees with the pole structure for $\calGold$ (given in Eq.~(7.52) of Ref.~\cite{Gehrmann-DeRidder:2005btv}), up to the rational part at $\order{1/\e^2}$. This is because the $G_3^0$ antenna differs at $\order{\e^0}$ from the three-particle antenna,$\Xold{G}$, used in Ref.~\cite{Gehrmann-DeRidder:2005btv}. Using as input the original three-particle antenna restores the agreement through to $\order{\e^0}$.

As for the $\Et (i_q^h,j_{\bar{Q}},k_g,l_{Q}^h)$ antenna, the sub-leading colour antenna $\Gt (i_g^h,j_{\bar{Q}},k_g,l_Q^h)$ only contains singularities associated with the $\bar{Q}gQ$ cluster. These are the triple-collinear limit, the soft-gluon limit, and the collinear limits for $\bar{Q}g$ and $gQ$. 
These limits are independent of the particle type of the first hard radiator $a$. The expression for $\Gt$ is therefore the same as for $\Et$ and is included in the auxiliary materials. $\Gt$ is symmetric under exchange of the quark-antiquark pair,
\begin{equation}
\Gt(i^h,j,k,l^h) = \Gt(i^h,l,k,j^h).
\end{equation}
Integrating over the antenna phase space, we find
\begin{eqnarray}
\label{eq:G40tint}
\calGt (s_{ijkl}) &=& \left( s_{ijkl} \right)^{-2\e}\Biggl [
-\frac{1}{6\e^3}
-\frac{13}{18\e^2}
+\frac{1}{\e} \left(
-\frac{80}{27}
+\frac{1}{4}\pi^2
\right)
\nonumber \\&& \hspace{2cm}
+ \left(
-\frac{7501}{648}
+\frac{13}{12}\pi^2
+\frac{40}{9}\zeta_3
\right)
 + \order{\e}\Biggr].
\end{eqnarray}
The combination $2\calGt$ agrees with the pole structure for $\calGtold$ (given in Eq.~(7.52) of Ref.~\cite{Gehrmann-DeRidder:2005btv}), up to the rational part at $\order {1/\e^2} $. This is as expected because the $G_3^0$ antenna differs at $\order{\e^0}$ from the three-particle antenna, $\Xold{G}$, used in Ref.~\cite{Gehrmann-DeRidder:2005btv}. Using as input the original three-particle antenna restores the agreement through to $\order{\e^0}$.

The $\bar{q}q\bar{Q}Q$ antenna, $\H$, contains no double-soft, no triple-collinear, and no single-soft limits.
It is composed entirely from the limits where $q$ and $\bar{q}$ and/or $Q$ and $\bar{Q}$ are collinear. $\H$ is symmetric under the exchange of $q$ and $\qb$ and/or $Q$ and $\Qb$, as well as the interchange of the quark pairs,
\begin{align}
&\H(i,j,k,l) = \H(j,i,k,l) = \H(i,j,l,k) = \H(j,i,l,k)\nonumber \\
=&\H(k,l,i,j) = \H(k,l,j,i) = \H(l,k,i,j) = \H(l,k,j,i).
\end{align}

The resulting expression for $\H$ is given by
\begin{equation}
\begin{split}
\label{eq:H40}
\H(i_{\qb},j_{q},k_{\Qb},l_{Q}) &=
\frac{1}{s_{ij}s_{k l}} 
+ \frac{2\left(s_{ik}s_{jl}+s_{il}s_{jk}\right)}{s_{ij}s_{kl}s_{ijkl}^2\ome^2} \\
&
- \frac{2 ((s_{ik}+s_{jl})(s_{il}+s_{jk}) + 2s_{ik}s_{jl} + 2
s_{il}s_{jk})}{s_{ij}s_{kl}s_{ijkl}^2\ome} \\
& -\frac{\left(s_{ijkl}+s_{ik}+s_{jk}+s_{il}+s_{jl}\right)}
{ s_{ij}s_{ijkl}^2}
\times
\frac{\left(s_{ik}s_{jl}+s_{il}s_{jk}\right)}{s_{ijk}s_{ijl}\ome} \\
& - \frac{\left(s_{ijkl}+s_{ik}+s_{jk}+s_{il}+s_{jl}\right)}
{s_{kl}s_{ijkl}^2}
\times
\frac{\left(s_{ik}s_{jl}+s_{il}s_{jk}\right)}{s_{jkl}s_{ikl}\ome }.
\end{split}
\end{equation}
Integrating over the antenna phase space, we find
\begin{eqnarray}
\label{eq:H40int}
\calH (s_{ijkl}) &=& \left( s_{ijkl} \right)^{-2\e}\Biggl [
+\frac{1}{9\e^2}
+\frac{1}{2\e}
+ \left(
\frac{283}{162}
-\frac{17}{108}\pi^2
\right)
 + \order{\e}\Biggr].
\end{eqnarray}
This agrees with the pole structure for $\calHold$ (given in Eq.~(7.59) of Ref.~\cite{Gehrmann-DeRidder:2005btv}), up to the rational part at $\order {1/\e}$. 
It is instructive to compare the antenna generated by our algorithm, Eq.~\eqref{eq:H40}, with the result given in Ref.~\cite{Gehrmann-DeRidder:2005btv} derived from matrix elements, 
\begin{align}
\label{eq:H40old}
\Hold(i_{\bar{q}},j_{q},k_{\bar{Q}},l_{Q}) &= 
\frac{1}{s_{ijkl}^2} \Bigg[ 
\frac{2\left(s_{ik}s_{j l}-s_{i l}s_{jk}\right)^2}{s_{ij}^2s_{k l}^2 \ome} + \frac{\left(s_{ik}+s_{il}+s_{jk}+s_{jl}\right)^2}{s_{ij}s_{k l}} + \frac{2}{\ome} \nonumber\\
&\hspace{1.5cm} 
- \frac{2 ((s_{ik}+s_{j l})(s_{i l}+s_{jk}) + 2s_{ik}s_{j l} + 2
s_{i l}s_{jk})}{s_{ij}s_{k l}\ome}\Bigg ] \, .
\end{align}
Comparing Eq.~\eqref{eq:H40} with Eq.~\eqref{eq:H40old}, we make the following observations
\begin{itemize}
    \item The absence of double poles in $s_{ij}$ and $s_{k l}$ in Eq.~\eqref{eq:H40}.  This is because angular-averaged splitting functions were used to construct $\H$, thereby ensuring that azimuthal correlations are not present in the antenna.
    \item The presence of triple invariants in $s_{ijk}$ and $s_{jkl}$ in the denominator.  This is a consequence of the $\PCup$ projector, which defines the momentum fraction with respect to one of the other two particles in the antenna.    For example, for the $\PCup_{ij}$ projector, $\xj\omxj \to s_{il}s_{jk}/(s_{ijk}s_{ijl})$. 
    For most antennae, this is a natural choice, and generates singular structures that are already present in the triple-collinear limit. In this particular case, there is no triple-collinear limit, and this looks unnatural.

\end{itemize}

To demonstrate how the constructed antenna is affected by the choice of $X_3^0$ antenna in describing the single unresolved limits, Eq.~\eqref{eq:H40OLDX3} shows the expression generated by the algorithm for $\H$ using the original $\Xold{G}$ antenna for the single unresolved limits,
\begin{equation}
\begin{split}
\H(i_{\qb},j_{q},k_{\Qb},l_{Q})[\Xold{G}] &=
\frac{1}{s_{ij}s_{k l}} 
+ \frac{2\left(s_{ik}s_{jl}+s_{il}s_{jk}\right)}{s_{ij}s_{kl}s_{ijkl}^2\ome^2} \\
&
- \frac{2 ((s_{ik}+s_{jl})(s_{il}+s_{jk}) + 2s_{ik}s_{jl} + 2
s_{il}s_{jk})}{s_{ij}s_{kl}s_{ijkl}^2\ome} \\
& -\frac{\left(s_{ijkl}+s_{ik}+s_{jk}+s_{il}+s_{jl}\right)}
{s_{ij}s_{ijkl}^2} \\
&
  -\frac{\left(s_{ijkl}+s_{ik}+s_{jk}+s_{il}+s_{jl}\right)}
{s_{kl}s_{ijkl}^2} \, .
\end{split}
\label{eq:H40OLDX3}
\end{equation}
We see that the double unresolved contributions in the first two lines of Eq.~\eqref{eq:H40OLDX3} are exactly the same as the double unresolved contribution in Eq.~\eqref{eq:H40}.  The single unresolved contributions (third and fourth lines) are however different because of the choice of single-real antenna.
In the $i||j$ limit, the difference between Eq.~\eqref{eq:H40OLDX3} and Eq.~\eqref{eq:H40}  is proportional to
\begin{equation}
    \frac{1}{s_{ij}} \Pqq(\xj) \left( \Xold{G}([i+j],k,l)-G_3^{0}([i+j],k,l)\right)
\end{equation}
with
\begin{equation}
\label{eq:G3diff}
    \left( \Xold{G}([i+j],k,l)-G_3^{0}([i+j],k,l)\right) \sim 
    -\frac{\left(s_{ijkl}+s_{ik}+s_{jk}+s_{il}+s_{jl}\right)}
{s_{ijkl}^2}
\end{equation} 
and
\begin{equation}
    \Pqq(x_j) = 1 - \frac{2\xj\omxj}{\ome} \stackrel{\PCup_{ij}}{\longrightarrow} 1 - \frac{\left(s_{ik}s_{jl}+s_{il}s_{jk}\right)}{s_{ijk}s_{ijl}\ome}.
\end{equation}
The dependence on the momentum fraction $\xj$ in Eq.~\eqref{eq:H40} is precisely cancelled, leading to the absence of triple invariants in the denominator of Eq.~\eqref{eq:H40OLDX3}.

\begin{figure}[t]
    \centering
    \includegraphics[width=0.49\textwidth,page=1]{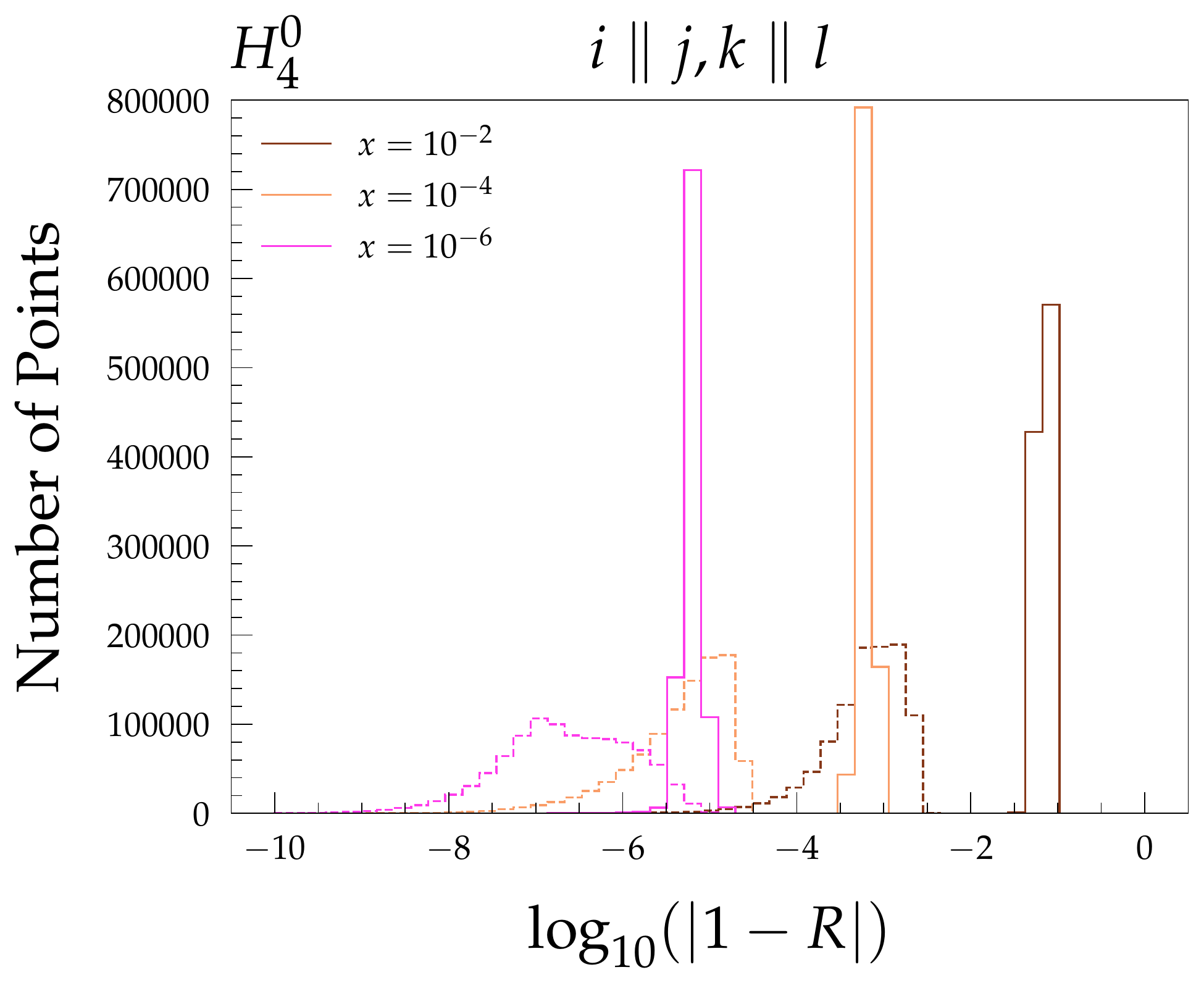}
    \includegraphics[width=0.49\textwidth,page=2]{H40n.pdf}
    \caption{Numerical tests of the new $\H$ against $\Hold$ in all relevant singular limits. Solid lines represent tests of $\H$ constructed with the new $G_3^0$ antennae and dashed lines represent tests with $\H$ constructed with $\Xold{G}$. For three different values of the scaling parameter $x$, the relative disagreement of the ratio $R = \H/\Hold$ is shown on a logarithmic axis.}
    \label{fig:spiketests_H40}
\end{figure}

In Fig.~\ref{fig:spiketests_H40}, we show numerical tests of $\H$ against $\Hold$ in the relevant double-unresolved and single-unresolved limits. The procedure is analogous to the one for Fig.~\ref{fig:spiketests_A40} and we show both versions of $\H$; the one constructed using the new form for $G_3^0$ in Eq.~\eqref{eq:H40} and the one using the original  $\Xold{G}$ antenna function in Eq.~\eqref{eq:H40OLDX3}.

In the double-collinear limit, both versions of $\H$ agree well with $\Hold$; as $x$ decreases they both describe the double-collinear limit increasingly correctly. We observe that the version constructed using the new $G_3^0$ antenna produces much sharper peaks at slightly larger values of $\log(|1-R|)$ compared to that based on the $\Xold{G}$ antenna. 

In the single-unresolved limits, we see much bigger differences; for the $\H$ using the $\Xold{G}$ antenna, there is good agreement, while there is very poor agreement for the $\H$ using the new $G_3^0$ antenna.  This is as expected!  In the single unresolved region, the $\Hold$ antennae function, which is based on $H \to q\qb Q\Qb$ matrix elements behaves as,
\begin{align}
    \Hold \sim &|\calM_4^0(i_{\qb},j_{q},k_{\Qb},l_{Q})|^2 \stackrel{i ||j}{\longrightarrow} \frac{1}{s_{ij}} \Pqq(\xj) |\calM_3^0([i+j]_g,k_{\Qb},l_{Q})|^2,
\end{align}
while by construction, the new $\H $ antenna behaves as,
\begin{align}
    \H \stackrel{i ||j}{\longrightarrow} \frac{1}{s_{ij}} \Pqq(\xj) G_3^0([i+j]_g,k_{\Qb},l_{Q}).
\end{align}
Since $\Xold{G}$ is constructed from the three particle matrix elements, $|\calM_3^0([i+j]_g,k_{Q},l_{\Qb})|^2$, the $\H$ antenna based on $\Xold{G}$ is guaranteed to have the correct limit. However, the finite $\order{\e^0}$ differences between $G_3^{0}$ and $\Xold{G}$ (see \eqref{eq:G3diff}) are multiplied by $1/s_{ij}$ and therefore lead to a different single unresolved limit for the $\H$ antenna using $G_3^{0}$. In fact, although these differences are sub-leading in the double unresolved region, they are responsible for the small numerical differences observed there.

Note that we do not require that the $\X$ antenna correctly describes the single-unresolved region; their role is to correctly describe the double unresolved limits.  The single-unresolved limits are always correctly described by other subtraction terms and when using an $\X$ antenna, one first has to subtract single-unresolved contributions. This leads to groups of subtraction terms like,
\begin{eqnarray}
\label{eq:H4subtraction}
   \H(i^h,j,k,l^h) - E_3^0(k,j,i)G_3^0(\widetilde{{ij}},\widetilde{{jk}},l)
        - E_3^0(j,k,l) G_3^0(\widetilde{{lk}},\widetilde{{kj}},i).
\end{eqnarray} 
Provided that the same $G_3^0$ is used for both the construction of $\H$ and in the analogues of Eq.~\eqref{eq:H4subtraction}, the single unresolved limits will be correctly described.

Additionally, we summarise the symmetry properties of the $\X$ constructed in this paper in Table~\ref{tab:X40sym}. 

\begin{table}[t]
\centering
\begin{tabular}{ccc}
$\X (1^h,2,3,4^h)$ & Symmetries \\
\hline
 $\A(1^h,2,3,4^h)$ & $(1,2,3,4) \leftrightarrow (4,3,2,1)$ \\
 $\At(1^h,2,3,4^h)$ & $(1,2,3,4) \leftrightarrow (4,3,2,1)$ and $(1,2,3,4) \leftrightarrow (1,3,2,4)$\\
 $\B(1^h,2,3,4^h)$ & $(1,2,3,4) \leftrightarrow (4,3,2,1)$ and $(1,2,3,4) \leftrightarrow (1,3,2,4)$ \\
 $\C(1^h,2,3,4^h)$ & $(1,2,3,4) \leftrightarrow (1,4,3,2)$ \\
$\D(1^h,2,3,4^h)$ & None  \\
 $\Dt(1^h,2,3,4^h)$ & $(1,2,3,4) \leftrightarrow (1,3,2,4)$   \\
$\Ea(1^h,2,3,4^h)$ & None  \\ 
 $\Eb(1^h,2,3,4^h)$ & None\\ 
 $\Et(1^h,2,3,4^h)$ & $(1,2,3,4) \leftrightarrow (1,4,3,2)$ \\ 
$\F(1^h,2,3,4^h)$ & $(1,2,3,4) \leftrightarrow (4,3,2,1)$   \\
$\Ft(1^h,2,3,4^h)$ & $(1,2,3,4) \leftrightarrow (4,3,2,1)$ and $(1,2,3,4) \leftrightarrow (1,3,2,4)$ \\
$\Ga(1^h,2,3,4^h)$ & $(1,2,3,4) \leftrightarrow (4,3,2,1)$ \\ 
 $\Gb(1^h,2,3,4^h)$ & None \\ 
 $\Gt(1^h,2,3,4^h)$ & $(1,2,3,4) \leftrightarrow (1,4,3,2)$\\ 
$\H(1^h,2,3,4^h)$ & $(1,2,3,4) \leftrightarrow (2,1,3,4)$ and $(1,2,3,4) \leftrightarrow (1,2,3,4)$ \\
& and $(1,2,3,4) \leftrightarrow (3,4,1,2)$ \\
\end{tabular}
\caption{Symmetries present for each $\X (1^h,2,3,4^h)$ antenna.}
\label{tab:X40sym}
\end{table}

\section{Outlook}
\label{sec:outlook}
We have proposed a general algorithm for the construction of real-radiation antenna functions directly from their desired unresolved limits.
The technique makes use of an iterative procedure to remove overlaps between different singular factors that are subsequently projected into the full phase space.
As the technique produces only denominators that match physical propagators, all antenna functions can straightforwardly be integrated analytically, which is a cornerstone of the antenna-subtraction method.

We have implemented our algorithm in an automated framework for the calculation of antenna functions and explicitly demonstrated that our technique can be used for single-real and double-real radiation antenna functions relevant to NLO and NNLO calculations, respectively.
In particular, we presented a full set of single-real and double-real tree-level antenna functions.
All of the new antenna functions we have presented here have been checked both analytically and numerically against the respective singular factors. 
As another strong validation, we have confirmed the correctness of their pole structure explicitly on the integrated level.
While the single-real antenna functions differ from previously used ones only in finite pieces, a residual dependence on the choice of single-real antennae is left in the construction of double-real antenna functions. 
This ambiguity is reflected in a difference in subleading poles starting from $1/\e^2$, while the deeper $1/\e^4$ and $1/\e^3$ poles correspond to the universal double-soft and triple-collinear behaviour. This is understood and poses no issues in application to the antenna-subtraction scheme, when used with the new single-real antenna functions.

Our work marks the first step towards a refined antenna-subtraction framework, as, for the first time, we have calculated a full set of double-real antenna functions that correspond to true antennae, meaning that they consist of exactly two hard radiators and contain no spurious singularities.
These features are vital for the subtraction scheme, as they eliminate the need for auxiliary subtraction terms to remove such spurious singularities.

Another avenue for future work might include the introduction of antenna functions with azimuthal correlations.  Here the main issue is matching the azimuthal correlations with the process-specific matrix elements under consideration.  For this reason, in this paper we removed the azimuthal correlations from the antennae. In principle the azimuthal terms can be directly removed from the matrix elements by using pairs (or multiple pairs) of correlated phase-space points.  Mass effects could also be included.  The unresolved limits are well known, however the integration of the antennae over the massive phase space is more involved.

To build a complete NNLO subtraction scheme, our work needs to be supplemented by an equivalent construction of one-loop single-real antenna functions that serve the purpose of real-virtual subtraction terms. This would require additional manipulation of explicit $\e$ poles and hypergeometric functions, alongside the single-real unresolved radiation. We leave this, as well as the description of a refined antenna-subtraction framework, for future work.

While we have focused on the application to single-real and double-real antenna functions, we wish to emphasise that the method can, in the future, be applied to build antenna functions for triple-real radiation as well, given that all singular factors with three unresolved particles are known~\cite{Catani:2019nqv,DelDuca:1999iql,DelDuca:2019ggv,DelDuca:2020vst,DelDuca:2022noh}.
This will form a substantial contribution to enable fully-differential calculations at N$^3$LO, where triple-real antenna functions can be used to subtract triple-unresolved singularities.

We believe that our new technique, as well as the antenna functions that we constructed here, will not only significantly simplify the antenna-subtraction framework, allowing eventually for a full automation, but also find their application in parton showers and their matching to NNLO calculations.

\acknowledgments
We thank Aude Gehrmann-De Ridder and Thomas Gehrmann for many fruitful discussions and help with the integrated antennae. We also thank Xuan Chen, Alexander Huss, Matteo Marcoli, and Giovanni Stagnitto for useful discussions.
CTP would like to thank John Campbell, Walter Giele, Stefan H{\"o}che, and Peter Skands for many interesting discussions.
NG thanks Albert Taormina and Maria Caleffi for their generous hospitality during the early stages of this work.
This research was supported in part by the UK Science and Technology Facilities Council under contract ST/T001011/1 and by the Swiss National Science Foundation under contract 200021-197130.

\appendix

\section{$\X$ Limits}
\label{app:X40limits}

In this appendix we list, for convenience, the limits of all $\X$ constructed using this paper's algorithm. If the result of an NNLO down projector is not given for a certain $\X$, then its result is $0$.

\subsection*{$\A (1_q^h,2_g,3_g,4^h_{\bar{q}} )$}
\begin{eqnarray} 
\PSdown_{23} A_4^0 (1^h,2,3,4^h) &=& \Sgg(1^h,2,3,4^h) \\
\PTCdown_{123} A_4^0 (1^h,2,3,4^h) &=& \Pqgg(1^h,2,3) \\
\PTCdown_{234} A_4^0 (1^h,2,3,4^h) &=& \Pqgg(4^h,3,2) \\
\PDCdown_{1234} A_4^0 (1^h,2,3,4^h) &=&  \Pqg(1^h,2) \Pqg(4^h,3)\\
\PSdown_{2} A_4^0 (1^h,2,3,4^h) &=& \frac{2 s_{13}}{s_{12} s_{23}} A_3^0(1,3,4)\\
\PSdown_{3} A_4^0 (1^h,2,3,4^h) &=& \frac{2 s_{24}}{s_{23} s_{34}} A_3^0(1,2,4)\\
\PCdown_{12} A_4^0 (1^h,2,3,4^h) &=&  \Pqg(1^h,2) A_3^0([1+2],3,4)\\
\PCdown_{23} A_4^0 (1^h,2,3,4^h) &=&  \Pgg(2,3) A_3^0(1,[2+3],4)\\
\PCdown_{34} A_4^0 (1^h,2,3,4^h) &=&  \Pqg(4^h,3) A_3^0(1,2,[3+4])\\
\PSCdown_{2;34} A_4^0 (1^h,2,3,4^h) &=& \frac{2 s_{134}}{s_{12} s_{234}}  \Pqg(4^h,3)\\
\PSCdown_{3;12} A_4^0 (1^h,2,3,4^h) &=& \frac{2 s_{124}}{s_{123} s_{34}}  \Pqg(1^h,2)
\end{eqnarray}

\subsection*{$\At (1_q^h,2_\gamma,3_\gamma,4_{\bar{q}}^h )$}
\begin{eqnarray}       
\PSdown_{23} \At (1^h,2,3,4^h) &=&  \Spp(1^h,2,3,4^h)\\
\PTCdown_{123} \At (1^h,2,3,4^h) &=& \Pqpp(1^h,2,3)\\
\PTCdown_{234} \At (1^h,2,3,4^h) &=& \Pqpp(4^h,3,2)\\
\PDCdown_{1234} \At (1^h,2,3,4^h) &=&  \Pqg(1^h,2) \Pqg(4^h,3)\\
\PDCdown_{1324} \At (1^h,2,3,4^h) &=&  \Pqg(1^h,3) \Pqg(4^,2)\\
\PSdown_{2} \At (1^h,2,3,4^h) &=& \frac{2 s_{14}}{s_{12} s_{24}} A_3^0(1,3,4)\\
\PSdown_{3} \At (1^h,2,3,4^h) &=& \frac{2 s_{14}}{s_{13} s_{34}} A_3^0(1,2,4)\\
\PCdown_{12} \At (1^h,2,3,4^h) &=&  \Pqg(1^h,2) A_3^0([1+2],3,4)\\
\PCdown_{13} \At (1^h,2,3,4^h) &=&  \Pqg(1^h,3) A_3^0([1+3],2,4)\\
\PCdown_{24} \At (1^h,2,3,4^h) &=&  \Pqg(4^h,2) A_3^0(1,3,[2+4])\\
\PCdown_{34} \At (1^h,2,3,4^h) &=&  \Pqg(4^h,3) A_3^0(1,2,[3+4])\\
\PSCdown_{2;13} \At (1^h,2,3,4^h) &=& \frac{2 s_{14}}{s_{12} s_{24}}  \Pqg(1^h,3)\\
\PSCdown_{2;34} \At (1^h,2,3,4^h) &=& \frac{2 s_{14}}{s_{12} s_{24}}  \Pqg(4^h,3)\\
\PSCdown_{3;12} \At (1^h,2,3,4^h) &=& \frac{2 s_{14}}{s_{13} s_{34}}  \Pqg(1^h,2)\\
\PSCdown_{3;24} \At (1^h,2,3,4^h) &=& \frac{2 s_{14}}{s_{13} s_{34}}  \Pqg(4^h,2)
\end{eqnarray}

\subsection*{$\B (1_q^h,2_{\bar{Q}},3_Q,4_{\bar{q}}^h)$}
\begin{eqnarray}
\PSdown_{23} \B (1^h,2,3,4^h) &=& \Sqq(1^h,2,3,4^h)\\
\PTCdown_{123} \B (1^h,2,3,4^h) &=& \PqQQ(1^h,2,3)\\
\PTCdown_{234} \B (1^h,2,3,4^h) &=& \PqQQ(4^h,3,2)\\
\PCdown_{23} \B (1^h,2,3,4^h) &=&  \Pqq(2,3) A_3^0(1,[2+3],4)
\end{eqnarray}

\subsection*{$\C (1_q^h,2_{\bar{q}},3_q,4_{\bar{q}}^h)$}
\begin{eqnarray}
\PTCdown_{234} \C (1^h,2,3,4^h) &=& \frac{1}{2} \Pqqq(2,3,4) 
\end{eqnarray}

\subsection*{$\D (1_q^h,2_g,3_g,4_g^h)$}
\begin{eqnarray}
\PSdown_{23} \D (1^h,2,3,4^h) &=& \Sgg(1^h,2,3,4^h)\\
\PTCdown_{123} \D (1^h,2,3,4^h) &=& \Pqgg(1^h,2,3)\\
\PTCdown_{234} \D (1^h,2,3,4^h) &=& \Pggg(4^h,3,2)\\
\PDCdown_{1234} \D (1^h,2,3,4^h) &=&  \Pqg(1^h,2) \Pgg(4^h,3)\\
\PSdown_{2} \D (1^h,2,3,4^h) &=&   \frac{2 s_{13}}{s_{12} s_{23}} D_3^0(1,3,4)\\
\PSdown_{3} \D (1^h,2,3,4^h) &=&   \frac{2 s_{24}}{s_{23} s_{34}} D_3^0(1,2,4)\\
\PCdown_{12} \D (1^h,2,3,4^h) &=&  \Pqg(1^h,2) D_3^0([1+2],3,4)\\
\PCdown_{23} \D (1^h,2,3,4^h) &=&  \Pgg(2,3) D_3^0(1,[2+3],4)\\ 
\PCdown_{34} \D (1^h,2,3,4^h) &=&  \Pgg(4^h,3) D_3^0(1,2,[3+4])\\
\PSCdown_{2;34} \D (1^h,2,3,4^h) &=& \frac{2 s_{134}}{s_{12} s_{234}}  \Pgg(4^h,3)\\
\PSCdown_{3;12} \D (1^h,2,3,4^h) &=& \frac{2 s_{124}}{s_{123} s_{34}}  \Pqg(1^h,2)
\end{eqnarray}

\subsection*{$\Dt (1_q^h,2_g,3_g,4_g^h)$}
\begin{eqnarray}  
\PSdown_{23} \Dt (1^h,2,3,4^h) &=& \Spp(1^h,2,3,4^h)\\
\PTCdown_{123} \Dt (1^h,2,3,4^h) &=& \Pqpp(1^h,2,3)\\
\PTCdown_{234} \Dt (1^h,2,3,4^h) &=& \Pggg(3,4^h,2)\\
\PDCdown_{1234} \Dt (1^h,2,3,4^h) &=&  \Pqg(1^h,2)  \Pgg(4^h,3)\\
\PDCdown_{1324} \Dt (1^h,2,3,4^h) &=&  \Pqg(1^h,2)  \Pgg(4^h,2)\\
\PSdown_{2} \Dt (1^h,2,3,4^h) &=&   \frac{2 s_{14}}{s_{12} s_{24}} D_3^0(1,3,4)\\
\PSdown_{3} \Dt (1^h,2,3,4^h) &=&   \frac{2 s_{14}}{s_{13} s_{34}} D_3^0(1,2,4)\\
\PCdown_{12} \Dt (1^h,2,3,4^h) &=&   \Pqg(1^h,2) D_3^0([1+2],3,4)\\
\PCdown_{13} \Dt (1^h,2,3,4^h) &=&   \Pqg(1^h,3) D_3^0([1+3],2,4)\\
\PCdown_{24} \Dt (1^h,2,3,4^h) &=&   \Pgg(4^h,2) D_3^0(1,3,[2+4])\\ 
\PCdown_{34} \Dt (1^h,2,3,4^h) &=&   \Pgg(4^h,3) D_3^0(1,2,[3+4])\\
\PSCdown_{2;34} \Dt (1^h,2,3,4^h) &=& \frac{2 s_{134}}{s_{12} s_{234}}  \Pgg(4^h,3)\\
\PSCdown_{3;24} \Dt (1^h,2,3,4^h) &=& \frac{2 s_{124}}{s_{13} s_{234}}  \Pgg(4^h,2)\\  
\PSCdown_{3;12} \Dt (1^h,2,3,4^h) &=& \frac{2 s_{124}}{s_{34} s_{123}}  \Pqg(1^h,2)\\
\PSCdown_{2;13} \Dt (1^h,2,3,4^h) &=& \frac{2 s_{134}}{s_{24} s_{123}}  \Pqg(1^h,3)
\end{eqnarray}

\subsection*{$\Ea (1_q^h,2_{\bar{Q}},3_Q,4_g^h)$}     
\begin{eqnarray}  
\PSdown_{23} \Ea (1^h,2,3,4^h) &=& \Sqq(1^h,2,3,4^h)\\
\PTCdown_{123} \Ea (1^h,2,3,4^h) &=& \PqQQ(1^h,2,3)\\
\PTCdown_{234} \Ea (1^h,2,3,4^h) &=& \Pgqbq(4^h,3,2)\\   
\PCdown_{23} \Ea (1^h,2,3,4^h) &=&  \Pqq(2,3) D_3^0(1,[2+3],4)
\end{eqnarray}
  
\subsection*{$\Eb (1_q^h,2_g,3_{\bar{Q}},4_Q^h)$}    
\begin{eqnarray}       
\PTCdown_{234} \Eb (1^h,2,3,4^h) &=& \Pgqbq(2,3,4^h)\\      
\PDCdown_{1234} \Eb (1^h,2,3,4^h) &=&  \Pqq(4^h,3) \Pqg(1^h,2)\\            
\PSdown_{2} \Eb (1^h,2,3,4^h) &=& \frac{2 s_{13}}{s_{12} s_{23}} E_3^0(1,3,4)\\       
\PCdown_{12} \Eb (1^h,2,3,4^h) &=&  \Pqg(1^h,2) E_3^0([1+2],3,4)\\
\PCdown_{23} \Eb (1^h,2,3,4^h) &=&  \Pqg(3,2) E_3^0(1,[2+3],4)\\
\PCdown_{34} \Eb (1^h,2,3,4^h) &=&  \Pqq(4^h,3) D_3^0(1,2,[3+4])\\  
\PSCdown_{2;34} \Eb (1^h,2,3,4^h) &=& \frac{2 s_{13}}{s_{12} s_{23}}  \Pqq(4^h,3)
\end{eqnarray}
         
\subsection*{$\Et (1_q^h,2_{\bar{Q}},3_g,4_Q^h)$} 
\begin{eqnarray}       
\PTCdown_{234} \Et (1^h,2,3,4^h) &=& \Ppqbq(4^h,3,2) \\     
\PSdown_{3} \Et (1^h,2,3,4^h) &=& \frac{2 s_{24}}{s_{23} s_{34}} E_3^0(1,2,4) \\
\PCdown_{23} \Et (1^h,2,3,4^h) &=&  \Pqg(2,3) E_3^0(1,[2+3],4) \\
\PCdown_{34} \Et (1^h,2,3,4^h) &=&  \Pqg(4^h,3) E_3^0(1,2,[3+4]) 
\end{eqnarray}    
         
\subsection*{$\F (1_g^h,2_g,3_g,4_g^h)$}      
\begin{eqnarray}  
\PSdown_{23} \F (1^h,2,3,4^h) &=& \Sgg(1^h,2,3,4^h)\\
\PTCdown_{123} \F (1^h,2,3,4^h) &=& \Pggg(1^h,2,3)\\
\PTCdown_{234} \F (1^h,2,3,4^h) &=& \Pggg(4^h,3,2)\\
\PDCdown_{1234} \F (1^h,2,3,4^h) &=&  \Pgg(1^h,2) \Pgg(4^h,3)\\  
\PSdown_{2} \F (1^h,2,3,4^h) &=& \frac{2 s_{13}}{s_{12} s_{23}} F_3^0(1,3,4)\\
\PSdown_{3} \F (1^h,2,3,4^h) &=& \frac{2 s_{24}}{s_{23} s_{34}} F_3^0(1,2,4)\\   
\PCdown_{12} \F (1^h,2,3,4^h) &=&  \Pgg(1^h,2) F_3^0([1+2],3,4)\\
\PCdown_{23} \F (1^h,2,3,4^h) &=&  \Pgg(2,3) F_3^0(1,[2+3],4)\\
\PCdown_{34} \F (1^h,2,3,4^h) &=&  \Pgg(4^h,3) F_3^0(1,2,[3+4])\\    
\PSCdown_{2;34} \F (1^h,2,3,4^h) &=& \frac{2 s_{134}}{s_{12} s_{234}}  \Pgg(4^h,3)\\
\PSCdown_{3;12} \F (1^h,2,3,4^h) &=& \frac{2 s_{124}}{s_{123} s_{34}}  \Pgg(1^h,2)
\end{eqnarray}
       
\subsection*{$\Ft (1_g^h,2_g,3_g,4_g^h)$}
\begin{eqnarray}  
\PSdown_{23} \Ft (1^h,2,3,4^h) &=& \Spp(1^h,2,3,4^h)\\ 
\PTCdown_{123} \Ft (1^h,2,3,4^h) &=& \Pggg(2,1^h,3)\\
\PTCdown_{234} \Ft (1^h,2,3,4^h) &=& \Pggg(3,4^h,2)\\
\PDCdown_{1234} \Ft (1^h,2,3,4^h) &=&  \Pgg(1^h,2)  \Pgg(4^h,3)\\
\PDCdown_{1324} \Ft (1^h,2,3,4^h) &=&  \Pgg(1^h,3)  \Pgg(4^h,2)\\
\PSdown_{2} \Ft (1^h,2,3,4^h) &=&   \frac{2 s_{14}}{s_{12} s_{24}} F_3^0(1,3,4)\\
\PSdown_{3} \Ft (1^h,2,3,4^h) &=&   \frac{2 s_{14}}{s_{13} s_{34}} F_3^0(1,2,4)\\ 
\PCdown_{12} \Ft (1^h,2,3,4^h) &=&   \Pgg(1^h,2) F_3^0([1+2],3,4)\\
\PCdown_{13} \Ft (1^h,2,3,4^h) &=&   \Pgg(1^h,3) F_3^0([1+3],2,4)\\
\PCdown_{24} \Ft (1^h,2,3,4^h) &=&   \Pgg(4^h,2) F_3^0(1,3,[2+4])\\ 
\PCdown_{34} \Ft (1^h,2,3,4^h) &=&   \Pgg(4^h,3) F_3^0(1,2,[3+4])\\
\PSCdown_{2;34} \Ft (1^h,2,3,4^h) &=& \frac{2 s_{134}}{s_{12} s_{234}}  \Pgg(4^h,3)\\
\PSCdown_{3;24} \Ft (1^h,2,3,4^h) &=& \frac{2 s_{124}}{s_{13} s_{234}}  \Pgg(4^h,2)\\
\PSCdown_{3;12} \Ft (1^h,2,3,4^h) &=& \frac{2 s_{124}}{s_{34} s_{123}}  \Pgg(1^h,2)\\
\PSCdown_{2;13} \Ft (1^h,2,3,4^h) &=& \frac{2 s_{134}}{s_{24} s_{123}}  \Pgg(1^h,3)
\end{eqnarray}

\subsection*{$\Ga (1_g^h,2_{\bar{Q}},3_Q,4_g^h)$} 
\begin{eqnarray}  
\PSdown_{23} \Ga (1^h,2,3,4^h) &=& \Sqq(1^h,2,3,4^h)\\
\PTCdown_{123} \Ga (1^h,2,3,4^h) &=& \Pgqbq(1^h,2,3)\\
\PTCdown_{234} \Ga (1^h,2,3,4^h) &=& \Pgqbq(4^h,3,2)\\
\PCdown_{23} \Ga (1^h,2,3,4^h) &=&  \Pqq(2,3) F_3^0(1,[2+3],4)
\end{eqnarray}
 
\subsection*{$\Gb (1_g^h,2_g,3_{\bar{Q}},4_Q^h)$}       
\begin{eqnarray}  
\PTCdown_{234} \Gb (1^h,2,3,4^h) &=& \Pgqbq(2,3,4^h)\\
\PDCdown_{1234} \Gb (1^h,2,3,4^h) &=&  \Pgg(1^h,2) \Pqq(4^h,3)\\ 
\PSdown_{2} \Gb (1^h,2,3,4^h) &=& \frac{2 s_{13}}{s_{12} s_{23}} G_3^0(1,3,4)\\ 
\PCdown_{12} \Gb (1^h,2,3,4^h) &=&  \Pgg(1^h,2) G_3^0([1+2],3,4)\\
\PCdown_{23} \Gb (1^h,2,3,4^h) &=&  \Pqg(3,2) G_3^0(1,[2+3],4)\\
\PCdown_{34} \Gb (1^h,2,3,4^h) &=&  \Pqq(4^h,3) F_3^0(1,2,[3+4])\\  
\PSCdown_{2;34} \Gb (1^h,2,3,4^h) &=& \frac{2 s_{13}}{s_{12} s_{23}}  \Pqq(4^h,3)
\end{eqnarray}

\subsection*{$\Gt (1_g^h,2_{\bar{Q}},3_g,4_Q^h)$}
\begin{eqnarray}
\PTCdown_{234} \Gt (1^h,2,3,4^h) &=& \Ppqbq(4^h,3,2) \\          
\PSdown_{3} \Gt (1^h,2,3,4^h) &=& \frac{2 s_{24}}{s_{23} s_{34}} G_3^0(1,2,4)\\    
\PCdown_{23} \Gt (1^h,2,3,4^h) &=&  \Pqg(2,3) G_3^0(1,[2+3],4) \\
\PCdown_{34} \Gt (1^h,2,3,4^h) &=&  \Pqg(4^h,3) G_3^0(1,2,[3+4]) 
\end{eqnarray}

\subsection*{$\H (1_{\bar{q}}^h,2_q,3_{\bar{Q}},4_Q^h)$}
\begin{eqnarray} 
\PDCdown_{1234} \H (1^h,2,3,4^h) &=&  \Pqq(1^h,2) \Pqq(4^h,3) \\
\PCdown_{12} \H (1^h,2,3,4^h) &=&  \Pqq(1^h,2) G_3^0([1+2],3,4) \\
\PCdown_{34} \H (1^h,2,3,4^h) &=&  \Pqq(4^h,3) G_3^0([3+4],1,2) 
\end{eqnarray}  

\section{Integrals of single-unresolved contributions}
\label{app:X30regionintegrations}
In this appendix we list the integration over the single-unresolved final-final phase space for the different contributions of the $X_3^0$ antennae.

The integration of the soft contribution $\Ssoft$ is simply the integral of the soft eikonal $\Sg$ and is the same for all antennae that contain a soft limit,
\begin{equation}
\begin{split}
\calSsoft(i^h,j_g,k^h)
&= (s_{ijk})^{-\e} \left(
\frac{1}{\e^2}
+\frac{2}{\e}
+6 - \frac{7}{12}\pi^2
+\e\left(18-\frac{25}{3}\zeta_3-\frac{7}{6}\pi^2\right) \right. \\
&\left. \hspace{2.2cm}
+\e^2\left(54-\frac{50}{3}\zeta_3-\frac{7}{2}\pi^2-\frac{71}{1440}\pi^4
\right) 
+{\cal O}(\e^3) 
\right) \, .
\end{split}
\end{equation}
The integrals of the three different collinear remainders $\Scol$ (with the soft contribution subtracted) are,
\begin{align}
\calScol(i_q^h,j_g;k^h)
&= (s_{ijk})^{-\e} \left(
-\frac{1}{4\e}
-\frac{5}{8}
+\e\left(-\frac{31}{16}+\frac{7}{48}\pi^2\right) \right.\\
& \left. \hspace{2.2cm}
+ \e^2\left(-
\frac{189}{32}+\frac{25}{12}\zeta_3+\frac{35}{96}\pi^2\right) 
+{\cal O}(\e^3) \right) \, , \nonumber\\
\calScol(i_g^h,j_g;k^h)
&= (s_{ijk})^{-\e} \left(
-\frac{1}{12\e}
-\frac{7}{24}
+\e\left(-\frac{15}{16}+\frac{7}{144}\pi^2\right) \right.\\
& \left. \hspace{2.2cm}
+ \e^2\left(-
\frac{93}{32}+\frac{25}{36}\zeta_3+\frac{49}{288}\pi^2\right) 
+{\cal O}(\e^3) \right) \, , \nonumber \\
\calScol(i_{Q}^h,j_{\Qb};k^h)
&=
\left(s_{ijk}\right)^{-\e}
\left( - \frac{1}{3\e}
 -\frac{3}{4}
+\left(-\frac{15}{8}+\frac{7\pi^2}{36}\right)\e
\right.  \\
& \left. \hspace{2.2cm}
+\left(-\frac{81}{16}+\frac{7\pi^2}{16}+\frac{25\zeta_3}{9} \right)\e^2
+ \order{\e^3} \right) \, . \nonumber
\end{align}

Note that in all of the above formulae arguments merely serve the purpose of identifying the particle content and the hard radiators. There is no dependence on the particle momenta in the integrated contributions besides the overall normalisation to the invariant mass $s_{ijk}$.

\section{Integrals of double-unresolved contributions}
\label{app:X40regionintegrations}
In this appendix we list the integration over the double-unresolved final-final phase space for the different contributions to the $X_4^0$ antennae. These contributions do not depend on the form chosen for the single-real antenna functions.

\begingroup
\allowdisplaybreaks
The integrals of the double-soft contribution $\Dsoft$ are simply given by the integration of the double-soft factors $\Sgg$, $\Spp$, and $\Sqq$,
\begin{align}
\calDsoft(i^h,j_g,k_g,l^h) &= 
\left(s_{ijkl} \right)^{-2\e}
\left[+\frac{3}{4\e^4}
+\frac{89}{24\e^3}
+\frac{1}{\e^2}\left(\frac{599}{36} - \pi^2\right)
\right.\nonumber\\
&\left.\hspace{2.2cm}
+ \frac{1}{\e}\left(\frac{7705}{108} - \frac{787}{144}\pi^2 - \frac{53}{4}\zeta_3\right)
\right. \\
& \left.\hspace{2.2cm}
+ \left(\frac{195547}{648} - \frac{2705}{108}\pi^2 - \frac{3371}{36}\zeta_3 + \frac{199}{480}\pi^4\right)
+ \order{\e}\right] \, , \nonumber\\
\calDsoft(i^h,j_\gamma,k_\gamma,l^h) &= 
\left(s_{ijkl}\right)^{-2\e}
\left[+\frac{1}{\e^4}
+\frac{4}{\e^3}
+\frac{1}{\e^2}\left(18 - \frac{3}{2}\pi^2\right)
+\frac{1}{\e}\left(76 - 6\pi^2 - \frac{74}{3}\zeta_3\right)
\right. \nonumber\\
& \left.\hspace{2.2cm}
+\left(312 - 27\pi^2 - \frac{308}{3}\zeta_3 + \frac{49}{120}\pi^4\right)
+ \order{\e}\right] \, , \\
\calDsoft(i^h,j_q,k_{\qb},l^h) &= 
\left(s_{ijkl}\right)^{-2\e}
\left[-\frac{1}{12\e^3}
-\frac{17}{36\e^2}
+\frac{1}{\e}\left(-\frac{277}{108} + \frac{11}{72}\pi^2\right)
\right. \nonumber\\
& \left.\hspace{2.2cm}
+\left(-\frac{4199}{324} + \frac{169}{216}\pi^2 + \frac{67}{18}\zeta_3\right)
+ \order{\e}\right] \, .
\end{align}

The integrals over the nine triple-collinear remainders (with the double-soft contribution subtracted) are,
\begin{align}
\calTcol(i_g^h,j_g,k_g;l^h) &= 
\left(s_{ijkl}\right)^{-2\e}
\left[-\frac{1}{4\e^3}
+\frac{1}{\e^2}\left(-\frac{499}{288} - \frac{1}{24}\pi^2\right)
+\frac{1}{\e}\left(-\frac{1757}{192} + \frac{5}{18}\pi^2 - 2\zeta_3\right)
\right. \nonumber\\
& \left.\hspace{2.2cm}
+\left(-\frac{440147}{10368} + \frac{1363}{576}\pi^2 + \frac{11}{12}\zeta_3 - \frac{1}{18}\pi^4\right)
+\order{\e}\right] \, , \\
\calTcol(i_g,j_g^h,k_g;l^h) &= \left( s_{ijkl} \right)^{-2\e}\left [-\frac{1}{6\e^3}-\frac{9}{16\e^2}+\frac{1}{\e} \left(\frac{413}{864}+\frac{1}{4}\pi^2-2\zeta_3\right) 
\right. \nonumber\\
& \left.\hspace{2.2cm}
+ \left(\frac{25565}{1728}+\frac{79}{96}\pi^2+\frac{4}{9}\zeta_3-\frac{1}{6}\pi^4\right) + \order{\e}\right] \, , \\
\calTcol(i_q^h,j_g,k_g;l^h) &=
\left(s_{ijkl}\right)^{-2\e} 
\left[-\frac{1}{2\e^3}
+\frac{1}{\e^2}\left(-\frac{187}{96} - \frac{1}{24}\pi^2\right)
+\frac{1}{\e}\left(-\frac{5185}{576} + \frac{37}{48}\pi^2 - \frac{9}{4}\zeta_3\right)
\right. \nonumber\\
& \left.\hspace{2.2cm}
+\left(-\frac{141871}{3456} + \frac{1685}{576}\pi^2 + \frac{347}{24}\zeta_3 - \frac{7}{90}\pi^4\right)
+\order{\e}\right] \, , \\
\calTcol(i_q^h,j_\gamma,k_\gamma;l^h) &=
\left(s_{ijkl}\right)^{-2\e}
\left[-\frac{1}{2\e^3}
-\frac{43}{16\e^2}
+\frac{1}{\e}\left(-\frac{377}{32} + \frac{3}{4}\pi^2 -\zeta_3\right)
\right. \nonumber\\
& \left.\hspace{2.2cm}
+\left(-\frac{3003}{64} + \frac{129}{32}\pi^2 + \frac{34}{3}\zeta_3 - \frac{1}{12}\pi^4\right)
+\order{\e}\right] \, , \\
\calTcol(i_q^h,j_g,k_{\qb};l^h) &= 
\left(s_{ijkl}\right)^{-2\e}
\left[-\frac{1}{6\e^3}
-\frac{35}{36\e^2}
+\frac{1}{\e}\left(-\frac{277}{54} + \frac{1}{4}\pi^2\right)
\right. \nonumber\\
& \left.\hspace{2.2cm}
+\left(-\frac{7967}{324} + \frac{35}{24}\pi^2 + \frac{40}{9}\zeta_3\right)
+\order{\e}\right]
\, , \\
\calTcol(i_g^h,j_q,k_{\qb};l^h) &= 
\left(s_{ijkl}\right)^{-2\e}
\left[+\frac{1}{72\e^2}
+\frac{1}{\e}\left(-\frac{23}{288} + \frac{1}{18}\pi^2\right)
\right. \nonumber\\
& \left.\hspace{2.2cm}
+\left(-\frac{15857}{5184} + \frac{23}{72}\pi^2+3\zeta_3\right)
+\order{\e}\right]
\, , \\
\calTcol(i_{\qb}^h,j_q,k_{g};l^h) &= 
\left(s_{ijkl}\right)^{-2\e}
\left[-\frac{1}{3\e^3}-\frac{9}{8\e^2}
+\frac{1}{\e} \left(-\frac{2927}{864}+\frac{1}{2}\pi^2\right)
\right. \nonumber\\
& \left.\hspace{2.2cm}
+ \left(-\frac{16127}{1728}+\frac{27}{16}\pi^2+\frac{80}{9}\zeta_3\right) 
+ \order{\e}\right]
\, , \\
\calTcol(i_q^h,j_{\Qb},k_{Q};l^h) &= 
\left(s_{ijkl}\right)^{-2\e}
\left[+\frac{1}{24\e^2}
+\frac{31}{144\e}
+\left(\frac{395}{432} - \frac{1}{18}\pi^2\right)
+\order{\e}\right] \, ,\\
\calTcol(i_q^h,j_{\qb},k_{q};l^h) &=
\left(s_{ijkl}\right)^{-2\e}
\left[+\frac{1}{\e}\left(-\frac{13}{32}
+\frac{1}{16}\pi^2
-\frac{1}{4}\zeta_3\right) 
\right. \nonumber\\
& \left.\hspace{2.2cm}
+\left(-\frac{73}{16}
+\frac{23}{96}\pi^2
+\frac{23}{8}\zeta_3
-\frac{1}{45}\pi^4\right)
+\order{\e}\right] \, .
\end{align}

The integrals over the nine double-collinear remainders (with the double-soft and triple-collinear contributions subtracted) are,
\begin{align}
\calDcol(i_q^h,j_{g};k_{g},l_{\qb}^h) &= 
\left( s_{ijkl} \right)^{-2\e}\left [-\frac{3}{16\e^2}-\frac{21}{16\e}+
\left(-\frac{55}{8}+\frac{9}{32}\pi^2\right) + \order{\e}\right],\\
\calDcol(i_q^h,j_{\gamma};k_{\gamma},l_{\qb}^h)  &=
\left( s_{ijkl} \right)^{-2\e}\left [+\frac{3}{16\e^2}+\frac{17}{16\e}+
\left(\frac{21}{4}-\frac{9}{32}\pi^2\right) + \order{\e}\right],\\
\calDcol(i_q^h,j_{g};k_{g},l_{g}^h) &= 
\left( s_{ijkl} \right)^{-2\e}\left
[+\frac{23}{72\e^2}+\frac{335}{144\e}+ \left(\frac{32573}{2592}-
\frac{23}{48}\pi^2\right) + \order{\e}\right],\\
\calDcol(i_q^h,j_{\tilde{g}};k_{\tilde{g}},l_{g}^h) &=
\left( s_{ijkl} \right)^{-2\e}\left
[+\frac{1}{8\e^2}+\frac{13}{16\e}+ \left(\frac{407}{96}-
\frac{3}{16}\pi^2\right) + \order{\e}\right],\\
\calDcol(i_q^h,j_{g};k_{\Qb},l_{Q}^h) &=
\left( s_{ijkl} \right)^{-2\e}\left [-\frac{1}{4\e}-\frac{95}{48} +
\order{\e}\right],\\
\calDcol(i_g^h,j_{g};k_{g},l_{g}^h)  &=
\left( s_{ijkl} \right)^{-2\e}\left
[+\frac{41}{48\e^2}+\frac{97}{16\e}+ \left(\frac{13997}{432}-
\frac{41}{32}\pi^2\right) + \order{\e}\right],\\
\calDcol(i_g^h,j_{\tilde{g}};k_{\tilde{g}},l_{g}^h)&=
\left( s_{ijkl} \right)^{-2\e}\left
[+\frac{13}{144\e^2}+\frac{95}{144\e}+ \left(\frac{4693}{1296}-
\frac{13}{96}\pi^2\right) + \order{\e}\right],\\
\calDcol(i_g^h,j_{g};k_{\Qb},l_{Q}^h) &= 
\left( s_{ijkl} \right)^{-2\e}\left [-\frac{1}{18\e^2}-
\frac{1}{2\e}+ \left(-\frac{3905}{1296}+\frac{1}{12}\pi^2\right) +
\order{\e}\right],\\
\calDcol(i_{\qb}^h,j_{q};k_{\Qb},l_{Q}^h) &=
\left( s_{ijkl} \right)^{-2\e}\left
[+\frac{1}{9\e^2}+\frac{13}{36\e}+ \left(\frac{139}{324}-
\frac{1}{6}\pi^2\right) + \order{\e}\right].
\end{align}

The remaining single-unresolved terms, $\Ssoft$ and $\Scol$, are not universal, since they depend on the choice of single-real antenna function. We do not list the integrals of these contributions.

Note that in all of the above formulae arguments merely serve the purpose of identifying the particle content and the hard radiators. There is no dependence on the particle momenta in the integrated contributions besides the overall normalisation to the invariant mass.
\endgroup

\section{Overlap of double-soft and triple-collinear contributions}
\label{app:TCprojectionsofDS}
The projections into the triple-collinear phase space for each of the three double-soft factors are given by,
\begin{align}
\PTCdown_{ijk} \Sgg(i^h,j,k,l^h) &= \frac{2 \ome W_{jk}}{\sjk^2\sijk^2\omxi^2}
+\frac{4\xi\xj\xk}{\sjk\sijk\omxi^3}
+\frac{2\xi^2}{\sij\sijk\xk\omxi} \\
& \qquad
+\frac{2\xi}{\sjk\sijk\xk}
-\frac{8\xi}{\sjk\sijk\omxi}
+\frac{2\xi}{\sij\sjk\xk}
+\frac{2\xi}{\sij\sjk\omxi} \, , \nonumber \\
\PTCdown_{ijk} \Spp(i^h,j,k,l^h) &= \frac{4\sjk\xi^2}{\sij\sik\sijk\xj\xk}
+\frac{4\xi^2}{\sij\sijk\xj\xk}
+\frac{4\xi^2}{\sik\sijk\xj\xk} \, , \\
\PTCdown_{ijk} \Sqq(i^h,j,k,l^h) &= -\frac{2 
W_{jk}}{\sjk^2\sijk^2\omxi^2}
-\frac{4\xi\xj\xk}{\sjk\sijk\omxi^3\ome} \\
& \qquad
+\frac{2\xi}{\sjk\sijk\omxi} \, , \nonumber
\end{align}
where 
\begin{align}
    W_{ij} &= (x_i s_{jk} - x_j s_{ik})^2 - \frac{2}{\ome}\frac{x_i x_j x_k}{(1-x_k)}s_{ij} s_{ijk} \, . \label{eq:Wdef} 
\end{align}
\section{Triple-collinear splitting functions}
\label{app:Pabc}
In this appendix, we list the complete set of triple-collinear splitting functions for the specific case where one of the particles is hard using the decomposition formulae given in Ref.~\cite{paper1}.

\subsection{Three collinear gluons}
For three collinear gluons, the triple-collinear splitting functions with one hard radiator are given by
\begin{align}
\Pggg(i^h,j,k) &= 
\frac{\PggS(x_k)}{s_{ijk}}  
\frac{\PggS\left(\frac{x_j}{1-x_k} \right)}{s_{ij}} +
\frac{\PggS(1-x_i)}{s_{ijk}}  
\frac{\Pgg\left(\frac{x_j}{1-x_i} \right)}{s_{jk}} \nonumber \\
& \qquad + 
\frac{1}{s_{ijk}^2} \Rgggsub(i,j,k) ,\\
\Pggg(i,j^h,k) &= 
\frac{\PggS(x_k)}{s_{ijk}}  
\frac{\PggS\left(\frac{x_i}{1-x_k} \right)}{s_{ij}}
+
\frac{\PggS(x_i)}{s_{ijk}}  
\frac{\PggS\left(\frac{x_k}{1-x_i} \right)}{s_{jk}} ,\\
\Pggg(i,j,k^h) &= \Pggg(k^h,j,i),
\end{align}
where
\begin{equation}
\begin{split}
\Rgggsub(i,j,k) &= \frac{2 \ome W_{jk}}{(1-x_i)^2 s_{jk}^2} 
+ \frac{4 \ome x_k}{(1-x_i)^2} \frac{ \Tr{i}{j}{k}{\ell} }{ s_{jk} } \\
& \qquad + \fa_0(x_i,x_j,x_k) + \fa(x_i,x_j,x_k) \frac{s_{ijk}\Tr{i}{j}{k}{\ell}}{s_{ij} s_{jk} } \, ,
\end{split}
\label{eq:Rggg}
\end{equation}
and
\begin{eqnarray}
\label{eq:fa0}
\fa_0(x_i,x_j,x_k) &=& \ome B_0(x_k,x_i)  ,\\
\label{eq:fa}
\fa(x_i,x_j,x_k) &=& -\frac{x_k \Pgg(x_k)}{x_j(1-x_i)} 
- \frac{\Pgg(x_j)}{x_k} + \frac{2}{x_j(1-x_k)}  -1 - \frac{1}{(1-x_i)(1-x_k)} .
\end{eqnarray}
We have also used that 
\begin{align}
    W_{ij} &= (x_i s_{jk} - x_j s_{ik})^2 - \frac{2}{\ome}\frac{x_i x_j x_k}{(1-x_k)}s_{ij} s_{ijk} \, , \label{eq:Wdef} \\
    \Tr{i}{j}{k}{\ell} &= x_k s_{ij} - x_j s_{ik} + x_i s_{jk} \, ,    
\end{align}
and
\begin{align}
    A_0(x,y) &= 1 - \frac{(1-x)}{(1-y)} \, , \label{eq:A0def}\\
    B_0(x,y) &= 1 + \frac{2x(x-2)}{(1-y)^2} + \frac{4x}{(1-y)} \, . \label{eq:B0def}
\end{align}

\subsection{Two gluons with a collinear quark or antiquark}

In the case where gluon $j$ is colour-connected to quark $i$ and gluon $k$, we have,
\begin{align}
\Pqgg(i^h,j,k) &= \frac{\Pqg(x_k)}{s_{ijk}} \frac{\Pqg\left(\frac{x_j}{1-x_k}\right)}{s_{ij}}
+ \frac{\Pqg(1-x_i)}{s_{ijk}}   \frac{\Pgg\left( \frac{x_j}{1-x_i} \right)}{s_{jk}} 
\nonumber \\
& \qquad + \frac{1}{s_{ijk}^2} \Rqgg(i,j,k) \, , \\
\Pqgg(i,j^h,k) &= 0 \, , \\
\Pqgg(i,j,k^h) &= 0 \, , 
\label{eq:Pqgg}
\end{align}
where
\begin{equation}
\begin{split}
\Rqgg (i,j,k) &= \frac{2 \ome}{(1-x_i)^2}  \frac{W_{jk}}{s_{jk}^2} 
+ \frac{4 \ome x_k }{(1-x_i)^{2}} \frac{\Tr{i}{j}{k}{\ell}}{s_{jk}} + \frac{\ome^2}{(1-x_k)}  \frac{\Tr{i}{j}{k}{\ell}}{s_{ij}} \\
& \qquad + \fb_0 (x_i,x_j,x_k) +  \fb(x_i,x_j,x_k) \frac{s_{ijk} \Tr{i}{j}{k}{\ell}}{s_{ij}s_{jk}} \, ,
\end{split}
\label{eq:Rqgg}
\end{equation}
and
\begin{align}
\fb_0 (x_i,x_j,x_k) &= \ome \left(B_0(x_k,x_i) -1 + \ome A_0(x_i,x_k)\right) \, , \label{eq:fb} \\
\fb(x_i,x_j,x_k) &=- \frac{x_j \Pqg(x_j)}{x_k (1-x_i)} - \frac{2 x_k\Pqg(x_k)}{x_j (1-x_i)} + \frac{4}{(1-x_i)} - 3 \ome \, .
\label{eq:fb0}
\end{align}

In the case where the gluons are abelianised ($\tilde{g}$) or two photons are collinear to the quark, then the splitting function is symmetric under the exchange of the two bosons ($j,k$).  We have,
\begin{align}
\Pqpp(i^h,j,k) &=
		\frac{\Pqg(x_k)}{s_{ijk}}  
		\frac{\Pqg\left(\frac{x_j}{1-x_k}\right)}{s_{ij}} 
        + \frac{\Pqg(x_j)}{s_{ijk}}  
		\frac{\Pqg\left(\frac{x_k}{1-x_j}\right)}{s_{ik}} \nonumber \\
        &\qquad	+ \frac{1}{s_{ijk}^2} \Rqpp (i,j,k) \, , \\ \label{app:Pqpp}
\Pqpp(i,j^h,k) &= 0 \, , \\
\Pqpp(i,j,k^h) &= 0 \, ,
\label{eq:Pqpp}
\end{align}
where
\begin{equation}
\begin{split}
    \Rqpp (i,j,k) &=
    - \frac{\ome^2}{(1-x_k)} \frac{\Tr{j}{i}{k}{\ell}}{s_{ij}} 
    - \frac{\ome^2}{(1-x_j)} \frac{\Tr{j}{i}{k}{\ell}}{s_{ik}} \\
    & \qquad + \fc_0(x_i,x_j,x_k) 
    + \fc(x_i,x_j,x_k) 
	\frac{s_{ijk} \Tr{j}{i}{k}{\ell}}{s_{ij}s_{ik}} \, ,
\end{split}
\label{eq:Rqpp}
\end{equation}
and
\begin{eqnarray}
\label{eq:fc0}
    \fc_0(x_i,x_j,x_k) &=& \ome\left(2 - \ome A_0(x_j,x_k) -\ome A_0(x_k,x_j)\right), \\
\label{eq:fc}
	\fc(x_i,x_j,x_k) &=& - \frac{x_j \Pqg(x_j)}{x_k (1-x_i)}  - \frac{x_k \Pqg(x_k)}{x_j (1-x_i)} + \frac{4}{(1-x_i)} - 4 \ome + \ome^2.
\end{eqnarray}

\subsection{Quark-antiquark pair with a collinear gluon}
There are also two distinct splitting functions representing the clustering of a gluon with a quark-antiquark pair into a parent gluon. 

When the gluon is colour-connected to the antiquark, we have, 
\begin{align}
\Pgqbq(i^h,j,k) &=
\frac{\PggS(1-x_i)}{s_{ijk}}
\frac{\Pqq\left(\frac{x_k}{1-x_i}\right)}{s_{jk}}
+ \frac{1}{s_{ijk}^2} \Rgqbq(i,j,k) \, , \\
\Pgqbq(i,j^h,k) &= 0 \, , \\
\Pgqbq(i,j,k^h) &= \frac{\Pqq(x_k)}{s_{ijk}} 
\frac{\Pqg\left(\frac{x_i}{1-x_k}\right)}{s_{ij}} 
+ \frac{\PggS(x_i)}{s_{ijk}} 
\frac{\Pqq\left(\frac{x_k}{1-x_i}\right)}{s_{jk}} \, ,
\label{eq:Pgqbq}
\end{align}
where
\begin{equation}
\begin{split}
\Rgqbq(i,j,k) &= 
- \frac{2}{(1-x_i)^2} \frac{W_{jk}}{s_{jk}^2} 
- \frac{\ome}{(1-x_k)} \frac{\Tr{i}{j}{k}{\ell}}{s_{ij}} 
- \frac{4 x_k}{(1-x_i)^2} \frac{ \Tr{i}{j}{k}{\ell}}{s_{jk}} \\
& \qquad
+ \fd_0 (x_i,x_j,x_k) 
+ \fd (x_i,x_j,x_k) \frac{s_{ijk} \Tr{i}{j}{k}{\ell}}{s_{ij}s_{jk}} \, ,
\end{split}
\label{eq:Rgqbq}
\end{equation}
and
\begin{align}
\label{eq:fd0}
    \fd_0 (x_i,x_j,x_k) &= - B_0(x_k,x_i) + 1 - \ome A_0(x_i,x_k) 
    \, , \\
\label{eq:fd}    
    \fd (x_i,x_j,x_k) &=
    - \frac{\Pqq (x_k)}{x_i (1-x_i)} + \frac{2}{(1-x_i)} + 1 - 2 x_i + \frac{2( x_j -x_k - 2x_j x_k)}{\ome (1-x_i)} \, .
\end{align}

The QED-like splitting, where the gluon, quark and antiquark form a photon-like colour singlet is given by,
\begin{align}
\Ppqbq(i^h,j,k) &=
\frac{\Pqq(1-x_k)}{s_{ijk}} 
\frac{\Pqg\left(\frac{x_j}{1-x_k}\right)}{s_{ij}} 
+
\frac{\Pqq(1-x_i)}{s_{ijk}} 
\frac{\Pqg\left(\frac{x_j}{1-x_i}\right)}{s_{jk}} \nonumber \\ 
& \qquad + \frac{1}{s_{ijk}^2} \Rpqbq(i,j,k) \, , \\
\Ppqbq(i,j^h,k) &= 0 \, , \\
\Ppqbq(i,j,k^h) &= \frac{\Pqq(1-x_k)}{s_{ijk}} 
\frac{\Pqg\left(\frac{x_j}{1-x_k}\right)}{s_{ij}} 
+
\frac{\Pqq(1-x_i)}{s_{ijk}} 
\frac{\Pqg\left(\frac{x_j}{1-x_i}\right)}{s_{jk}} \nonumber \\ 
& \qquad + \frac{1}{s_{ijk}^2} \Rpqbq(i,j,k) \, , 
\label{eq:Ppqbq}
\end{align}
where
\begin{equation}
\begin{split}
\Rpqbq(i,j,k) &=
 \frac{\ome}{(1-x_k)} \frac{\Tr{i}{j}{k}{\ell}}{s_{ij}} 
+ \frac{\ome}{(1-x_i)} \frac{\Tr{i}{j}{k}{\ell}}{s_{jk}} \\
& \qquad
+ \fe_0 (x_i,x_j,x_k) 
+ \fe (x_i,x_j,x_k) \frac{s_{ijk} \Tr{i}{j}{k}{\ell}}{s_{ij}s_{jk}} \, , 
\end{split}
\label{eq:Rpqbq}
\end{equation}
and
\begin{align}
    \fe_0 (x_i,x_j,x_k) &=
    -2 + \ome A_0(x_i,x_k) + \ome A_0(x_k,x_i) \, , \label{eq:fe0} \\
    \fe (x_i,x_j,x_k) &= 
    -\frac{\Pqq(x_i)}{x_j}-\frac{\Pqq(x_k)}{x_j} + \frac{2 \epsilon}{\ome} x_j \, . \label{eq:fe}
\end{align}

\subsection{Quark-antiquark pair  with a collinear quark or antiquark}
Finally, we consider the clustering of a quark-antiquark pair ($Q\bar{Q}$) and a quark $q$ to form a parent quark with the same flavour as q. There are two splitting functions, one where the quark flavours are different and one where the quarks have the same flavour.  

For distinct quarks, we have
\begin{align}
\PqQQ(i^h,j,k) &=
\frac{\Pqg(1-x_i)}{s_{ijk}} \frac{\Pqq\left( \frac{x_j}{1-x_i} \right)}{s_{jk}} 
+ \frac{1}{s_{ijk}^2} \RqQQ(i,j,k) \, , \\
\PqQQ(i,j^h,k) &= 0 \, ,\\
\PqQQ(i,j,k^h) &= 0 \, ,
\label{eq:PqQQ}
\end{align}
where
\begin{equation}
\begin{split}
\RqQQ(i,j,k) &=  - \frac{2}{\omxi^2} \frac{W_{jk}}{s_{jk}^2} 
 - \frac{2 x_k}{\omxi^2} \frac{\Tr{i}{j}{k}{\ell}}{ s_{jk}} - \frac{2 x_j}{\omxi^2} \frac{\Tr{i}{k}{j}{\ell}}{ s_{jk}}  \\
&\qquad + \ff_0 (x_i,x_j,x_k) \, , 
\end{split}
\label{eq:RqQQ}
\end{equation}
and
\begin{equation}
\label{eq:ff0}
\ff_0 (x_i,x_j,x_k) = - \frac{1}{2} ( B_0(x_j,x_i) + B_0(x_k,x_i)) + 1 +\epsilon \, .
\end{equation}

Finally, for identical quarks, we have 
\begin{align}
\Pqqq(i^h,j,k) &= \frac{1}{s_{ijk}^2} \Rqqq(i,j,k) \, , \\
\Pqqq(i,j^h,k) &= 0 \, , \\
\Pqqq(i,j,k^h) &= \frac{1}{s_{ijk}^2} \Rqqq(i,j,k) \, , 
\label{eq:Pqqq}
\end{align}
where
\begin{equation}
\begin{split}
\Rqqq(i,j,k) &=
-  \frac{2\ome}{\omxi} \frac{\Tr{i}{j}{k}{\ell}}{s_{jk}} 
-  \frac{2\ome}{\omxk} \frac{\Tr{i}{j}{k}{\ell}}{s_{ij}}  \\
& \qquad + \fg_0(x_i,x_j,x_k) + \fg(x_i,x_j,x_k) 
\frac{s_{ijk}\Tr{i}{j}{k}{\ell}}{s_{ij} s_{jk}} \, ,
\end{split}
\label{eq:Rqqq}
\end{equation}
and
\begin{align}
\fg_0 (x_i,x_j,x_k) &=
 -2 \ome (\epsilon + A_0(x_i,x_k) + A_0(x_k,x_i)) \, , \label{eq:fg0} \\
\fg(x_i,x_j,x_k) &=
-\frac{2x_j}{\omxi\omxk}
+\ome \left(\frac{\omxk}{\omxi} + \frac{\omxi}{\omxk}
+2+\epsilon \right) \, . \label{eq:fg}
\end{align}

\section{Integrals of $\X$ antennae derived using the $X_3^0$ of Ref.~\cite{Gehrmann-DeRidder:2005btv}}
\label{app:intX4oldX3}
\begingroup
\allowdisplaybreaks
In this appendix, we list the integrals over the antenna phase space of the $\X$ antennae constructed using the $X_3^0$ antennae of Ref.~\cite{Gehrmann-DeRidder:2005btv}:
\begin{eqnarray}
\label{eq:A40intoldX3}
\calA (s_{ijkl}) &=& \left( s_{ijkl} \right)^{-2\e}\Biggl [
+\frac{3}{4\e^4}
+\frac{65}{24\e^3}
+\frac{1}{\e^2} \left(
\frac{217}{18}
-\frac{13}{12}\pi^2
\right)
+\frac{1}{\e} \left(
\frac{43223}{864}
-\frac{589}{144}\pi^2
-\frac{71}{4}\zeta_3
\right)
\nonumber \\&& \hspace{2cm}
 + \left(
\frac{1094807}{5184}
-\frac{8117}{432}\pi^2
-\frac{1327}{18}\zeta_3
+\frac{373}{1440}\pi^4
\right)
 + \order{\e}\Biggr], \\
\label{eq:A40tintoldX3}
\calAt (s_{ijkl}) &=& \left( s_{ijkl} \right)^{-2\e}\Biggl [
+\frac{1}{\e^4}
+\frac{3}{\e^3}
+\frac{1}{\e^2} \left(
13
-\frac{3}{2}\pi^2
\right)
+\frac{1}{\e} \left(
\frac{845}{16}
-\frac{9}{2}\pi^2
-\frac{80}{3}\zeta_3
\right)
\nonumber \\&& \hspace{2cm}
 + \left(
\frac{6865}{32}
-\frac{39}{2}\pi^2
-80\zeta_3
+\frac{29}{120}\pi^4
\right)
 + \order{\e}\Biggr], \\
\label{eq:B40intoldX3}
\calB (s_{ijkl}) &=& \left( s_{ijkl} \right)^{-2\e}\Biggl [
-\frac{1}{12\e^3}
-\frac{7}{18\e^2}
+\frac{1}{\e} \left(
-\frac{407}{216}
+\frac{11}{72}\pi^2
\right)
\nonumber \\&& \hspace{2cm}
+ \left(
-\frac{5809}{648}
+\frac{145}{216}\pi^2
+\frac{67}{18}\zeta_3
\right)
 + \order{\e}\Biggr], \\
\label{eq:C40intoldX3}
\calC (s_{ijkl}) &=& \left( s_{ijkl} \right)^{-2\e}\Biggl [
+\frac{1}{\e} \left(
-\frac{13}{32}
+\frac{1}{16}\pi^2
-\frac{1}{4}\zeta_3
\right)
\nonumber \\&& \hspace{2cm}
 + \left(
-\frac{73}{16}
+\frac{23}{96}\pi^2
+\frac{23}{8}\zeta_3
-\frac{1}{45}\pi^4
\right)
 + \order{\e}\Biggr], \\
\label{eq:D40intoldX3}
\calD (s_{ijkl}) &=& \left( s_{ijkl} \right)^{-2\e}\Biggl [
+\frac{3}{4\e^4}
+\frac{71}{24\e^3}
+\frac{1}{\e^2} \left(
\frac{257}{18}
-\frac{13}{12}\pi^2
\right)
+\frac{1}{\e} \left(
\frac{13661}{216}
-\frac{35}{8}\pi^2
-\frac{35}{2}\zeta_3
\right)
\nonumber \\&& \hspace{2cm}
 + \left(
\frac{22286}{81}
-\frac{9335}{432}\pi^2
-\frac{5473}{72}\zeta_3
+\frac{9}{32}\pi^4
\right)
 + \order{\e}\Biggr], \\
\label{eq:D40tintoldX3}
\calDt (s_{ijkl}) &=& \left( s_{ijkl} \right)^{-2\e}\Biggl [
+\frac{1}{\e^4}
+\frac{10}{3\e^3}
+\frac{1}{\e^2} \left(
\frac{47}{3}
-\frac{3}{2}\pi^2
\right)
+\frac{1}{\e} \left(
\frac{30313}{432}
-5\pi^2
-\frac{83}{3}\zeta_3
\right)
\nonumber \\&& \hspace{2cm}
 + \left(
\frac{132451}{432}
-\frac{1129}{48}\pi^2
-\frac{818}{9}\zeta_3
+\frac{19}{120}\pi^4
\right)
 + \order{\e}\Biggr], \\
\label{eq:E40aintoldX3}
\calEa (s_{ijkl}) &=& \left( s_{ijkl} \right)^{-2\e}\Biggl [
-\frac{1}{12\e^3}
-\frac{5}{12\e^2}
+\frac{1}{\e} \left(
-\frac{1631}{864}
+\frac{1}{8}\pi^2
\right)
\nonumber \\&& \hspace{2cm}
+ \left(
-\frac{46315}{5184}
+\frac{77}{108}\pi^2
+\frac{20}{9}\zeta_3
\right)
 + \order{\e}\Biggr], \\
\label{eq:E40bintoldX3}
\calEb (s_{ijkl}) &=& \left( s_{ijkl} \right)^{-2\e}\Biggl [
-\frac{1}{3\e^3}
-\frac{41}{24\e^2}
+\frac{1}{\e} \left(
-\frac{7325}{864}
+\frac{1}{2}\pi^2
\right)
\nonumber \\&& \hspace{2cm}
+ \left(
-\frac{22745}{576}
+\frac{41}{16}\pi^2
+\frac{80}{9}\zeta_3
\right)
 + \order{\e}\Biggr], \\
\label{eq:E40tintoldX3}
\calEt (s_{ijkl}) &=& \left( s_{ijkl} \right)^{-2\e}\Biggl [
-\frac{1}{6\e^3}
-\frac{35}{36\e^2}
+\frac{1}{\e} \left(
-\frac{1045}{216}
+\frac{1}{4}\pi^2
\right)
\nonumber \\&& \hspace{2cm}
+ \left(
-\frac{28529}{1296}
+\frac{35}{24}\pi^2
+\frac{40}{9}\zeta_3
\right)
 + \order{\e}\Biggr], \\
\label{eq:F40intoldX3}
\calF (s_{ijkl}) &=& \left( s_{ijkl} \right)^{-2\e}\Biggl [
+\frac{3}{4\e^4}
+\frac{77}{24\e^3}
+\frac{1}{\e^2} \left(
\frac{559}{36}
-\frac{13}{12}\pi^2
\right)
+\frac{1}{\e} \left(
\frac{59249}{864}
-\frac{671}{144}\pi^2
-\frac{69}{4}\zeta_3
\right)
\nonumber \\&& \hspace{2cm}
 + \left(
\frac{508343}{1728}
-\frac{9923}{432}\pi^2
-\frac{2819}{36}\zeta_3
+\frac{437}{1440}\pi^4
\right)
 + \order{\e}\Biggr], \\
\label{eq:F40tintoldX3}
\calFt (s_{ijkl}) &=& \left( s_{ijkl} \right)^{-2\e}\Biggl [
+\frac{1}{\e^4}
+\frac{11}{3\e^3}
+\frac{1}{\e^2} \left(
\frac{313}{18}
-\frac{3}{2}\pi^2
\right)
+\frac{1}{\e} \left(
\frac{34571}{432}
-\frac{11}{2}\pi^2
-\frac{86}{3}\zeta_3
\right)
\nonumber \\&& \hspace{2cm}
 + \left(
\frac{924559}{2592}
-\frac{209}{8}\pi^2
-\frac{916}{9}\zeta_3
+\frac{3}{40}\pi^4
\right)
 + \order{\e}\Biggr], \\
\label{eq:G40aintoldX3}
\calGa (s_{ijkl}) &=& \left( s_{ijkl} \right)^{-2\e}\Biggl [
-\frac{1}{12\e^3}
-\frac{4}{9\e^2}
+\frac{1}{\e} \left(
-\frac{745}{432}
+\frac{7}{72}\pi^2
\right)
\nonumber \\&& \hspace{2cm}
+ \left(
-\frac{6431}{864}
+\frac{163}{216}\pi^2
+\frac{13}{18}\zeta_3
\right)
 + \order{\e}\Biggr], \\
\label{eq:G40bintoldX3}
\calGb (s_{ijkl}) &=& \left( s_{ijkl} \right)^{-2\e}\Biggl [
-\frac{1}{3\e^3}
-\frac{139}{72\e^2}
+\frac{1}{\e} \left(
-\frac{8669}{864}
+\frac{1}{2}\pi^2
\right)
\nonumber \\&& \hspace{2cm}
+ \left(
-\frac{248495}{5184}
+\frac{139}{48}\pi^2
+\frac{80}{9}\zeta_3
\right)
 + \order{\e}\Biggr], \\
\label{eq:G40tintoldX3}
\calGt (s_{ijkl}) &=& \left( s_{ijkl} \right)^{-2\e}\Biggl [
-\frac{1}{6\e^3}
-\frac{41}{36\e^2}
+\frac{1}{\e} \left(
-\frac{1327}{216}
+\frac{1}{4}\pi^2
\right)
\nonumber \\&& \hspace{2cm}
+ \left(
-\frac{38291}{1296}
+\frac{41}{24}\pi^2
+\frac{40}{9}\zeta_3
\right)
 + \order{\e}\Biggr], \\
\label{eq:H40intoldX3}
\calH (s_{ijkl}) &=& \left( s_{ijkl} \right)^{-2\e}\Biggl [
+\frac{1}{9\e^2}
+\frac{7}{9\e}
+ \left(
\frac{1345}{324}
-\frac{1}{6}\pi^2
\right)
 + \order{\e}\Biggr].
\end{eqnarray}
In all cases, we find agreement with the analogous integrated antenna given in Ref.~\cite{Gehrmann-DeRidder:2005btv} through to $\order{\e^0}$.
\endgroup

\bibliographystyle{jhep}
\bibliography{bib2}{}
\end{document}